\documentclass[12pt]{article}
\usepackage{multirow}
\usepackage{booktabs}
\usepackage{mathrsfs}
\usepackage{fullpage}
\usepackage{amsmath}
\usepackage{graphicx}%
\usepackage{amsfonts}%
\usepackage{amssymb}
\usepackage{afterpage}
\usepackage{url}
\usepackage{fancyhdr}
\usepackage{isomath}
\usepackage{amsmath}
\usepackage{amsbsy}
\usepackage{amssymb}
\usepackage{amscd}
\usepackage{amsfonts}
\usepackage{graphicx}

\usepackage{verbatim}
\usepackage{euscript}
\usepackage{alltt}
\usepackage{stmaryrd}
\usepackage{float}
\usepackage[titletoc,title]{appendix}
\usepackage{caption}
\usepackage[english]{babel}
\usepackage{enumitem}
\usepackage{subcaption} 
\usepackage{layout}
\usepackage{longtable}
\usepackage{tikz}

\input{temp-definitions.inp}

\usepackage[margin=1in]{geometry}
\usepackage{titlesec}

\titleformat{\paragraph}
{\normalfont\normalsize\bfseries}{\theparagraph}{1em}{}
\titlespacing*{\paragraph}
{0pt}{3.25ex plus 1ex minus .2ex}{1.5ex plus .2ex}

\PassOptionsToPackage{normalem}{ulem}
\usepackage{ulem}

\begin{document}
	
\pagenumbering{roman}

\title{A unif\mbox{}icat\mbox{}ion of f\mbox{}inite deformat\mbox{}ion $J_2$ Von-Mises plasticity and quantitative dislocation mechanics}

\author{Rajat Arora\thanks{Dept.~of Civil \& Environmental Engineering, Carnegie Mellon University, Pittsburgh, PA, 15213.  Currently:  Ansys, Inc. rajat.arora9464@gmail.com.}    \qquad Amit Acharya\thanks{Dept.~of Civil \& Environmental Engineering, and Center for Nonlinear Analysis, Carnegie Mellon University, Pittsburgh, PA, 15213. acharyaamit@cmu.edu.}
}

\date{}

\maketitle

\begin{abstract}

\noindent We present a framework which unifies classical phenomenological $J_2$ and crystal plasticity theories with quantitative dislocation mechanics. The theory allows the computation of stress fields of arbitrary dislocation distributions and, coupled with minimally modified classical ($J_2$ and crystal plasticity)  models for the plastic strain rate of statistical dislocations, results in a versatile model of finite deformation mesoscale plasticity. We demonstrate some capabilities of  the framework by solving two outstanding challenge problems in mesoscale plasticity: 1) recover the experimentally observed power-law scaling of stress-strain behavior in constrained simple shear of thin metallic films inferred from micropillar experiments which all strain gradient plasticity models overestimate and fail to predict; 2) predict the finite deformation stress and energy density fields of a sequence of dislocation distributions representing a progressively dense dislocation wall in a finite body, as might arise in the process of polygonization when viewed macroscopically, with one consequence being the demonstration of the inapplicability of current mathematical results based on $\mathrm{\Gamma}$-convergence for this physically relevant situation. Our calculations in this case expose a possible `phase transition' - like behavior for further theoretical study.  We also provide a  quantitative solution to the fundamental question of the volume change induced by dislocations in a finite deformation theory, as well as show the massive non-uniqueness in the solution for the (inverse) deformation map of a body inherent in a model of finite strain dislocation mechanics, when approached as a problem in classical finite elasticity.

\end{abstract}

\pagenumbering{arabic}

\section{Introduction}
\label{sec:introduction}

It is by now an accepted fact that the plastic deformation of metallic materials is primarily an outcome of the motion of dislocation line defects and that the evolving distribution of these defects, i.e., microstructure, plays a pivotal role in determining the strength and mechanical properties of such materials. In particular, there appears to be scientific consensus that the accumulation of Ashby's \cite{ashby1970deformation} `Geometrically Necessary Dislocations' (GNDs) leads to the phenomenon of \textit{size-effect} \cite{ashby1970deformation, fleck1994strain, 
stelmashenko1993microindentations, ma1995size} in micron-scale bodies, in that a stronger overall stress-strain response is observed with a decrease in size of the specimen.  Dislocations also form intricate microstructures under the action of their mutual interactions and applied loads such as dislocation cells  \cite{mughrabi1976asymmetry, mughrabi1979persistent, mughrabi1981cyclic, hughes2000microstructure} and labyrinths \cite{jin1984dislocation}, often with dipolar dislocation walls, and mosaics \cite{theyssier1995mosaic}. The presence of such dislocation microstructures, in particular their ‘cell size’ and orientation, has a strong influence on the macroscopic response of materials \cite{humphreys2012recrystallization, reed_2006}. While such dislocation mediated features of mechanical behavior from small to large length scales abound, commensurate theory and computational tools for understanding and predictive design are lacking, especially at finite deformation.

Several candidates for continuum scale gradient plasticity models at finite deformation  \cite{acharya2000grain, arsenlis2004evolution, tang2004effects,evers2004non,kuroda2008finite,ma2006dislocation,niordson2004size,niordson2005instabilities,lynggaard2019finite,niordson2019homogenized, ma2006dislocation,po2019continuum}, including inertia \cite{kuroda2019nonuniform}, are available with the goal of accurately modeling elastic plastic response of materials. These models can predict length-scale effects and some even account for some version of dislocation transport. However, to our knowledge, there is no continuum formulation that takes into account the stress field of signed dislocation density and its transport at finite deformation
. Current versions of Dislocation Dynamics models \cite{deshpande2003finite, irani2015finite} accounting for some features of finite deformation, reviewed in \cite{arora_acharya_ijss},  are also not capable of computing the finite deformation stress and energy density fields of dislocation distributions. Appendix \ref{sec:literature_review_jmps} presents a brief review of the vast literature on strain gradient plasticity (SGP) theories, as well as some models for computing static finite deformation fields of dislocations within a couple stress theory where dislocations are modeled as singular force distributions.


This paper reports on the further assessment of a modeling approach grounded in the idea of continuously distributed dislocations to understand  plasticity in solids at finite deformation, as it arises from dislocation motion, interaction, and nucleation within the material. Based on  partial differential equations (pde), the microscopic model -- Field Dislocation Mechanics (FDM) \cite{acharya2001model, acharya2004constitutive} -- takes into account the stress field and energy distribution of signed dislocation density along with its spatio-temporal evolution at finite strains. The model is capable of dealing with several features of defects in the crystal lattice at the atomic scale and, through a commonly used filtering approach in multiphase flows \cite{babic1997average}, provides a formal pathway for posing a space-time averaged pde model at the meso and macro scales \cite{acharya2006size, acharya2007jump, acharya2011microcanonical} termed as Mesoscale Field Dislocation Mechanics (MFDM). Built on rigorous  kinematical and thermodynamical ideas, MFDM allows for a study of finite deformation mesoscale plasticity rooted firmly in quantitatively identifiable links to the mechanics of dislocations.

A finite element based parallel computational tool based on the MFDM framework was verified and validated in \cite{arora_xia_acharya_cmame} and assessed in \cite{arora_acharya_ijss}. Here, we use this computational framework to study the various problems outlined in the abstract of this paper. In addition, we also demonstrate strong Bauschinger effects in our extension of $J_2$ plasticity theory as an outcome of inhomogeneity induced by boundary constraints to plastic flow, without the introduction of any ad-hoc model beyond an isotropic model of work hardening.

This paper is organized as follows:  after introducing notation and terminology in the remainder of this Introduction, Sec.~\ref{sec:theoritical_framework} presents the governing equations of finite deformation MFDM. The staggered computational algorithm for the quasi-static framework is discussed in Sec.~\ref{sec:fem}.  Section \ref{sec:results} presents the results of the three illustrative problems mentioned in the abstract. We end with some concluding remarks in Sec.~\ref{sec:conclusion}.

Significant portions of the description of the theory in Sec.~\ref{sec:theoritical_framework} are common with \cite{arora_xia_acharya_cmame}, a paper developed concurrently with this work. We include this content here for the sake of being self-contained, and because the theory being discussed is quite recent in the literature and not commonly known.

\section*{Notation and terminology}
\label{sec:notation}
Vectors and tensors  are represented by bold face lower and upper-case letters, respectively. The action of a second order tensor $\bfA$ on a vector $\bfb$ is denoted by $\bfA\bfb$. The inner product of two vectors is denoted by $\bfa\cdot\bfb$ and the inner product of two second order tensors is denoted by $\bfA:\bfB$. A superposed dot denotes a material time derivative. A rectangular Cartesian coordinate system is invoked for ambient space and all (vector) tensor components are expressed with respect to the basis of this coordinate system. $(\cdot)_{,i}$ denotes the partial derivative of the quantity $(\cdot)$ w.r.t.~the $x_i$ coordinate direction of this coordinate system. $\bfe_{i}$ denotes the unit vector in the $x_i$ direction. Einstein's summation convention is always implied unless mentioned otherwise. All indices span the range $1$-$3$ unless stated otherwise. The condition that any quantity (scalar, vector, or tensor) $a$ is defined to be $b$ is indicated by the statement $a := b$ (or $b =: a$).  $tr(\bfA)$ and $det(\bfA)$ denote the  trace and the determinant of the second order tensor $\bfA$, respectively. The symbol $|(\cdot)|$ represents the magnitude of the quantity $(\cdot)$.  The symbol $a\,en$  in figures  denotes $a \times 10^n$.

The current configuration and its external boundary is denoted by $\mOmega$ and $\partial \mOmega$, respectively. $\bfn$ denotes the unit outward normal field on $\partial \mOmega$. The symbols $grad$, $div$, and $curl$ denote the gradient, divergence, and curl on the current configuration. For a second order tensor $\bfA$, vectors $\bfv$, $\bfa$, and $\bfc$, and a spatially constant vector field $\bfb$, the operations of $div$, $curl$, and cross product of a tensor ($\times$) with a vector are defined as follows:
\begin{align*}
	(div\bfA)\cdot\bfb &= div(\bfA^T \bfb), ~~~~~~~~~ \forall ~ \bfb \\
	\bfb\cdot(curl\bfA)\bfc &=  \left[curl(\bfA^T \bfb)\right]\cdot \bfc, ~~~ \forall ~ \bfb, \bfc \\
	\bfc\cdot(\bfA\times\bfv)\bfa &= \left[(\bfA^T \bfc)\times \bfv \right]\cdot\bfa ~~~~\forall ~ \bfa, \bfc.
\end{align*}
In rectangular Cartesian coordinates, these are denoted by
\begin{align*}
	(div\bfA)_i =  A_{ij,j},\\
	(curl\bfA)_{ri} =  \varepsilon_{ijk}A_{rk,j}, \\
	(\bfA \times \bfv)_{ri} = \varepsilon_{ijk}A_{rj}v_k,
\end{align*} where $\varepsilon_{ijk}$ are the components of the third order alternating tensor $\bfX$.  The corresponding operations on the reference configuration are denoted by the symbols $Grad$, $Div$, and $Curl$. $\bfI$ is the second order Identity tensor whose components w.r.t.~any orthonormal basis are denoted by $\delta_{ij}$. The vector $\bfX(\bfA\bfB)$ is defined by $\left[\bfX(\bfA\bfB)\right]_i =  \varepsilon_{ijk}A_{jr}B_{rk}$.  The following list describes some of the mathematical symbols used in this paper.\newline
$\bbC$ : Fourth order elasticity tensor, assumed to be positive definite on the space of second order symmetric tensors\\
$E$ : Young's modulus\\
$\mu $: Shear modulus\\
$\nu$ : Poisson's ratio\\
$\bfC^e$ : Elastic Right Cauchy-Green deformation tensor\\
$I_1(\bfC^e)$ : First invariant of $\bfC^e$\\
$\phi$ : Elastic energy density of the material\\
$\rho$ : Mass density of the current configuration\\
$\rho^*$ : Mass density of the pure, unstretched lattice\\
$(\cdot)_{sym}$ : Symmetric part of $(\cdot)$\\
$m$ : Material rate sensitivity\\
$\hat{\gamma}_0$ : Reference strain rate\\
$\hat{\gamma}$ : Magnitude of slipping rate due to statistical dislocations for the $J_2$ plasticity model\\
$\hat{\gamma}^k$ : Magnitude of slipping rate due to statistical dislocations on the $k^{th}$ slip system for the crystal plasticity model\\
$n_{sl}$ : Number of slip systems\\ 
$\tau^k$ : Resolved shear stress on $k^{th}$ slip system\\
$sgn(\tau^k)$ : Sign of the scalar $\tau^k$\\
$\bfm^{k}$, $\bfn^k$ : Slip direction and the slip plane normal for the $k^{th}$ slip system in the current configuration\\
$\bfm^k_0$, $\bfn^k_0$ : Slip direction and the slip plane normal for the $k^{th}$ slip system in the pure, unstretched lattice\\
$g_0$ : Initial strength (Initial yield stress in shear)\\ 
$g_s$ : Saturation strength\\
$g$ : Material strength \\
$\mTheta_0$ : Stage $2$ hardening rate\\
$k_0$: Material constant incorporating the effect of GNDs on evolution of material strength\\
$\eta$: Non-dimensional material constant in empirical Taylor relationship ($\sim \frac{1}{3}$) for macroscopic strength vs. dislocation density\\
$\epsilon$ : Material constant, with dimensions of $stress \times length^2$, used in the description of dislocation core energy density\\
$b$ : Burgers vector magnitude of a full dislocation in the crystalline material\\
$h$ : Length of the smallest edge of an element in the finite element mesh under consideration

\section{Theory}
\label{sec:theoritical_framework}

This section summarizes the governing equations, initial and boundary conditions, and constitutive assumptions of finite deformation Mesoscale Field Dislocation Mechanics (details in Appendix \ref{app:MFDM}), the model that is computationally implemented, verified, and assessed in \cite{arora2019computational,arora_xia_acharya_cmame, arora_acharya_ijss} and further evaluated in this paper. These governing equations are:
\begin{subequations}
	\begin{align}
		&~\mathring{\bfalpha}\equiv (div\,\bfv)\bfalpha+\dot{\bfalpha}-\bfalpha\bfL^T = -curl\left(\bfalpha\times \bfV + \bfL^p\right)\label{eq:mfdm_a}\\[1mm]
		&~\bfW = \bfchi+grad\bff \nonumber\\[1.25mm]
		&\left.\begin{aligned}
			&curl{{\bfW}} = curl{\bfchi} = -\bfalpha\\
			&div{\bfchi} = \bf0
			\label{eq:mfdm_chi} 
		\end{aligned}\right\}\\[1.25mm]
		&~div\left(grad\dot{\bff}\right) = div\left(\bfalpha\times \bfV + \bfL^p - \dot{\bfchi}-\bfchi\bfL\right)\label{eq:mfdm_fevol}\\
		&~div\,[\bfT(\bfW)]  = \begin{cases}
			\bf 0 & \text{quasistatic} \\
			\rho\,\dot{\bfv} & \text{dynamic}.\\
		\end{cases}
		\label{eq:mfdm_f}
	\end{align}
	\label{eq:mfdm}
\end{subequations}
In \eqref{eq:mfdm}, $\bfW$ is the inverse-elastic distortion tensor, ${\bfchi}$ is the incompatible part of $\bfW$, $\bff$ is the plastic position vector, $grad\bff$ represents the compatible part of $\bfW$, $\bfalpha$ is the dislocation density tensor, $\bfv$ represents the material velocity field, $\bfL=grad\bfv$ is the velocity gradient, $\bfT$ is the (symmetric) Cauchy stress tensor, and $\bfV$ is the dislocation velocity field. The upshot of the development in Appendix \ref{app:MFDM} is that if $\bfL^p = \bfzero$ then the system \eqref{eq:mfdm} refers to the governing equations of FDM theory; otherwise, it represents the MFDM model. FDM applies to understanding the mechanics of small collections of dislocations, resolved at the scale of individual dislocations. MFDM is a model for  mesoscale plasticity with clear connections to  microscopic FDM.  The fields involved in the MFDM model are space-time averaged counterparts of the fields of FDM \eqref{eq:fdm}, with $\bfL^p$ being an emergent additional mesoscale field.

\subsection{Boundary conditions}
\label{sec:boundary_conditions}
The $\bfalpha$ evolution equation \eqref{eq:mfdm_a},  the incompatibility equation for $\bfchi$ \eqref{eq:mfdm_chi},  the $\bff$ evolution equation \eqref{eq:mfdm_fevol}, and the equilibrium equation \eqref{eq:mfdm_f} require specification of boundary conditions at all times.

The $\bfalpha$ evolution equation  \eqref{eq:mfdm_a} admits a `convective' boundary condition of the form $(\bfalpha \times \bfV + \bfL^p) \times \bfn=\bfPhi$ where $\bfPhi$ is a second order tensor valued function of time and position on the boundary characterizing the flux of dislocations at the surface satisfying the constraint $\bfPhi\bfn=\bf0$.  The boundary condition is specified in one of the following two ways:
\begin{itemize}
	\item \textit{Constrained}: $\bfPhi(\bfx, t) = \bf0$, for $\bfx$ on the boundary and for all times $t$. This makes the body plastically constrained at the boundary point $\bfx$ which means that dislocations cannot exit the body through that point, while only being allowed to move parallel to the boundary at $\bfx$.

	\item \textit{Unconstrained}: A less restrictive boundary condition where $\hat\bfL^p \times \bfn$ \eqref{tab:const_Lp} is specified on the boundary, along with the specification of dislocation flux $\bfalpha (\bfV\cdot\bfn)$ on the inflow part of the boundary (where $\bfV\cdot\bfn < 0$). In addition to this, for non-zero $l$, specification of $l^2\hat{\gamma}_{sd} (curl\bfalpha \times \bfn)$ on the boundary is also required, where $\hat{\gamma}_{sd}$  is defined in \eqref{eq:gaama_hat_ssd}. 
\end{itemize}

The incompatibility equation \eqref{eq:mfdm_chi} admits a boundary condition of the form $	\bfchi \bfn = \bf0 $ on the external boundary $\partial\mOmega$ of the domain. This condition along with the system \eqref{eq:mfdm_chi} ensures vanishing of $\bfchi$ when the dislocation density field vanishes on the domain, $\bfalpha \equiv \bf0$. The $\bff$ evolution equation \eqref{eq:mfdm_fevol} requires a Neumann boundary condition of the form
\begin{align*}
	(grad \,\dot{\bff})\bfn  = (\bfalpha\times \bfV + \bfL^p - \dot{\bfchi} - \bfchi\bfL)\bfn
\end{align*}
on the external boundary of the domain. The equilibrium equation \eqref{eq:mfdm_f} requires specification of standard displacement/velocity and/or statically admissible tractions on complementary parts of the boundary of the domain.

\subsection{Initial conditions}
\label{sec:initial_conditions}
The evolution equations for $\bfalpha$  and  $\bff$, \eqref{eq:mfdm_a} and \eqref{eq:mfdm_fevol}, require specification of initial conditions on the domain.

For the $\bfalpha$ equation, an initial condition of the form $\bfalpha(\bfx, t = 0) = \bfalpha_0(\bfx)$ is required.  To obtain  the initial condition on $\bff$, the problem can be more generally posed as follows: determine the $\bff$ and $\bfT$ fields on a given configuration with a known  dislocation density $\bfalpha$. This problem is solved by solving for $\bfchi$ from the incompatibility equation and then $\bff$ from the equilibrium equation as described by the system
\begin{eqnarray}
	\left.\begin{aligned}
		curl\bfchi  &= -\bfalpha \\
		div\bfchi &= \bf0\\
		div\left[\bfT(grad\!\bff, \bfchi)\right] &= \bf0\\
	\end{aligned}\quad\right\} \text{on } \mOmega  \label{eq:ECDD}
	\\
	\left.\begin{aligned}
		\bfchi\bfn &= \bf0\\
		\bfT\bfn &= \bft\\
	\end{aligned}\qquad \right\} \text{on }\partial\mOmega,
	\label{eq:ECDD_bc}
\end{eqnarray}
where $\bft$ denotes a statically admissible, prescribed traction field on the boundary. This determination of $\bfchi$, $\bff$, and $\bfT$ for a given dislocation density $\bfalpha$ and $\bft$ on  any known configuration will be referred to as the ECDD solve on that configuration.  Hence, we do the ECDD solve on the `as-received' configuration, i.e.~the current configuration at $t = 0$,  to determine the initial value of $\bff$ which determines the stress field at $t = 0$. For the dynamic case, an initial condition on material velocity field $\bfv(\bfx, t = 0)$ is required.

The model admits an arbitrary specification of $\dot{\bff}$ at a point to uniquely evolve $\bff$ from  Eq.~\eqref{eq:mfdm_fevol}  in time and we prescribe it to be $\dot{\bff} = \bf0$.

\subsection{Constitutive equations for $\bfT$, $\bfL^p$, and $\bfV$} 
\label{sec:dissipation}
MFDM requires constitutive statements for the stress $\bfT$, the plastic distortion rate $\bfL^p$, and the dislocation velocity $\bfV$. The details of the thermodynamically consistent constitutive formulations are presented in \cite[Sec.~3.1]{arora_acharya_ijss}.  This constitutive structure is summarized below.

	{\renewcommand{\arraystretch}{1.2}
\begin{table}
	\centering
	\begin{tabular}{|p{3.5cm}|p{6.5cm}|p{5cm}|}
		\hline
		~~\newline
		Saint-Venant-Kirchhoff Material
		& \begin{align*}\phi(\bfW) = \dfrac{1}{2\rho^*} \bfE^e:\bbC:\bfE^e\end{align*} & 		\begin{align}	\bfT = \bfF^e\left[\bbC:\bfE^e\right]\bfF^{eT}\label{eq:stress_svk}\end{align}\\  
		
		\hline
		Core energy density & \multicolumn{2}{|c|}{ $~~~~~~~\mUpsilon(\bfalpha) :=  \dfrac{1}{2\rho^*}{\epsilon} \,\bfalpha:\bfalpha$} \\
		\hline
		
	\end{tabular}
	\caption{Constitutive choices for elastic energy density, Cauchy stress, and core energy density.}
	\label{tab:const_T}
\end{table}
}

Table \ref{tab:const_T} presents the Cauchy stress expressions for the Saint-Venant-Kirchhoff 
material . It also presents the assumed constitutive form of the  mesoscopic core energy density (per unit mass) for the material. Tables \ref{tab:const_Lp} and \ref{tab:const_V} present the constitutive assumptions for $\bfL^p$ and $\bfV$, respectively, for the crystal and $J_2$ plasticity models.  Table \ref{tab:const_g} presents the governing equation for the evolution of material strength $g$ for the two models.  The use of $\hat{\gamma}_{sd}$ in \eqref{eq:softening} stems from the fact that isotropic (or Taylor) hardening is used for the  evolution of strength on every slip system with equal initial values, i.e.,
\begin{align}\nonumber
\hat{\gamma}^k &= sgn(\tau^k)\, \hat{\gamma_0}^k\left(\frac{|\tau^k|}{g^k} \right)^{\frac{1}{m}},\qquad k = 1,\ldots,n_{sl} \nonumber\\
\dot{g}_{kj} &= h(\bfalpha,g)\left(\left|\bfF^e\bfalpha\times\bfV\right| + \sum_{j = 1}^{n_{sl}}  [q + (1-q)\delta_{kj}]\left| \hat{\gamma}^j \right| \right), \qquad 1 \leq q \leq 1.4, \qquad k,j = 1,\ldots,n_{sl}, \label{eqn:latent_hardening}
\end{align}
where the function $h$ is defined in \eqref{eq:softening} and \eqref{eqn:latent_hardening} is a simple modification of standard latent hardening phenomenology assumed in classical crystal plasticity (see, e.g., \cite{peirce1983material}). In this paper, we focus on constitutive assumptions using the $J_2$ plasticity model while the crystal plasticity counterparts are presented for completeness. Also, isotropic hardening is not a necessary condition for the formulation.

\begin{table}
	
	\centering
	\begin{tabular}{|p{4cm}|p{11.5cm}|}
		\hline
		~ & {\begin{align}
				\hat{\bfL}^p&= \bfW\, \left(\sum_k^{n_{sl}} \hat{\gamma}^k \, \bfm^k\otimes\bfn^k \right)_{sym} \\
				\bfL^p &= \hat{\bfL}^p + \quad  \left( \dfrac{l^2}{n_{sl}} \sum_k^{n_{sl}} |\hat\gamma^k| \right) \,curl \bfalpha \label{eq:Lp_crystal}\\
				\hat{\gamma}^k &= sgn(\tau^k)\, \hat{\gamma_0}^k\left(\frac{|\tau^k|}{g} \right)^{\frac{1}{m}}\end{align}}\\
		\vspace{-3cm} Crystal plasticity & \vspace{-1cm}{\begin{align*}
				\tau^k = \bfm^k\cdot\bfT\bfn^ k  ;~~
				\bfm^k = {\bfF^e\bfm_0^k}  ; ~~
				\bfn^k = {{\bfF^e}^{-T}\bfn_0^k} 
		\end{align*}}\\[-2em] \hline 
		\vspace{1.5cm}			$J_2$ plasticity &  {\begin{align}\hat{\bfL}^p &=\, \hat{\gamma}\bfW\frac{\bfT^{'}}{|\bfT^{'}|}  ;  ~~ \hat{\gamma} = \hat{\gamma_0} \left(\dfrac{|\bfT'|}{\sqrt{2} \,g}\right)^{\frac{1}{m}}\nonumber\\[.1cm]
				\bfL^p &=\hat{\bfL}^p +  l^2  \hat\gamma \, curl\bfalpha \label{eq:Lp_j2}  
		\end{align}} \\ \hline
	\end{tabular}
	\caption{Constitutive choices for plastic strain rate due to SDs $\bfL^p$.}
	\label{tab:const_Lp}
\end{table}

\begin{table}
	\centering
	\begin{tabular}{|p{15cm} |}
		\hline 
		{\begin{align}
				T'_{ij} = T_{ij} - \dfrac{T_{mm}}{3}\delta_{ij}; ~~~~ a_i &:= \dfrac{1}{3}T_{mm} \varepsilon_{ijk}{F^e}_{jp}\alpha_{pk}; ~~~~			c_i := \varepsilon_{ijk}T'_{jr}{F^e}_{rp}\alpha_{pk}\nonumber\\
				\bfd = \bfc - \left(\bfc\cdot \frac{\bfa}{|\bfa|} \right) \frac{\bfa}{|\bfa|}; ~~&~~			\hat{\gamma}_{avg} =
				\begin{cases} 
					\hat{\gamma} & J_2~\textrm{plasticity} \\
					\dfrac{1}{n_{sl}}\sum_k^{\,n_{sl}} |\hat{\gamma}^k| & \textrm{Crystal plasticity}.
				\end{cases}\nonumber\\
				\bfV = \zeta & \frac{\bfd}{|\bfd|} ~~ ; ~~\zeta =  \left(\dfrac{\mu}{g}\right)^2\,\eta^2\, b  \, \hat{\gamma}_{avg}\label{eq:V_lpj2} 
		\end{align}}\\
		\hline
	\end{tabular}
	\caption{Constitutive choices for dislocation velocity $\bfV$.}
	\label{tab:const_V}
\end{table}

\begin{table}
	\centering
	\begin{tabular}{|p{15cm} |}
		\hline 
		{\begin{align}
				\hat{\gamma}_{sd} &=
				\begin{cases} 
					\hat{\gamma} & J_2~\textrm{plasticity} \\
					\sum_k^{\,n_{sl}} |\hat{\gamma}^k| & \textrm{Crystal plasticity}.
				\end{cases}
				\label{eq:gaama_hat_ssd}
		\end{align}}
		{
			\begin{align}
				\dot{g} = h(\bfalpha, g) \left(\left|\bfF^e\bfalpha\times\bfV\right|+ \hat{\gamma}_{sd} \right); \qquad h(\bfalpha, g) = \frac{\mu^2\eta^2b}{2(g-g_0)}k_0 \left|\bfalpha\right|+ \mTheta_0 \left(\frac{g_s-g}{g_s-g_0}\right)
				\label{eq:softening}
			\end{align} 
		}\\
		\hline
	\end{tabular}
	\caption{Evolution equation for material strength $g$.}
	\label{tab:const_g}
\end{table}

All material parameters, except $k_0$ and $l$, are part of the constitutive structure of well-accepted models of classical plasticity theory. Our model requires these  two extra material parameters beyond the requirements of classical theory. The length-scale $l$ (with physical dimensions of length) sets the scale for the mesoscopic core energy to be effective,  and $k_0$ (non-dimensional) characterizes the hardening rate due to GNDs.

We mention here that $l$, introduced in \eqref{eq:Lp_crystal} or \eqref{eq:Lp_j2}  as a dimensional consequence of including the core energy, is not responsible for producing enhanced size effects and microstructure in MFDM. Rather, the `smaller is harder' size effect  decreases with increasing magnitude of $l$ since its presence reduces the magnitude of the $\bfalpha$ field and, consequently, reduces hardening \eqref{eq:softening}.

\section{Numerical algorithm} 
\label{sec:fem}

The finite element implementation of the system of equations given in \eqref{eq:mfdm} has been discussed in \cite{arora_xia_acharya_cmame} where detailed numerical algorithms, verification, and validation exercises are provided. Here, we briefly describe the general flow of the algorithm for the sake of being self-contained.

Modeling material behavior through the use of MFDM requires the concurrent solution to the  coupled nonlinear system of pdes \eqref{eq:mfdm}. To efficiently solve the system \eqref{eq:mfdm} using a staggered scheme in each time increment as shown in \cite{roy2005finite, roy2006size} for the small deformation case, it is augmented with the rate form of the equilibrium equation to obtain the material velocity field in the domain. In the absence of body forces and inertia, the rate form of the equilibrium equation is given as 
\begin{align}
	div\left(div(\bfv) \bfT + \dot{\bfT} - \bfT\bfL^T\right) = \bf0.
	\label{eq:rate_form}
\end{align}
The material velocity, $\bfv$, obtained by solving \eqref{eq:rate_form}, is used to discretely update the geometry at each time increment. 

The augmented system (\eqref{eq:mfdm} and \eqref{eq:rate_form}) is solved numerically using the following discretization methods: the Galerkin FEM for the equilibrium equation  \eqref{eq:mfdm_f}, its rate form \eqref{eq:rate_form}, and the evolution equation \eqref{eq:mfdm_fevol} for the compatible part of the inverse  elastic distortion; the Least-Suares FEM for the incompatibility equation \eqref{eq:mfdm_chi}; and the Galerkin-Least-Squares FEM for the dislocation evolution equation \eqref{eq:mfdm_a}.  We now summarize the algorithm for the quasistatic case.

\subsection{Algorithm for quasistatic loading case}
\label{sec:algorithm}
The augmented system of equations (\eqref{eq:mfdm} and  \eqref{eq:rate_form}) is discretely evolved in time. An efficient time-stepping criteria and  an intricate `cut-back' algorithm are implemented to ensure stable and accurate evolution of state variables. The algorithm used to solve the augmented MFDM system for the quasistatic case is as follows:

\begin{itemize}
	\item Given the material parameters and initial conditions on $\bfalpha$ and the prescribed traction $\bft$ (most often vanishing), ECDD is used to  obtain $\bff$, $\bfchi$, and $\bfT$ on the configuration at $t = 0$.
	
	\item Given the geometry and state variables ($\bff^t, \bfchi^t, \bfalpha^t, \bfV^n, \bfL^p$, and $ g^n$) at time $t^n$, $\Delta t^n := t^{n+1} - t^n$ is then calculated based on the time stepping criteria explained in \cite[Sec.\,4]{arora_xia_acharya_cmame}. The following steps are used to evolve the system in a time increment $[t^n, t^{n+1}]$:
	
	\begin{enumerate}
		\item The rate form of the equilibrium equation \eqref{eq:rate_form} is solved on the configuration $\bfOmega^n$ to obtain the material velocity field $\bfv^n$. This velocity field $\bfv^n$ is used to (discretely) evolve the geometry to obtain the configuration at time $\bfOmega^{n+1}$. 
		
		\item The dislocation evolution equation \eqref{eq:mfdm_a} is solved on $\bfOmega^n$ to obtain $\bfalpha^{n+1}$ on the configuration $\bfOmega^{n+1}$.
		
		\item $\bfchi^{n+1}$ on $\bfOmega^{n+1}$ is obtained by solving \eqref{eq:mfdm_chi} with $\bfalpha^{n+1}$ as data. \label{three}
		
		\item $\bff^{n+1}$ is determined as follows:
		
		\begin{enumerate}
			\item $\bff$ is evolved from \eqref{eq:mfdm_fevol} to obtain $\bff^{n+1}$ on the configuration $\bfOmega^{n+1}$. \label{4a}
			
			\item In alternate increments, the equilibrium equation \eqref{eq:mfdm_f} is solved on the configuration $\bfOmega^{n+1}$ which is now posed as a traction boundary value problem (with rigid modes eliminated by kinematic constraints). The statically admissible nodal (reaction) forces, on the part of the boundary with Dirichlet boundary conditions on material velocity, are computed following the discussion in \cite[Sec.~3.1.1]{arora_xia_acharya_cmame}. \label{4b}
			
		\end{enumerate}
		
		\item Once the state at time $t^{n+1}$ is accepted after checking through the cut-back criterion \cite[Sec.\,4]{arora_xia_acharya_cmame}, the above algorithm is repeated to obtain the new state at $t^{n+2}$.
		
	\end{enumerate}
	
\end{itemize}

The fields obtained from  Steps \ref{three}.\! and \ref{4a}.\! above suffice to define an approximation for the stress field at $t^{n+1}$, using the hyperelastic constitutive equation. However, this may not satisfy (discrete) balance of forces at each time step (because the current geometry obtained from the rate-form of equilibrium and the stress field under discussion need not necessarily be `consistent' with each other in the sense of discretely satisfying force balance on $\Omega^{n+1}$); therefore we periodically use the discrete  equilibrium equation to correct for force balance (Step \ref{4b}. above) . 

The algorithm above is described in greater detail in \cite{arora2019computational,arora_xia_acharya_cmame}, and has been implemented  in an MPI(Message Passing Interface)-accelerated C++ code utilizing various comprehensive state of the art libraries such as Deal.ii \cite{dealII85}, P4est \cite{BursteddeWilcoxGhattas11}, MUMPS \cite{MUMPS:1}, and PetSc \cite{petsc-web-page} . 

\section{Results and discussion}
\label{sec:results}
We use the parallel computational framework of MFDM developed in \cite{arora_xia_acharya_cmame} to solve three fundamental problems in finite deformation dislocation mechanics and small-scale plasticity.

Before proceeding to discuss results,  we mention some details pertaining to our calculations. For all the results presented in this work, the input flux $\bfalpha (\bfV\cdot\bfn)$ and $curl\bfalpha \times \bfn$ are assumed to be  $\bf0$ on the boundary. Also, $\hat\bfL^p$ is directly evaluated at the boundary to calculate $\hat\bfL^p \times \bfn$. All fields are interpolated using bilinear/trilinear elements in $2$-d/$3$-d, unless otherwise stated.  The Burgers vector content of an area patch $A$ with normal $\bfn$ is given by $	\bfb_A = \int_A \bfalpha \bfn \, dA,$ where $\bfalpha$ denotes the dislocation density field in the domain. When the dislocation distribution $\bfalpha$ is localized in a cylinder around a space curve as its axis and this cylinder threads the area patch $A$,  then we denote its Burgers vector by $\bfb$. It must be noted that $\bfb$ is independent of any area patch $A$ for which  the intersection of the `core' cylinder and the patch is entirely contained within the patch, this being a consequence of the fact that $div\, \bfalpha = \bfzero$.  We refer to the magnitude of the Burgers vector, $|\bfb|$, as the strength of the dislocation. 
We define  a  measure of  magnitude of the GND density as \cite{arora_acharya_ijss}
\begin{align*}
\rho_g(\bfx, t) :=  \frac{|{\bfalpha}(\bfx,t)|}{b}. 
\end{align*}

All algorithms in this paper have been verified to reproduce classical plasticity solutions for imposed homogeneous deformation histories by comparison with solutions obtained by integrating the evolution equation \eqref{eq:conventional_Fedot} for the elastic distortion tensor $\bfF^e$  to determine the Cauchy stress response for an imposed spatially homogeneous velocity gradient history, $\bfL$:
\begin{align}
	\begin{split}
		& \dot\bfF^e = \bfL\bfF^e - \bfF^e\bfL^p\bfF^e =: \tilde\bff(\bfF^e, g),\\
		&\dot g = \tilde{g}(\bfF^e, g),
	\end{split}
	\label{eq:conventional_Fedot}
\end{align}
where $\bfL^p$ is defined from Eq.~\eqref{eq:Lp_j2} or \eqref{eq:Lp_crystal}  with $l = 0$, and $\tilde{g}$ is given by \eqref{eq:softening} with $k_0 = 0$.

\begin{figure}
	\centering
	{\includegraphics[width=.65\linewidth]{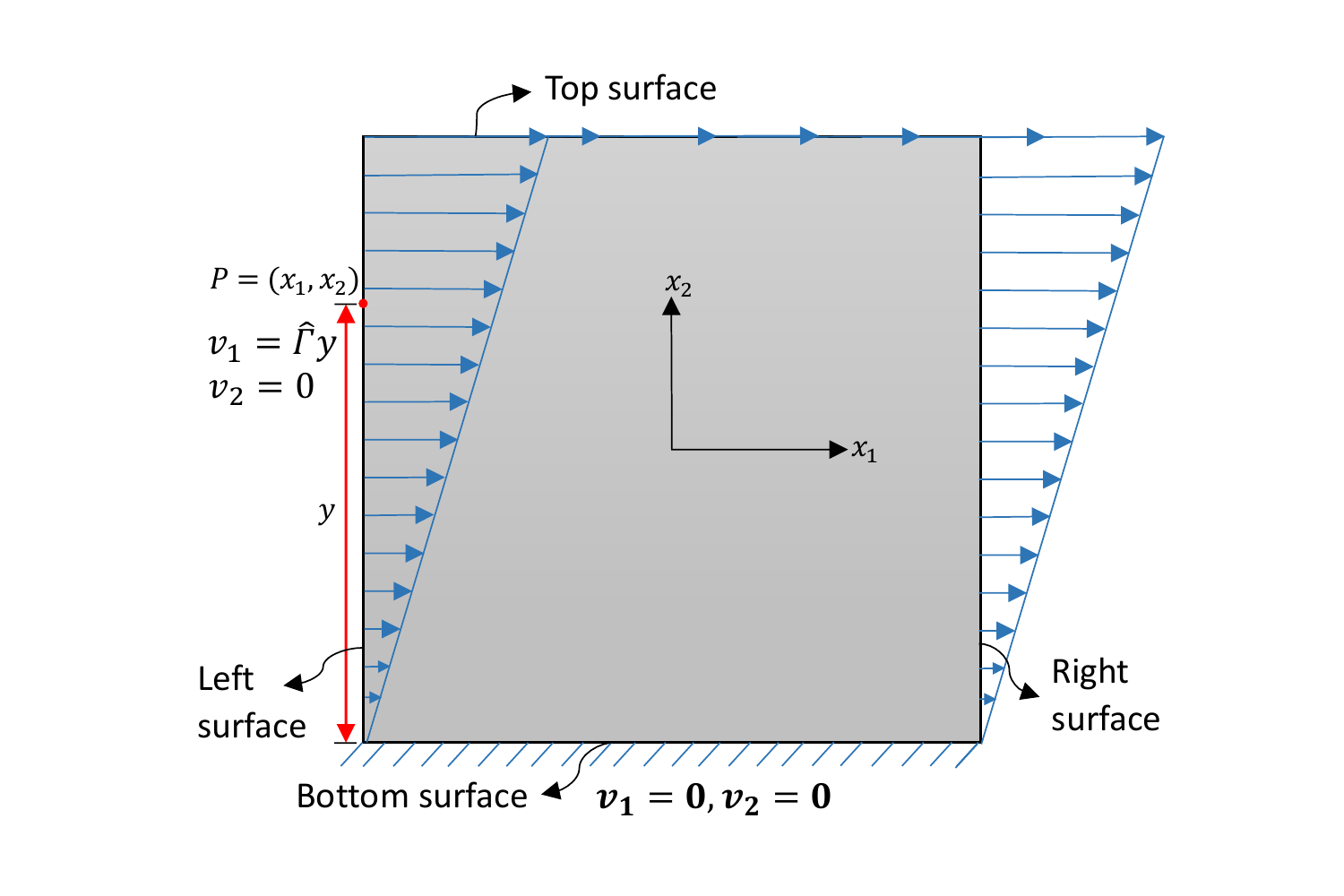}}
	\caption{Typical schematic of a rectangular body under shear loading.}
	\label{fig:se_schematic}
\end{figure}

A typical schematic of the basic geometry used in most problems (further details are mentioned as required) is shown in Fig.\,\ref{fig:se_schematic}. The averaged $T_{12}$ component of the stress tensor on the top surface is denoted by $\tau$. It is calculated by  summing the tangential components of the nodal reaction force on the top surface and then dividing by the current area (line length) of the surface. $\hat\mGamma$ represents the applied  strain rate. At any time $t$, $\mGamma$ denotes the engineering shear strain and is calculated as $\hat\mGamma t$. 

\subsection{Plastic flow in thin confined layers: comparison with experiment}
\label{sec:res_thin} 

Recent experiments by Mu et al.~ \cite{mu2014micro, mu2014thickness, mu2016dependence} report results of micropillars subjected to axial compression that contain a thin layer of ductile copper sandwiched within a stiff and brittle ceramic pillar. Confined simple shear loading conditions are generated when the thin Cu layer is inclined at $45^{\circ}$ to the pillar axis \cite{mu2016dependence}. The interfaces between the Cu layer and the ceramic bulk may be assumed to be plastically constrained as the ceramic is brittle. The experimental results demonstrate a power-law relationship with negative exponent between the inferred applied shear stress on the Cu layer versus its thickness.  This exponent was found to be  $\sim -0.2$ for the  as-deposited material \cite[Fig. 4a]{mu2016dependence} and $\sim -0.7$ for the annealed samples.

As noted in \cite[pp. 5--6]{mu2016dependence} and by Hutchinson \cite{hutch_sgp_scaling}, and posed as a fundamental challenge to \emph{all} higher-order strain gradient plasticity theories which predict a size effect due to interface constraints, such models inevitably predict a scaling exponent $ \leq -1$, representing a much stiffer response than observed in experiment.

The situation above has prompted a further formulation of strain gradient plasticity {\cite{dahlberg2019fractional} introducing a new fitting parameter and adapted to this single experiment, as well as reformulations \cite{kuroda_needle_sgp_bc} of existing SGP models wherein ad-hoc relaxation of boundary constraints dictated by the basic theory is suggested to accommodate the observed behavior with the justification that ``there is a maximum value of the magnitude of the plastic strain gradient at the layer boundaries that can be supported before plastic straining starts at the boundary.'' Moreover, it is understood from the kinematics of dislocation motion that plastic shearing parallel to a boundary is not constrained by constrained plastic flow boundary conditions \cite{gurtin2005boundary} \cite[Sec. \!2.2.1]{acharya2006size}.

	In contrast, here (Sec.~\ref{sec:res_thin}) we use the finite deformation $J_2$ MFDM computational framework to model the constrained simple shear of a thin polycrystalline metallic strip without any adjustment to the basic model to accommodate this specific experiment. We focus on the result for the as-deposited material and correspondingly use a reasonable value of the initial yield stress to reflect the presence of an initial statistical dislocation density in the material.
	
	The scaling exponent, denoted by $\beta$, is evaluated by assuming a power-law relationship
	\begin{align}
		\left(\frac{\tau - g_0}{g_0}\right) =  c\,H^\beta,
		\label{eq:thin_tau}
	\end{align}
	where $c$ is a scalar which is constant for a given strain and boundary conditions, and $H$ denotes
	the film thickness. Moreover, since MFDM accounts for the finite deformation stress fields for the GNDs and their spatio-temporal evolution coupled to the underlying kinematics, we also present the microstructure evolution during the deformation.

	{\renewcommand{\arraystretch}{1.2}
	\begin{table}
		\centering
		\begin{minipage}[b]{.54\linewidth}
			\centering{
				\begin{tabular}{ | c | c | }
					\hline
					Size $(\mu m \times \mu m)$ & Mesh  \\ 			
					\hline
					$5 \times .50 $ & $100 \times 10$	 \\ 
					$5 \times .65 $ & $100 \times 13$	 \\ 
					$5 \times .80 $ & $100 \times 16$	 \\ 
					$5 \times 1.0 $ & $100 \times 20$	 \\ \hline
			\end{tabular}}
		
			\caption{Mesh details for  different domain sizes.}
			\label{tab:thin_mesh}
			
				\centering{
				{\includegraphics[width=\linewidth]{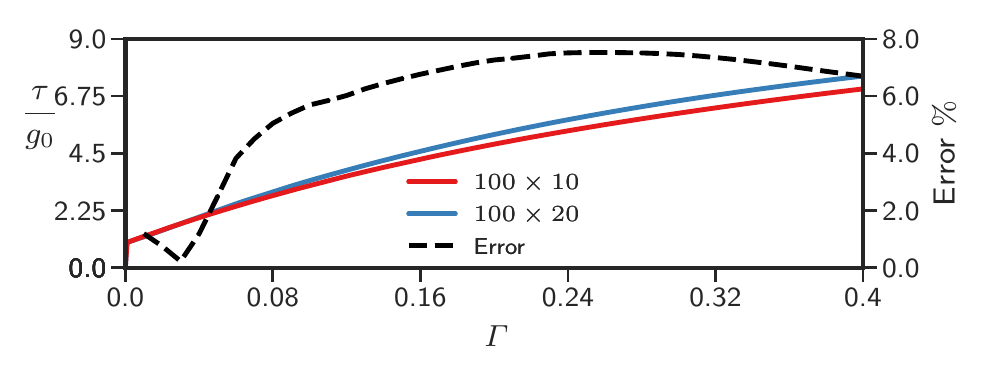}}
				\vspace*{-11mm}
				  \captionof{subfigure}{$H = 0.5 \mu m$}
				\label{fig:se_ss_conv_p5}}
			
			\centering{
				{\includegraphics[width=\linewidth]{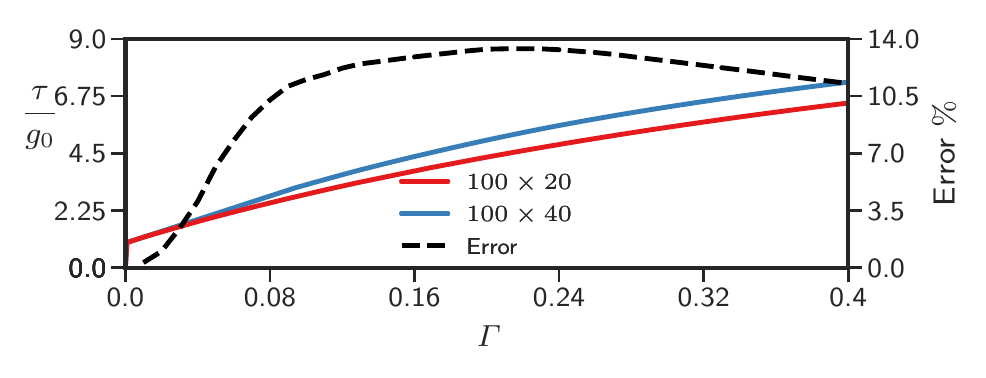}}
				\vspace*{-11mm}
	\captionof{subfigure}{$H = 1 \mu m$}
				\label{fig:se_ss_conv_1}}
			
			{
				\captionof{figure}{Convergence of stress-strain responses.}
				\label{fig:se_ss_conv_p5_1}
			}
		\end{minipage}\hfill
		\begin{minipage}[b]{.42\linewidth}
			\centering{
				\begin{tabular}{ | c | c | }
					\hline
					Parameter &   Value  \\ 			\hline
					$\hat\gamma_0$ &  $0.001s^{-1}$\\     
					$m$ &  $0.03$\\     
					$\eta$ & $\dfrac{1}{3}$\\[.2cm]
					$b$ &  $4.05 {\mbox{\normalfont\AA}}$\\  
					$g_0$ &  $.0173$ GPa\\     
					$g_s$ &  $.161$ GPa\\     
					$\mTheta_0$ &  $.3925$ GPa\\     
					$k_0$ &  $20$\\     
					$l$ &  $\sqrt{3}\times 0.1\,\mu m$\\     
					$E$ &  $62.78$ GPa \\     
					$\nu$ &  $.3647$ \\ \hline	
			\end{tabular}}
			\caption{Parameter values used to model the effect of thickness on the flow stress of thin metal films.}
			\label{tab:thin_data}
		\end{minipage}
	\end{table}
}

	Simulations are performed on rectangular domains of width $5\,\mu m$ and varying thicknesses, $H$, of  $0.5\,\mu m$, $0.65\,\mu m$, $0.8\,\mu m$, and $1\,\mu m$ (the pillar diameter in the experiments was $5\, \mu m$). Problems are set up in a $2$-d plane strain setting as follows:  velocity boundary conditions corresponding to the simple shear deformation for plane strain condition are imposed. At any point $\bfx \equiv (x_1, x_2)$ on the boundary denoted by $P$, $v_1 = \hat{\mGamma}y(x_2)$ and $v_2 = 0$ are imposed where $y(x_2)$ is the height of the point $P$ from the bottom surface as shown in Fig. \ref{fig:se_schematic}.  Table \ref{tab:thin_data} presents the values of the material constants used in this section. The applied strain rate $\hat{\mGamma} = .001s^{-1}$. The top and bottom surfaces are treated as (plastically) constrained whereas the other two lateral surfaces are treated as (plastically) unconstrained, as defined in Sec.~\ref{sec:boundary_conditions}}. These boundary conditions mimic the metallic sample sandwiched between two non-deforming brittle substrates while undergoing shear deformation, and the possibility of free exit of dislocations from the sides of the strip sandwiched within the micropillar. Section \ref{sec:res_se_j2}  explores the effect when all the four surfaces are treated as  plastically constrained. Figure \ref{fig:se_ss_conv_p5_1} shows the convergence of the stress strain response for the domains with thickness $0.5 \mu m$ and $1\mu m$ for two mesh sizes. We use the coarser mesh for our numerical computations, with details of the meshes for all the domain sizes given in Table \ref{tab:thin_mesh}.
	
%
%
	
	\begin{figure}
		\centering
		\begin{subfigure}[b]{.8\linewidth}
			\centering
			{\includegraphics[width=0.95\linewidth]{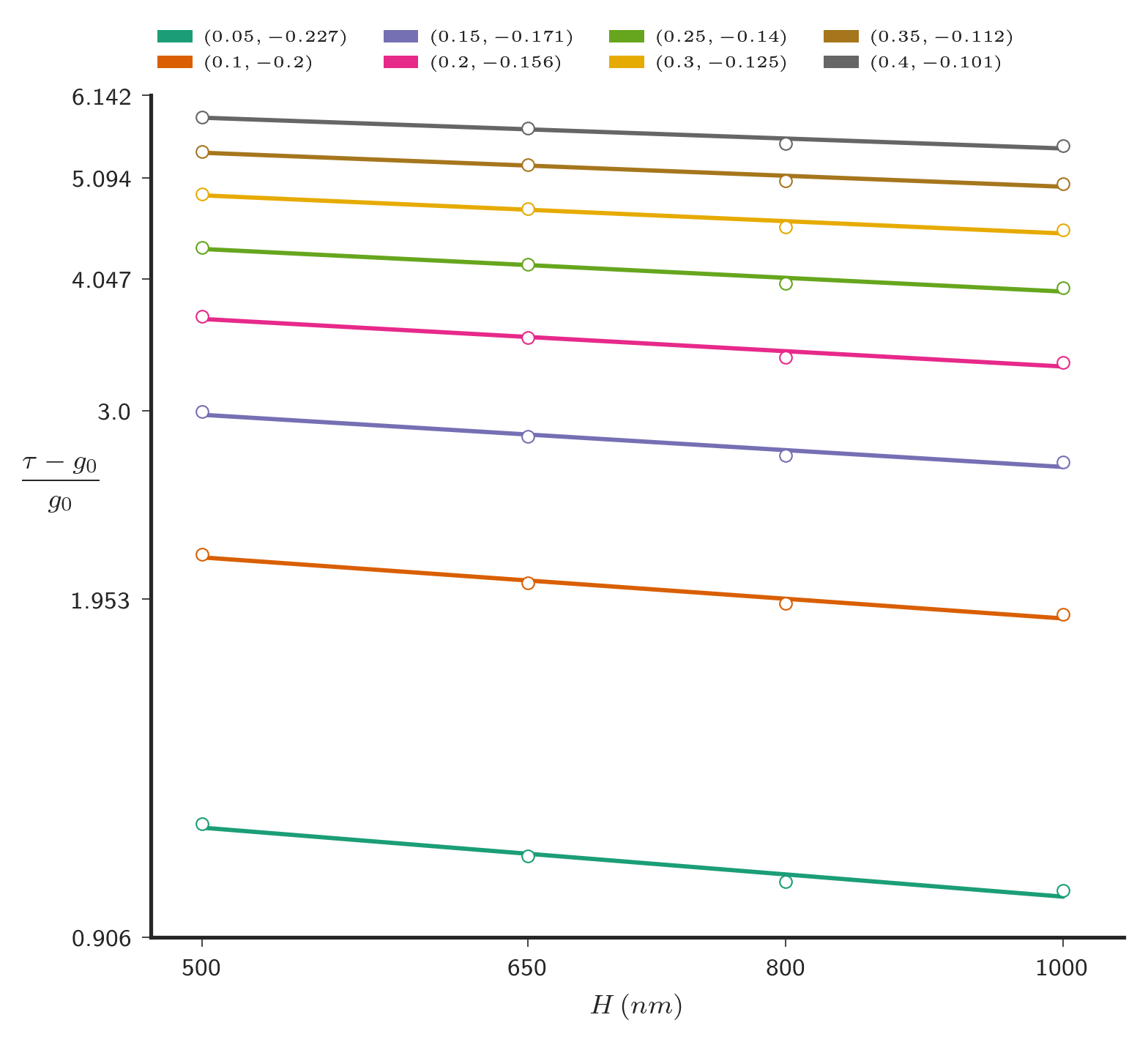}}
		\end{subfigure}
		\caption{Scaling exponent $\beta$ at different strains. For each color, the pair in the legend represents the strain and the value of $\beta$ (the slope) at that strain. The data obtained from MFDM is marked by `o' and the straight line represents the best fit curve on a $\log$-$\log$ scale.}
		\label{fig:thin_beta}
	\end{figure}

	Figure \ref{fig:thin_beta} shows the scaling exponent $\beta$ at different applied shear strains. The value of $\beta$ predicted by MFDM is between $0$ and $-1$ and very close to the value  observed in experiments on the as-deposited samples in \cite{mu2016dependence}. We predict a clear decreasing trend in the values of $|\beta|$ with increasing strain. We note that the predicted values of $\beta$ are a pure outcome of the theory unlike \cite{dahlberg2019fractional} wherein the scaling exponent is exactly equal to the fractional order of the discrete plastic-strain derivatives embedded in their theory, but undetermined by it.
	
\begin{figure}
		\centering
		\begin{subfigure}[b]{.495\linewidth}
			\centering
			\includegraphics[width=0.95\linewidth]{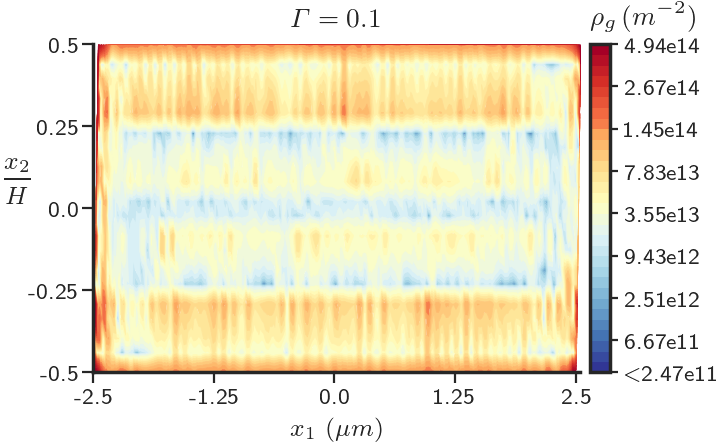}
			\caption{}
		\end{subfigure}
		\begin{subfigure}[b]{.495\linewidth}
			\centering
			\includegraphics[width=0.95\linewidth]{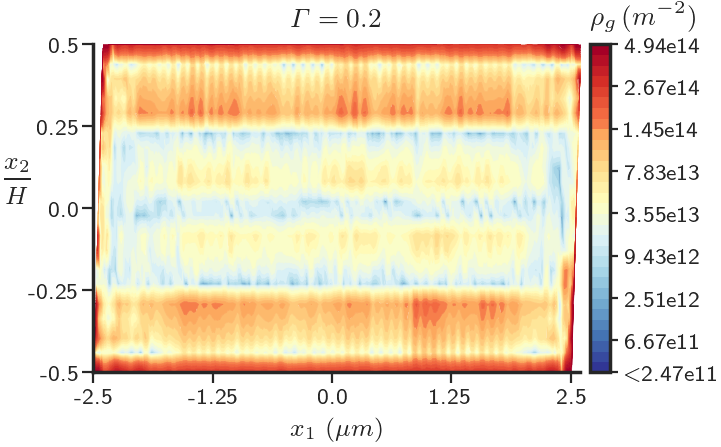}
			\caption{}
		\end{subfigure}\\[5mm]
		\begin{subfigure}[b]{.495\linewidth}
			\centering
			\includegraphics[width=0.95\linewidth]{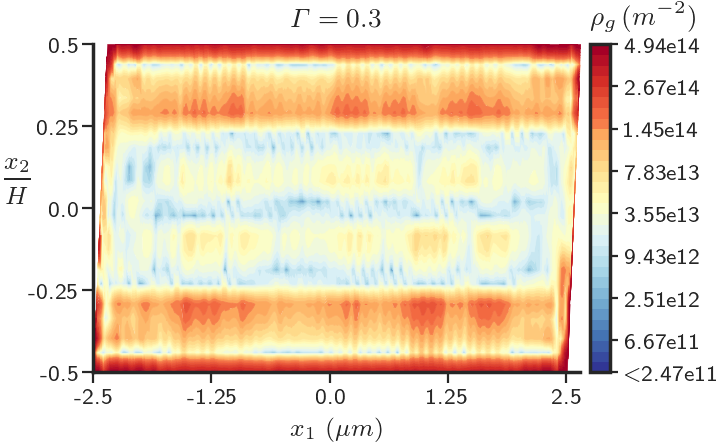}		\caption{}
		\end{subfigure}
		\begin{subfigure}[b]{.495\linewidth}
			\centering
			\includegraphics[width=0.95\linewidth]{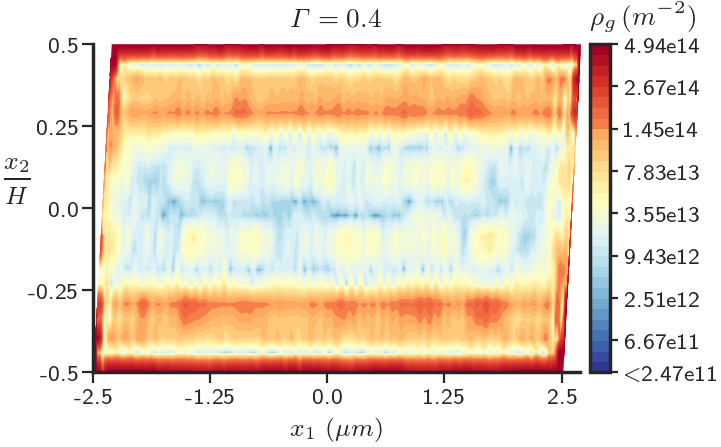}
			\caption{}
		\end{subfigure}\\
		\caption{$\rho_g (m^{-2})$ at different strains for the sample with $H = 0.5 \mu m$.}
		\label{fig:thin_anorm2}
	\end{figure}

	Figure \ref{fig:thin_anorm2} shows the $\rho_g$ distribution at different strains for the sample with $H = 0.5\, \mu m$. In the figure, the $x_2$ axis is scaled by the thickness $H$ of the sample. The constrained boundary condition on the top and bottom surfaces induce gradients in $\bfL^p$ near these boundaries. The gradients, in the $x_2$ direction, of $L^p_{21}$ - arising from the presence of the $T_{21}$ stress component -  leads to the continuous generation of $\alpha_{23}$ near the top and bottom boundaries.

\begin{figure}
		\centering
		\begin{subfigure}[b]{.495\linewidth}
			\centering
			\includegraphics[width=0.95\linewidth]{./figures/Thin/ANp5-40p0pcnt.png}
			\caption{}
		\end{subfigure}
		\begin{subfigure}[b]{.495\linewidth}
			\centering
			\includegraphics[width=0.95\linewidth]{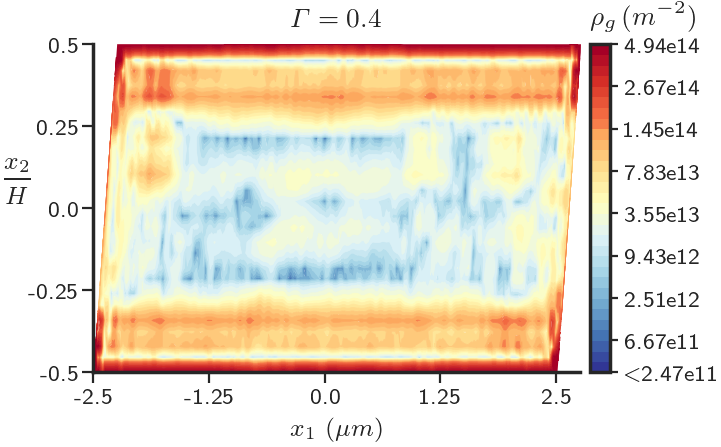}
			\caption{}
		\end{subfigure}\\[5mm]
		\begin{subfigure}[b]{.495\linewidth}
			\centering
			\includegraphics[width=0.95\linewidth]{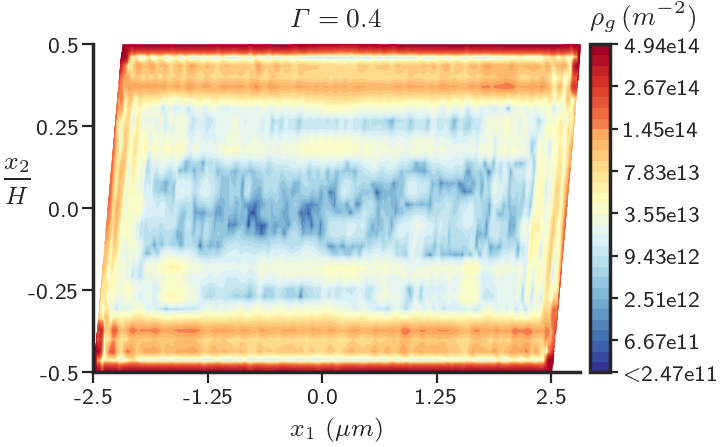}
			\caption{}
		\end{subfigure}
		\begin{subfigure}[b]{.495\linewidth}
			\centering
			\includegraphics[width=0.95\linewidth]{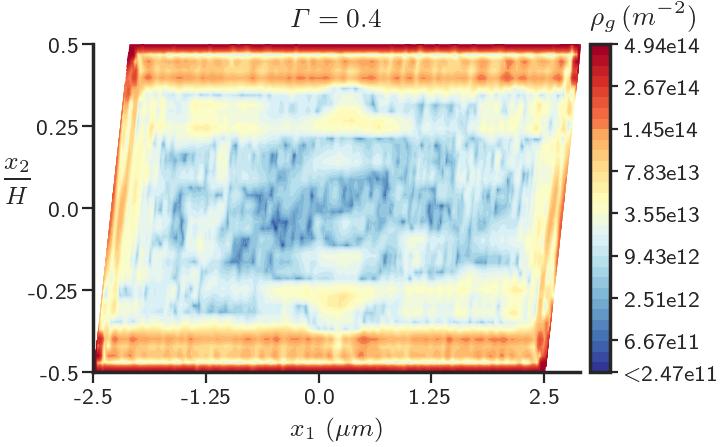}
			\caption{}
		\end{subfigure}
		\caption{$\rho_g$ at $40\%$ strain for different sample sizes. a) $H = 0.5 \mu m$ b) $H = 0.65 \mu m$ c) $H = 0.8 \mu m$ d) $H = 1 \mu m$.}
		\label{fig:thin_anorm}
	\end{figure}

	Figure \ref{fig:thin_anorm} shows the $\rho_g$ distribution at $\mGamma = 0.40$ for the various sample sizes. The $x_2$ axis is scaled by the respective thickness of the sample.  The figure shows that the non-dimensional dislocation layer width decreases with the increase in the height of the sample, an effect not predictable by classical theory on dimensional grounds \cite{roy2006size}.

In closing, we note that MFDM has not been shown to predict a size-effect at initial yield in small deformation analysis (at finite deformation, there is a  variation of initial yield stress w.r.t the initial GND distribution field to be theoretically expected that needs to be assessed in future simulation work). Even though in the modeling of \cite{dahlberg2019fractional, kuroda_needle_sgp_bc} it is implicitly assumed that the entire size effect is related to phenomena at initial yield, it is not entirely clear to us from the experimental load-displacement data in axial compression in \cite[Fig. 4b]{mu2016dependence} that the plateaus there, representing axial displacements up to shear-off of the sample, do not include hardening in the stress-strain response. The results we have presented are directly related to the size-effect scaling in work hardening response. Our computational infrastructure is ideally suited for independent prediction of the finite-deformation load-displacement response of the experiment, which is future work in progress.

\subsubsection{Effect of fully constrained boundaries and Bauschinger effect}
\label{sec:res_se_j2}
\begin{figure}
	\centering
	\begin{subfigure}[b]{.40\linewidth}
	{\includegraphics[width=\linewidth]{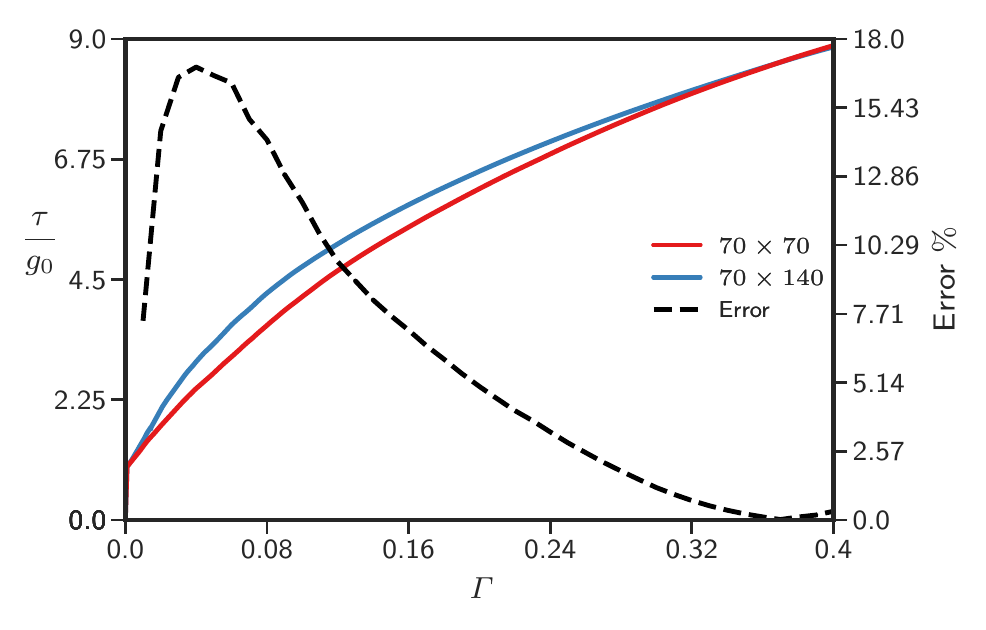}}
\caption{Convergence of stress-strain response for the $\mum{1}$ domain.}
\label{fig:se_ss_conv}
	\end{subfigure}\hfill
\begin{subfigure}[b]{.58\linewidth}
	\centering
	{\includegraphics[width=.99\linewidth]{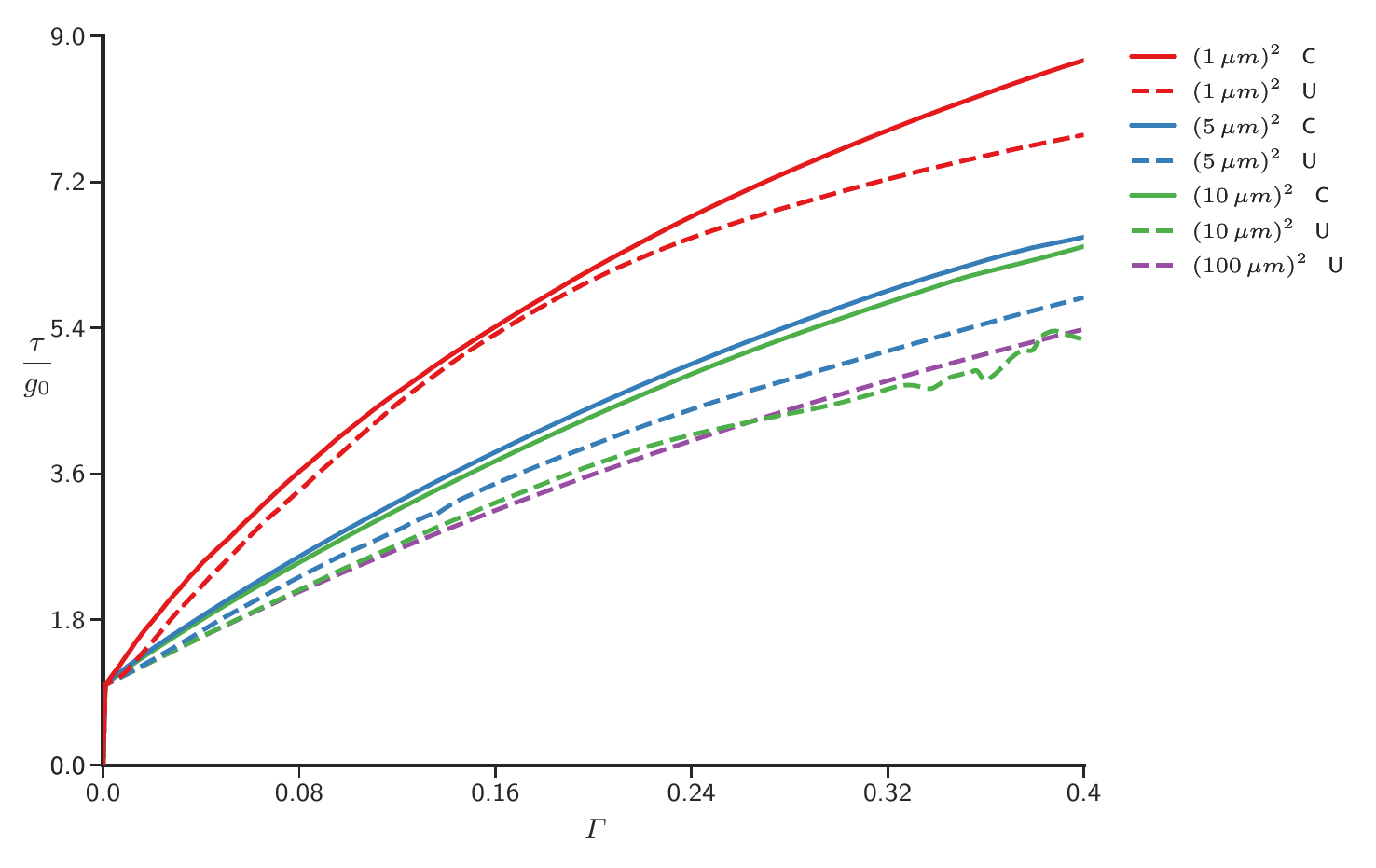}}
	\caption{Size effect under simple shear. C: Constrained boundaries U: Unconstrained boundaries.}
	\label{fig:se_ss}
\end{subfigure}
\end{figure}

Simulations are performed on square domains of sizes $\mum{1}$, $\mum{2}$, $\mum{5}$, $ \mum{10}$, and $\mum{100}$ with a typical schematic of the  geometry shown in Fig.~\ref{fig:se_schematic}. The boundary conditions are as for the simple shear case just discussed in this Sec.~\ref{sec:res_thin} except that the lateral boundaries are also considered to be plastically constrained. Table \ref{tab:thin_data} presents the values of the material parameters used in this section. An applied strain rate of $\hat{\mGamma} = .001 s^{-1}$ is used. Figure \ref{fig:se_ss_conv} shows the convergence of stress-strain curves for the $\mum{1}$ sample size for  two meshes with $70\times 70$ and $140\times 140$ elements. We use the coarser mesh of $70 \times 70$ elements for all the domain sizes. The unconstrained case represents a conservative simulation scenario with smaller gradients than the constrained case, and hence these mesh sizes suffice for the unconstrained case as well.

Figure \ref{fig:se_ss} shows the overall stress-strain response for all the domain sizes and both the boundary conditions (plastically constrained and unconstrained), demonstrating the `smaller is harder' size effect.  

\begin{figure}
	\centering
	\begin{subfigure}[b]{.8\linewidth}
		\centering
		{\includegraphics[width=0.95\linewidth]{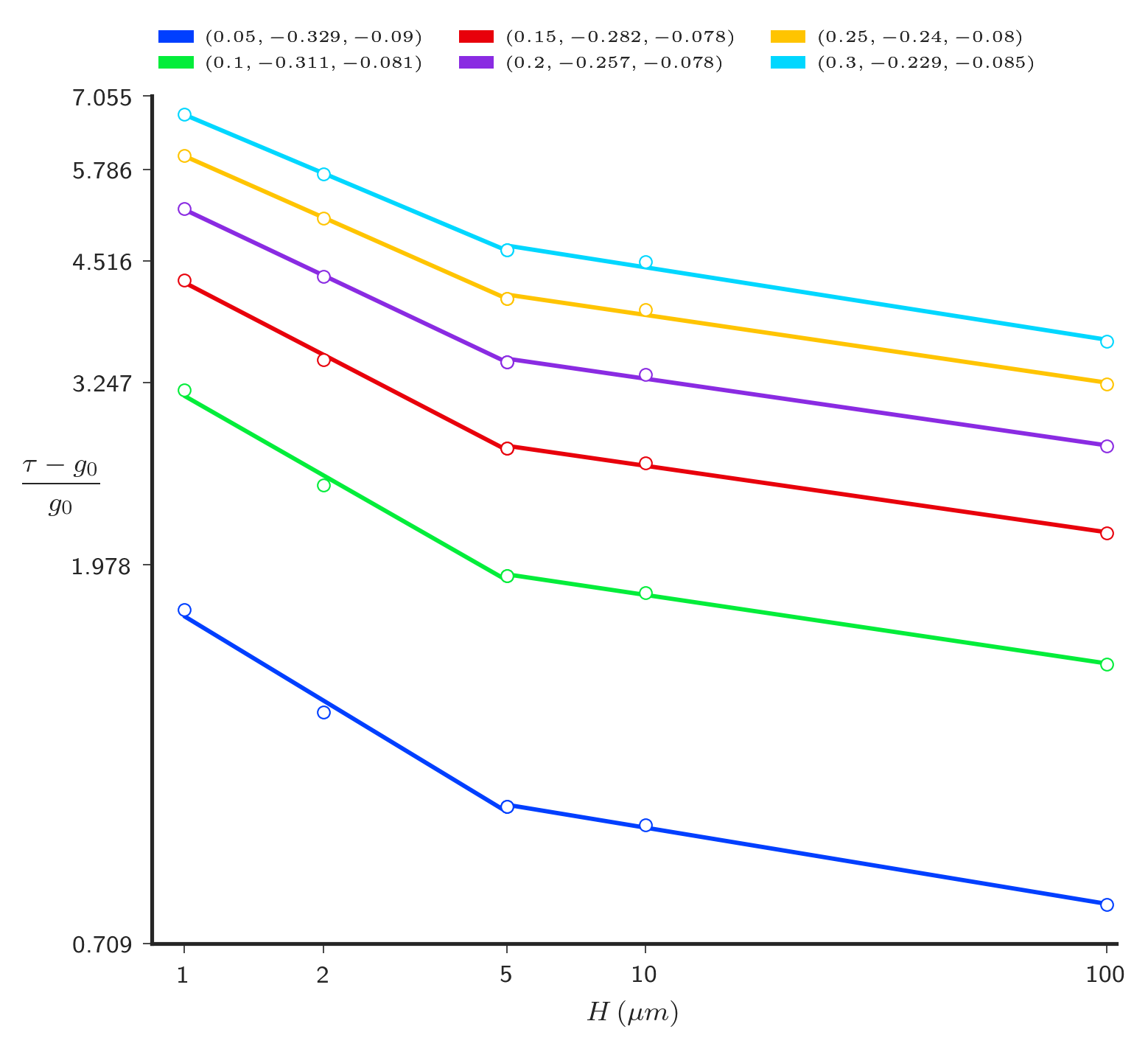}}
	\end{subfigure}
	\caption{Scaling exponent $\beta$ at different strains. For each curve, the trio in the legend represents the strain followed by the  values of the slope, $\beta$, from the left, of the two straight lines comprising the curve.  The data obtained from MFDM is marked by `o' and the straight portions of the curves represent the best fit lines of the data on a $\log$-$\log$ scale.}
	\label{fig:thin_beta_C}
\end{figure}

Figure \ref{fig:thin_beta_C} presents the scaling exponent $\beta$ at different strains. The values of $\beta$ lie between $0$ and $-1$ with magnitudes less than the case studied in Figure \ref{fig:thin_beta}. The data over the whole size range does not fit a single power law expression; as shown, two power-laws appear to provide a reasonable fit.

\begin{figure}
	\begin{subfigure}[b]{.495\linewidth}
		\centering
		{\includegraphics[width=\linewidth]{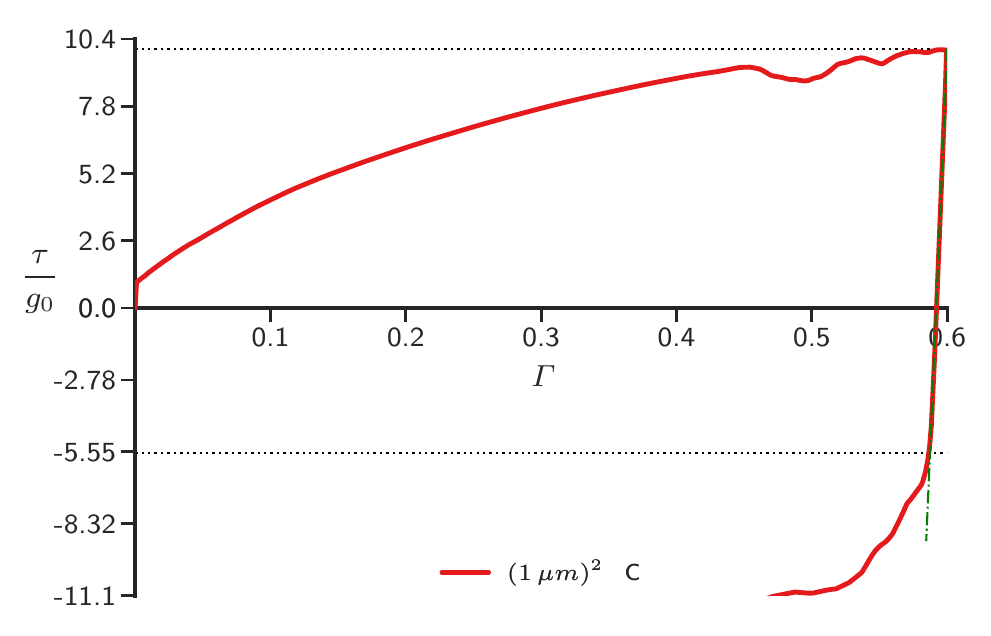}}
		\caption{}
		\label{fig:se_1mc_unloading_ss}
	\end{subfigure}\hfill
	\begin{subfigure}[b]{.495\linewidth}
		\centering
		{\includegraphics[width=\linewidth]{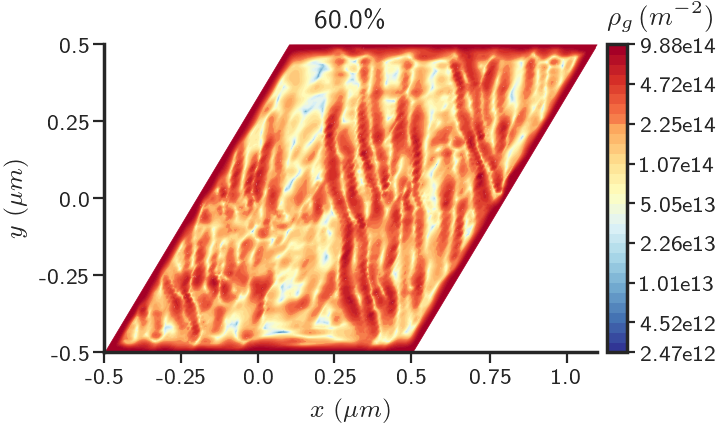}}
		\caption{}
		\label{fig:se_1mc_an_loaded60}
	\end{subfigure}\\
	\begin{subfigure}[b]{.495\linewidth}
		\centering
		{\includegraphics[width=\linewidth]{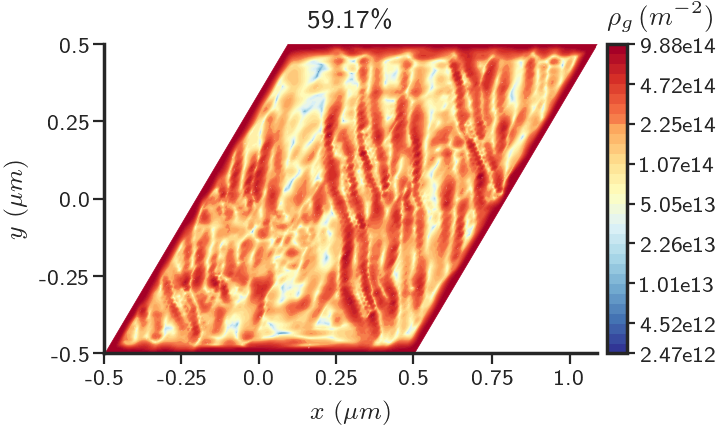}}
		\caption{}
		\label{fig:se_1mc_an_unloaded0}
	\end{subfigure}%
	\begin{subfigure}[b]{.495\linewidth}
		\centering
		{\includegraphics[width=\linewidth]{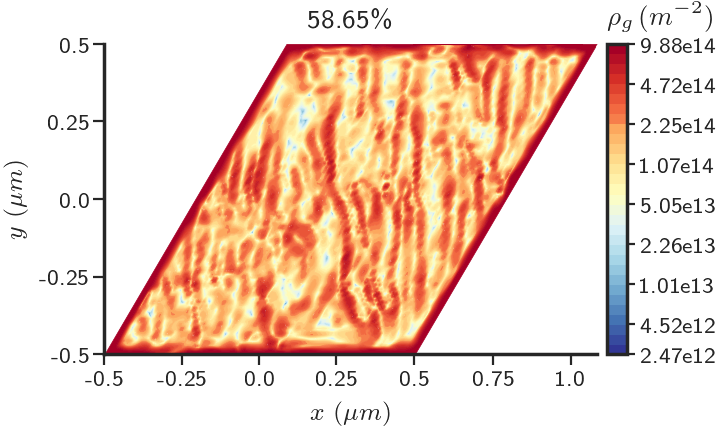}}
		\caption{}
		\label{fig:se_1mc_an_loadedr}
	\end{subfigure}%
	\caption{a) Stress strain response for $\mum{1}$ fully constrained (C) sample demonstrating Bauschinger effect. The green dotted line shows the elastic unloading curve. b) $\rho_g$ distribution at $60\%$ strain c) $\rho_g$ distribution at $0$ averaged load on the top surface d) $\rho_g$ distribution at $58.65\%$ strain when plastic deformation starts again after the loading direction is reversed.}
	\label{fig:se_1mc_an_unloadedp}
\end{figure}

Figure \ref{fig:se_1mc_unloading_ss} shows the stress strain plot for the $\mum{1}$ sample size with constrained boundaries up to $60\%$ strain and Figure \ref{fig:se_1mc_an_loaded60} shows the $\rho_g$ distribution at  that strain. The loading direction is then reversed and the body starts unloading elastically. Fig. \ref{fig:se_1mc_an_unloaded0} shows the $\rho_g$ distribution in the domain at $59.17\%$ strain when the averaged load on the top surface ($\tau$) is close to $0$. As the reverse loading continues,  the body starts deforming plastically again around $58.65\%$ strain displaying a strong Bauschinger effect, presumably due to the internal stresses of the $\bfalpha$ distribution. Figure \ref{fig:se_1mc_an_loadedr} show the $\rho_g$ distribution at $58.65\%$ strain when the plastic deformation initiates again.

\begin{figure}
	\begin{subfigure}[b]{.495\linewidth}
		\centering
		{\includegraphics[width=\linewidth]{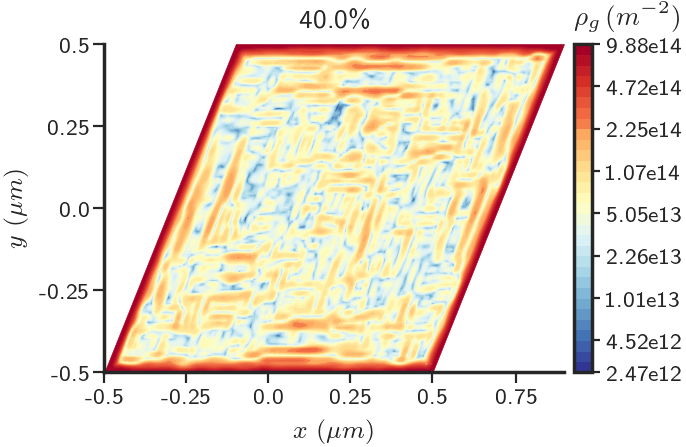}}
		\caption{}
	\end{subfigure}%
	\begin{subfigure}[b]{.495\linewidth}
		\centering
		{\includegraphics[width=\linewidth]{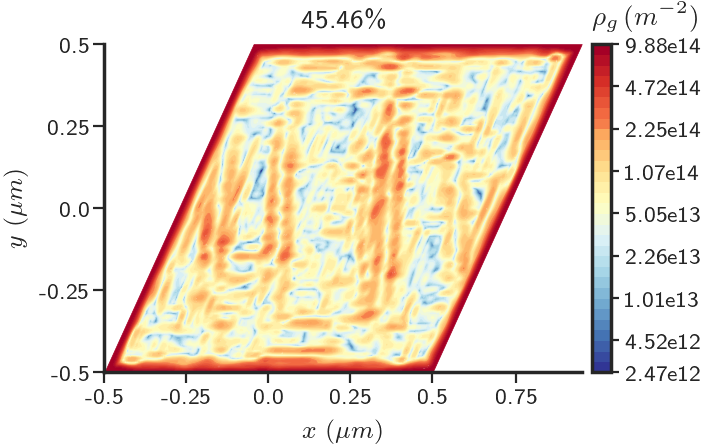}}
		\caption{}
	\end{subfigure}\\
	\centering
	\begin{subfigure}[b]{.495\linewidth}
		\centering
		{\includegraphics[width=\linewidth]{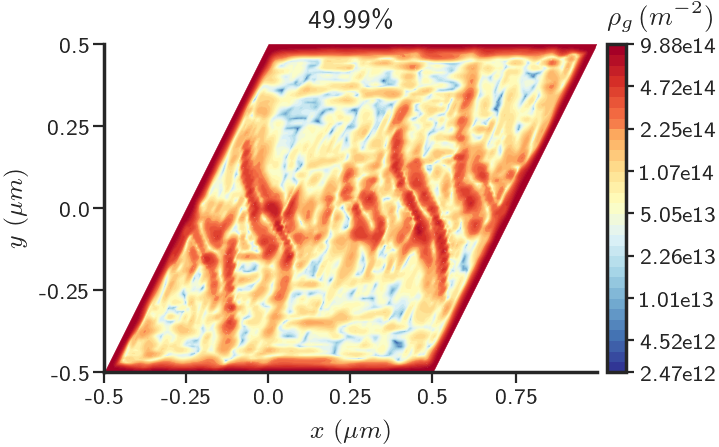}}
		\caption{}
	\end{subfigure}
	\caption{$\rho_g$ distribution in the domain at different strains around the onset of `waviness' in stress-strain curve Fig.\,\ref{fig:se_1mc_unloading_ss} during loading.}
	\label{fig:alpha_wavy}
\end{figure}

Figure \ref{fig:alpha_wavy} shows the $\rho_g$ distribution in the same problem at three different strains of $40\%$, $45.46\%$, and $49.99\%$ during the forward loading. There is considerable development of wall-like microstructure in the interior of the domain between $40\%$-$50\%$ strain. This interior  microstructure pattern formation roughly coincides with the `wavy' signature in the stress-strain response beyond $40\%$ strain in Fig. \ref{fig:se_1mc_unloading_ss}.

\subsection{Finite elastic f\mbox{}ields in polygonization}
\label{sec:res_poly}

We use finite deformation FDM to study the stress and energy density fields of a sequence of dislocation distributions whose limit is a through-dislocation wall, as observed in the physical process of polygonization \cite{gilman1955structure, nabarro1967theory}. After presenting our results, we make contact with available mathematical work \cite{DW:muller, ginsterb} on the limit energy functionals for \emph{nonlinear} elastic deformations with dislocations, show that the current state-of-the-art of mathematical work in this direction is inadequate for the problem of polygonization, and discuss our perspective on the problem.

\begin{figure}[h]
	\centering
	\begin{minipage}[b]{.6\linewidth}
		\centering
		{\includegraphics[width=\linewidth]{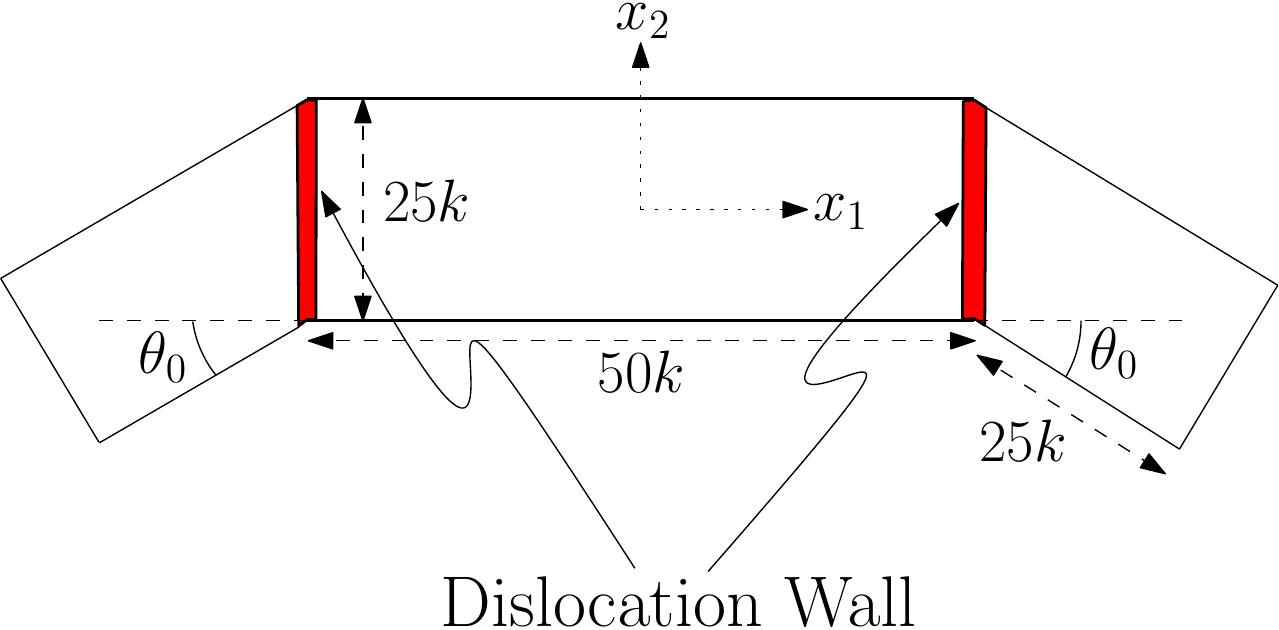}}
		\caption{Schematic layout of the polygonized domain.}
		\label{fig:poly_schematic}
	\end{minipage}\hfill
	\begin{minipage}[b]{.38\linewidth}
		\centering
		{\includegraphics[width=\linewidth]{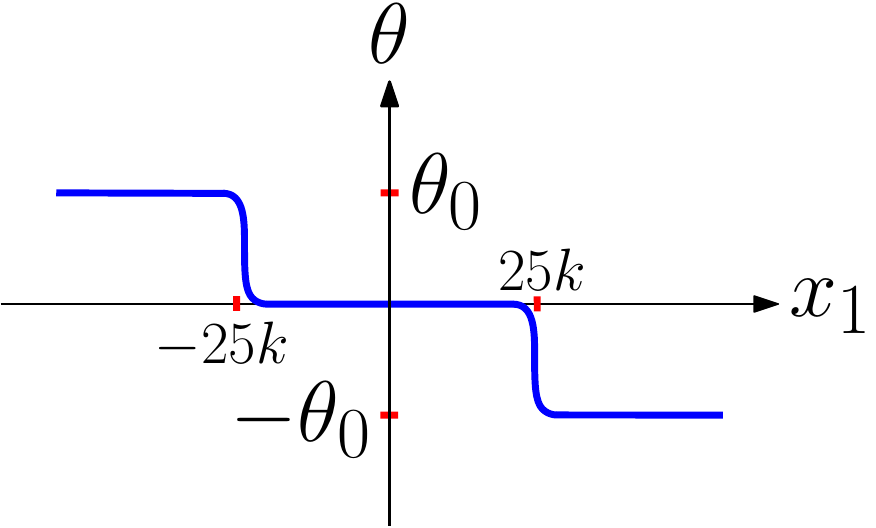}}
		\caption{Variation of $\theta$ in the polygonized domain for dislocation walls centered at $x_1 = \pm25k$.}
		\label{fig:theta_schematic}
	\end{minipage}\hfill
\end{figure}

We compute the stress field of a special sequence of dislocations, wherein the dislocation cores stack up to form a dislocation wall in the limit. The result is a tilt grain boundary consistent with a piecewise uniform finite rotation field resulting in zero-stresses in the domain.  Before analyzing the stress fields for the sequence of dislocation distributions, we  first study the limiting case, i.e., a polygonized domain with two dislocation walls as shown in Fig.~\ref{fig:poly_schematic}.  The height of the dislocation walls is chosen as $25k$ where $k$ denotes the width of the walls.

In the limiting case, the orientation of the lattice on one side of the wall differs from the other side by a rotation angle denoted by $\mis_0$. To construct the dislocation density field describing the two walls in the domain for a given  $\mis_0$, we first approximate this piecewise constant rotation angle field by a continuous field 
\begin{align}
	\theta(\bfx) &= -\dfrac{\theta_0}{2} \left[\,\tanh\left(\frac{x_1 + 25k}{a\,k}\right) + \,\tanh\left(\frac{x_1 - 25k}{a\,k}\right) \right],
	\label{eq:dw_poly_alphaeq}
\end{align}  
with the walls centered at $x_1 = \pm 25k$, and $a$ is a dimensionless scalar chosen to be $0.238$ which ensures  that the widths of the dislocation walls are $k$. Figure \ref{fig:theta_schematic} shows the variation of $\mis(\bfx)$ along $x_1$ in the polygonized domain. The corresponding elastic distortion $\bfF^e$ is then given by the rotation tensor field
\begin{align}
	\begin{split}
		\bfF^e(\bfx) &= \begin{bmatrix}
			\cos(\theta(\bfx)) & - \sin(\theta(\bfx)) & 0 \\
			\sin(\theta(\bfx)) &   \cos(\theta(\bfx)) & 0 \\
			0 & 0 & 1
		\end{bmatrix},
		\label{eq:dw_Fe_alpha}
	\end{split}
\end{align}
with the corresponding dislocation density field in the domain given by $\bfalpha = -curl \,\bfW$.

A non-uniform mesh, highly refined close to the dislocation walls, is used to discretize the polygonized domain comprising an isotropic Saint-Venant-Kirchhoff elastic material with $E = 200$GPa and $\nu = 0.3$.  The stress fields are calculated for $\theta_0 = \mdeg{45}$ in a $2$-d plane strain setting. Since $\theta$ does not vary in the $x_2$ direction
\begin{equation*}
	\alpha_{13} = -\epsilon_{312} \left(\bfW \right)_{12,1} = \cos(\theta)\dfrac{d\theta}{dx}; \qquad 
	\alpha_{23} = -\epsilon_{312} \left(\bfW \right)_{22,1} = \sin(\theta)\dfrac{d\theta}{dx}.
\end{equation*}
Thus, $\alpha_{23} \approx 0$ is a reasonable approximation for low-angle grain boundaries.  The solution for the nonlinear elastic stress field of the chosen $\bfF^e$ field is of course trivial - it is $\bfzero$ everywhere. However, the stress field of a progressively forming dislocation wall whose elastic distortion approaches a rotation tensor field everywhere is non-trivial, as we show in the following. Since the latter is our main interest, and the knowledge of the trivial solution does not help in solving the latter problem, we devise a unified strategy to solve the entire class of problems. This requires first to solve the `full-wall' problem with the trivial solution by a non-trivial method that then finds crucial use in solving the full range of `sparse' to `full' wall problems. In what follows, we first describe how the full-wall problem is dealt with.

With the dislocation density field and vanishing applied tractions specified, the ECDD system \eqref{eq:ECDD}-\eqref{eq:ECDD_bc} is solved in the polygonized domain to determine the stress field. The numerical solution of the nonlinear ECDD problem requires a good initial guess. This guess is obtained by solving for $\bff$ by using the Least-Squares finite element method with objective functional
$\frac{1}{2} \int_\Omega || grad\bff + \bfchi - \bfW||^2 \, dV$ with $\bfW$ and $\bfchi$ specified. The specified data is generated from $\bfW$, the inverse elastic distortion field corresponding to the prescribed $\bfF^e$  field given by \eqref{eq:dw_Fe_alpha}, and by solving for $\bfchi$ from the $div$-$curl$ system \eqref{eq:mfdm_chi} with $\bfalpha = - curl \, \bfW$ specified.
The weak form (to solve for $\bff$) is given by 
\begin{equation}
	\int_\Omega grad\!\bff:grad\delta\!\bff \, dV = \int_\Omega (\bfW - \bfchi):grad\delta\!\bff \, dV.
	\label{eq:fguess_R}
\end{equation}
$\bff$ is fixed at one point in the domain to obtain a unique solution. Solving for $\bff$ from the least squares method implies $(\bfW - \bfchi)\,\bfn = (grad\!\bff)\bfn$ holds weakly  on the external boundary.

\begin{figure}
	\centering
	\begin{subfigure}[b]{0.49\textwidth}
		\centering
		{\includegraphics[width=\linewidth]{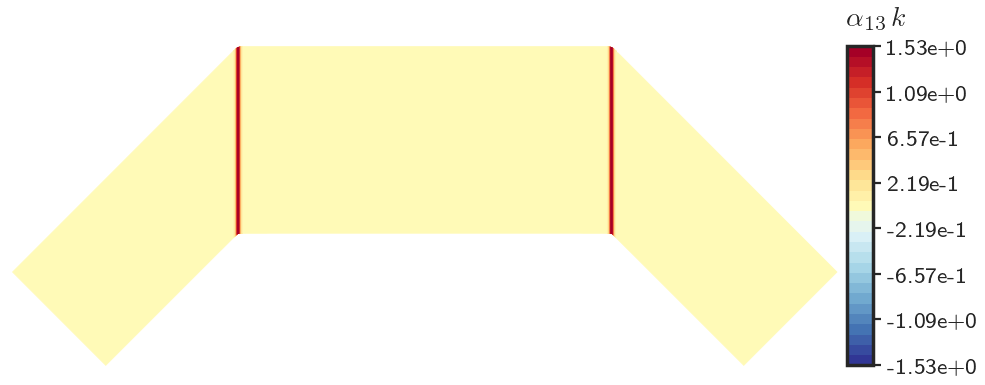}}
		\caption{}
		\label{fig:dw_alpha13_45_poly}
	\end{subfigure}\hfill
	\begin{subfigure}[b]{0.49\textwidth}
		\centering
		{\includegraphics[width=\linewidth]{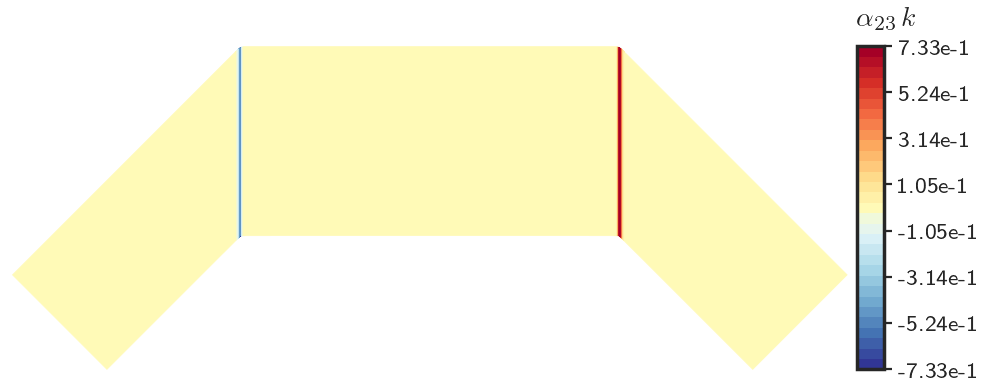}}
		\caption{}
		\label{fig:dw_alpha23_45_poly}
	\end{subfigure}
	\caption{ a) $\alpha_{13}$ b) $\alpha_{23}$ in the polygonized domain with $\mis_0 = \mdeg{45}$.}
	\label{fig:dw_alpha_45_poly}
\end{figure}

\begin{figure}
	\centering
	\includegraphics[width=.495\linewidth]{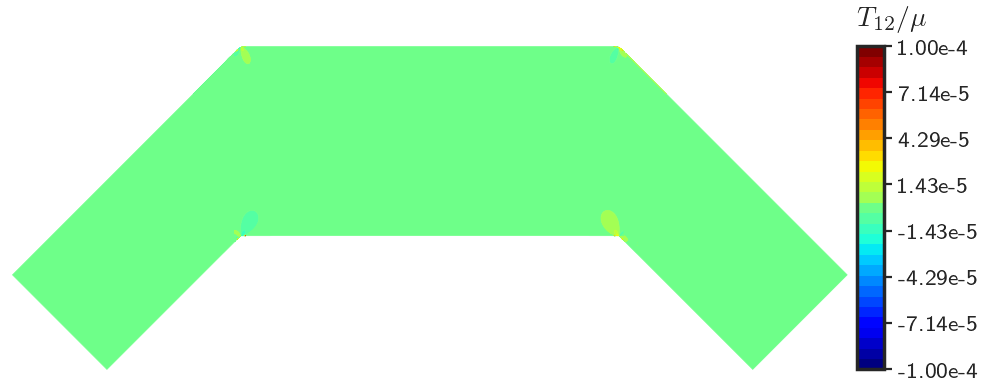}
	\caption{$T_{12}$ in the polygonized domain with $\mis_0 = \mdeg{45}$.}
	\label{fig:dw_stress_fields_45_poly}
\end{figure}

Figure \ref{fig:dw_stress_fields_45_poly} shows the distribution of the $T_{12}$ stress component in the domain, with the dislocation density fields shown in Figure \ref{fig:dw_alpha_45_poly}. As can be seen, the shear stresses are negligible in the entire domain and further decrease upon refinement. To verify that the values of all the components of the stress tensor $\bfT$ are negligible, we define two non-dimensional measures of strain energy density in the domain as follows:
\begin{align}
	\bar{\psi}_{fd} &= \dfrac{1}{2\mu}\bfE^e:\bbC:\bfE^e;  \qquad \bfE^e = \dfrac{1}{2}(\bfF^{eT}\bfF^{e} - \bfI)\label{eq:dw_psi_fd}\\
	\bar{\psi}_{sd} &= \dfrac{1}{2\mu} (\bfF^{e} - \bfI):\bbC:(\bfF^{e} - \bfI). \label{eq:dw_psi_sd}
\end{align}

\begin{figure}
	\centering	
	\begin{minipage}[t]{.48\linewidth}
		\centering
		{\includegraphics[width=\linewidth]{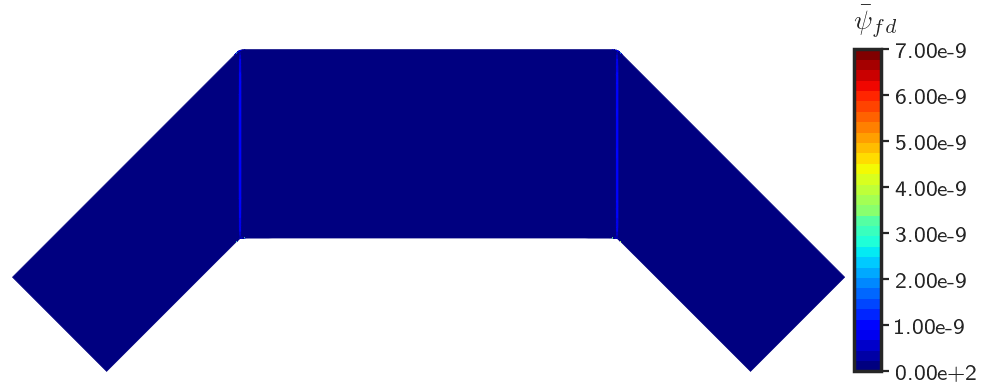}}
		\caption{Finite deformation strain energy density $\bar{\psi}_{fd}$ in the polygonized domain.}
		\label{fig:dw_sefd_45_poly}
	\end{minipage}
	\hfill
	\begin{minipage}[t]{.48\linewidth}	
		\centering
		{\includegraphics[width=\linewidth]{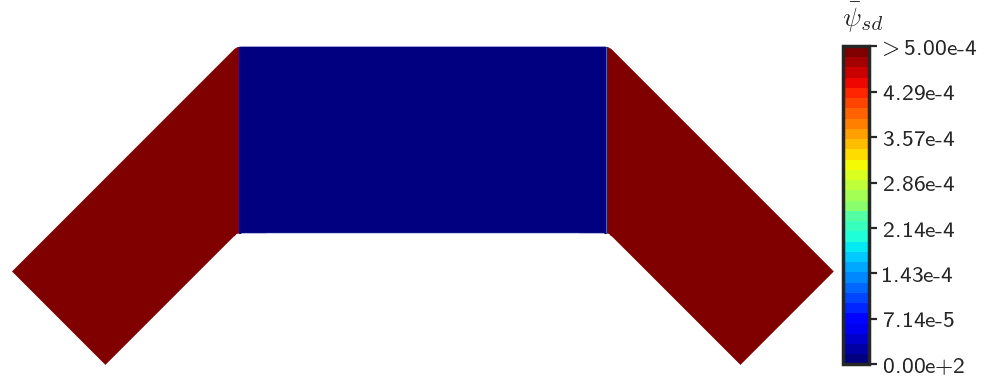}}
		\caption{Small deformation strain energy density $\bar{\psi}_{sd}$ in the polygonized domain. For the region where the normalized energy density $\bar{\psi}_{sd} > 5\times 10^{-4}$, the value is $0.4$.}
		\label{fig:dw_sesd_45_poly}
	\end{minipage}
\end{figure}

\noindent In the above, $\bar{\psi}_{fd}$ and $\bar{\psi}_{sd}$ denote the finite and small deformation non-dimensional strain energy densities, respectively. Figure \ref{fig:dw_sefd_45_poly} shows the distribution of  finite deformation strain energy density  $\bar{\psi}_{fd}$ in the domain. The negligible magnitude of the $\bar{\psi}_{fd}$ distribution in the body demonstrates that the body is stress-free. However, Figure  \ref{fig:dw_sesd_45_poly} demonstrates that the \emph{small deformation theory predicts a non-zero strain energy density profile even when the values of the elastic distortion field is a rotation  tensor everywhere}. Moreover, the expression for the strain energy density for the small deformation (linear) theory is not invariant under superposed rigid body motions.

\begin{figure}
	\centering
	\begin{subfigure}[b]{0.33\linewidth}
		\centering
		{\includegraphics[width=\linewidth]{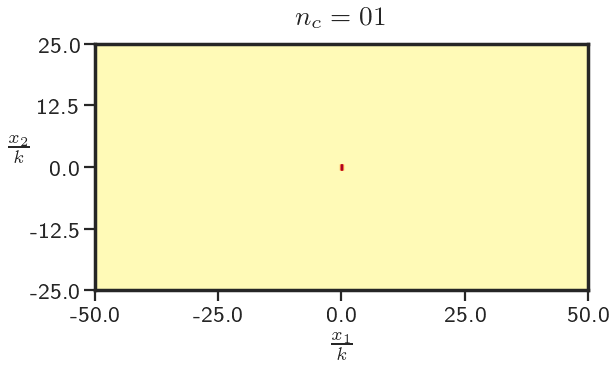}}
	\end{subfigure}%
	\begin{subfigure}[b]{0.33\linewidth}
		\centering
		{\includegraphics[width=\linewidth]{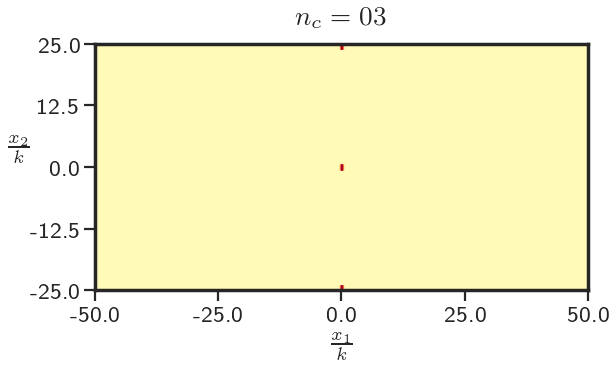}}
	\end{subfigure}
	\begin{subfigure}[b]{0.33\linewidth}
		\centering
		{\includegraphics[width=\linewidth]{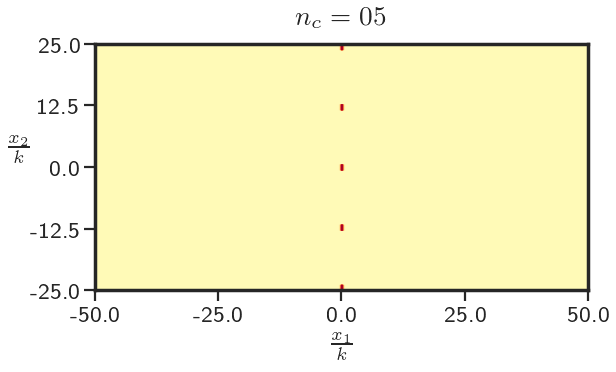}}
	\end{subfigure}\\
	\begin{subfigure}[b]{0.33\linewidth}
		\centering
		{\includegraphics[width=\linewidth]{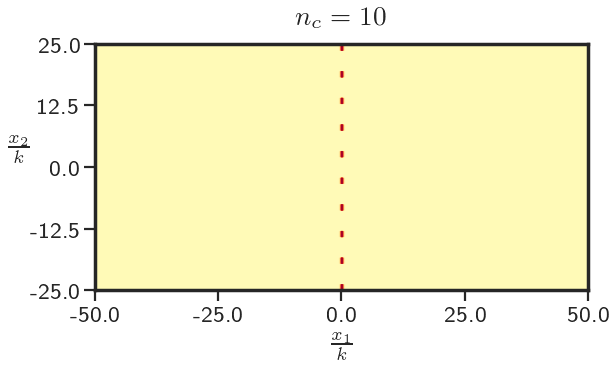}}
	\end{subfigure}%
	\begin{subfigure}[b]{0.33\linewidth}
		\centering
		{\includegraphics[width=\linewidth]{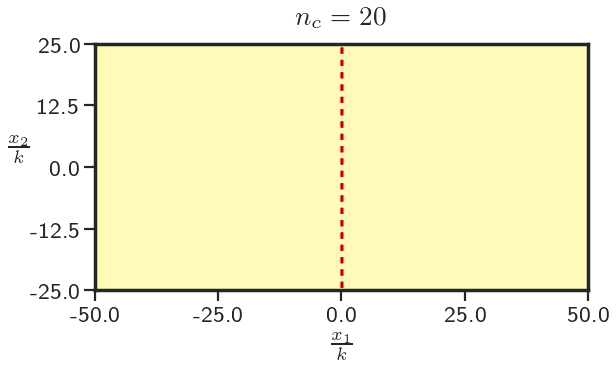}}
	\end{subfigure}
	\begin{subfigure}[b]{0.33\linewidth}
		\centering
		{\includegraphics[width=\linewidth]{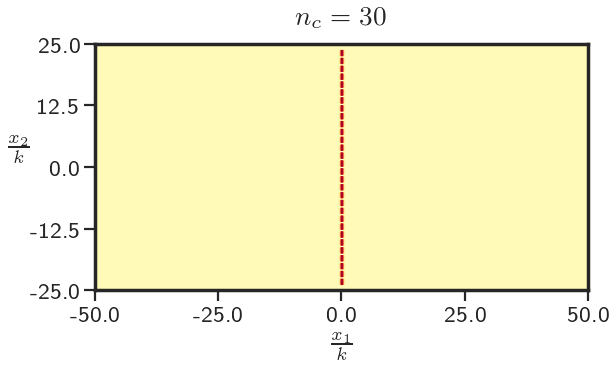}}
	\end{subfigure}\\
	\begin{subfigure}[b]{0.33\linewidth}
		\centering
		{\includegraphics[width=\linewidth]{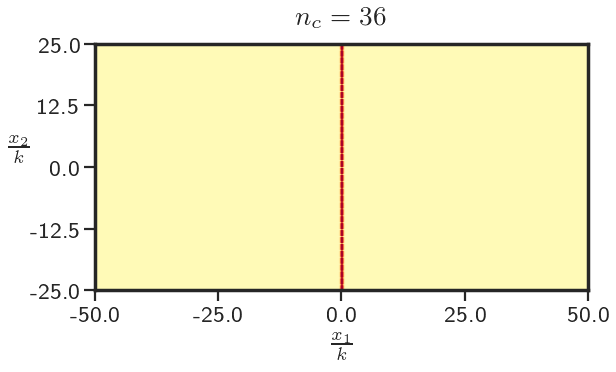}}
	\end{subfigure}%
	\begin{subfigure}[b]{0.33\linewidth}
		\centering
		{\includegraphics[width=\linewidth]{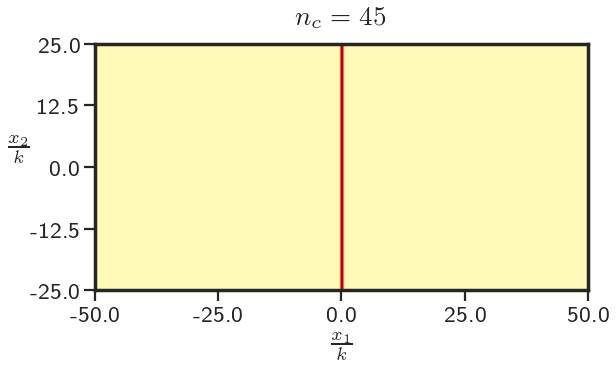}}
	\end{subfigure}%
	\begin{subfigure}[b]{0.33\linewidth}
		\centering
		{\includegraphics[width=\linewidth]{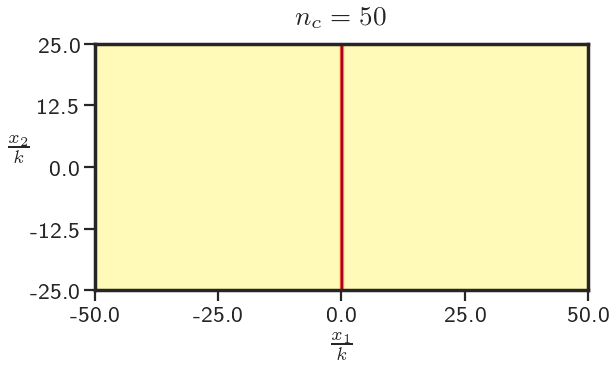}}
	\end{subfigure}\\
	\begin{subfigure}[b]{0.6\linewidth}
		\centering
		{\includegraphics[width=.6\linewidth]{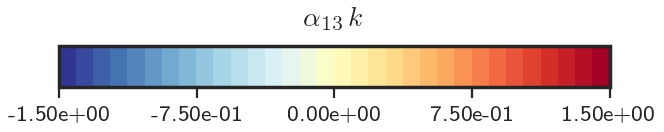}}
	\end{subfigure}
	\caption{Position of dislocation cores comprising the progressively developing dislocation wall.}
	\label{fig:dw_alpha13_45}
\end{figure}

We now demonstrate the stress and energy density field paths induced by a sequence of dislocation distributions, going from one core to a full wall, in a rectangular domain of size $100k \times 50k$. The $\mis(\bfx)$ corresponding to a dislocation wall centered at $x_1 = 0$ is given by 
\begin{align}
	\theta(\bfx) &= \dfrac{\theta_0}{2} \,\left[1 - \,\tanh\left(\frac{x_1}{a\,k}\right)\right],
	\label{fig:dw_rect_alphaeq}
\end{align}
where $\mis_0$ is the rotation angle, equal to the difference in the orientations of the lattice across the wall. Equation \eqref{eq:dw_Fe_alpha} gives the elastic distortion tensor field in the domain. The values of $a$ and $\theta_0$ are chosen to be same as in the previous section, i.e.~$0.238$ and $\mdeg{45}$, respectively. As before, the negative $curl$ of the inverse elastic distortion field gives the dislocation density comprising the full dislocation wall. The dislocation density field corresponding to a single core amounts to isolating the dislocation density distribution in a region of dimension $k \times k$ from the full wall, with value $\bf0$ everywhere else in the domain. The dislocation density corresponding to multiple  cores in a wall configuration can then be prescribed by using the dislocation density field corresponding to the single core with a vertical shift and aligning the so-obtained core with the previous ones. For any given number of dislocation cores in the rectangular body, denoted by $n_c$, the ECDD system is solved to obtain the stress field in the body. The chosen sequence of positions and number (of cores) for the $\alpha_{13}$ component of the dislocation density is shown in Figure \ref{fig:dw_alpha13_45}. $\alpha_{23}$ is similarly placed and its values are accordingly assigned. The Burgers vector for a single dislocation core is computed to be $\bfb = .778k\, \bfe_1 -.322k\, \bfe_2$. A uniform mesh of $(0.1k)^2$ sized elements is used to discretize the rectangular domain for all cases except for the case of $n_c = 50$. For $n_c = 50$, the mesh is  further refined near the dislocation wall.

For the simulations with $n_c \le 30$ the initial guess to the Newton-Raphson method is obtained from the small deformation equilibrium problem  solved on the current configuration  as discussed in \cite{zhang2018finite,arora2019computational}; in solving problems of finite deformation dislocation 
fields with sparsely distributed individual dislocations, this is found to be essential.
However, for the simulations with $n_c \geq 36$, the initial guess is obtained by solving \eqref{eq:fguess_R} (corresponding to a full dislocation wall), and this is also found to be essential to obtain a solution for the stress fields of these `dense' distributions of individual dislocations. For $n_c = 35$, the numerical solution did not converge with initial guess coming from either of the two approaches mentioned above. For $n_c = 32$, a solution, using the initial guess from the small-deformation theory, can be obtained. The same procedure does not succeed for  $n_c = 33, 34$.

\begin{figure}
	\centering
	\begin{subfigure}[b]{0.33\textwidth}
		\centering
		{\includegraphics[width=\linewidth]{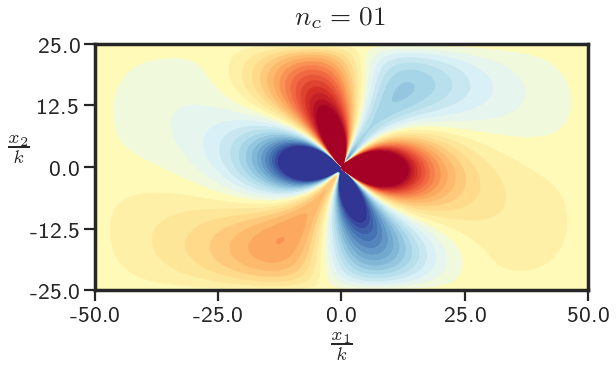}}
	\end{subfigure}%
	\begin{subfigure}[b]{0.33\textwidth}
		\centering
		{\includegraphics[width=\linewidth]{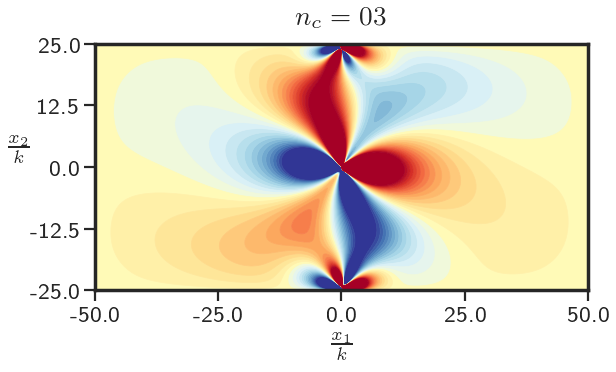}}
	\end{subfigure}%
	\begin{subfigure}[b]{0.33\textwidth}
		\centering
		{\includegraphics[width=\linewidth]{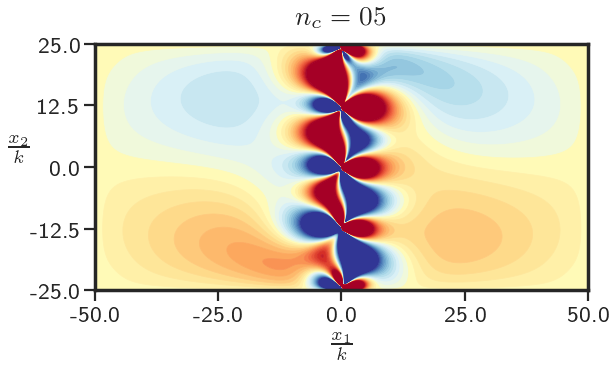}}
	\end{subfigure}\\
	\begin{subfigure}[b]{0.33\textwidth}
		\centering
		{\includegraphics[width=\linewidth]{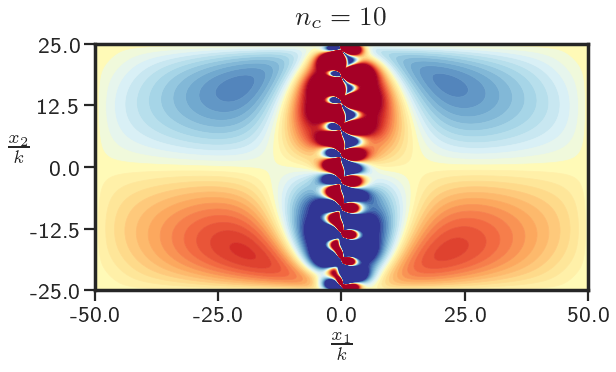}}
	\end{subfigure}%
	\begin{subfigure}[b]{0.33\textwidth}
		\centering
		{\includegraphics[width=\linewidth]{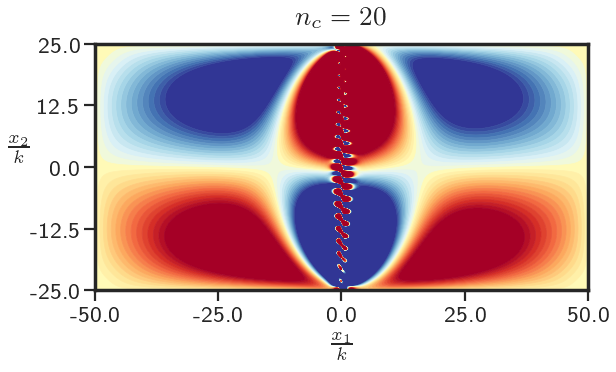}}
	\end{subfigure}%
	\begin{subfigure}[b]{0.33\textwidth}
		\centering
		{\includegraphics[width=\linewidth]{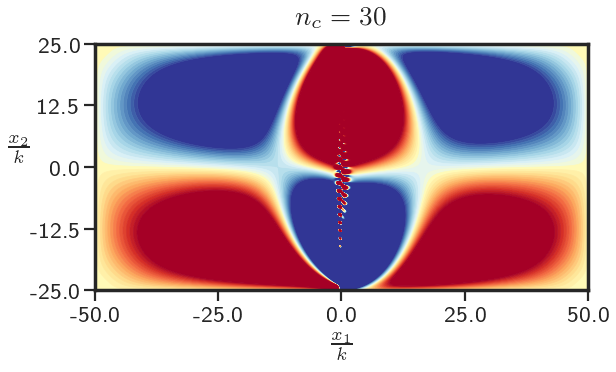}}
	\end{subfigure}\\
	\begin{subfigure}[b]{0.33\textwidth}
		\centering
		{\includegraphics[width=\linewidth]{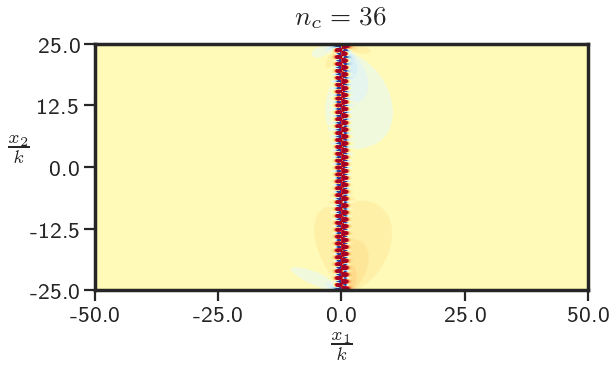}}
	\end{subfigure}%
	\begin{subfigure}[b]{0.33\textwidth}
		\centering
		{\includegraphics[width=\linewidth]{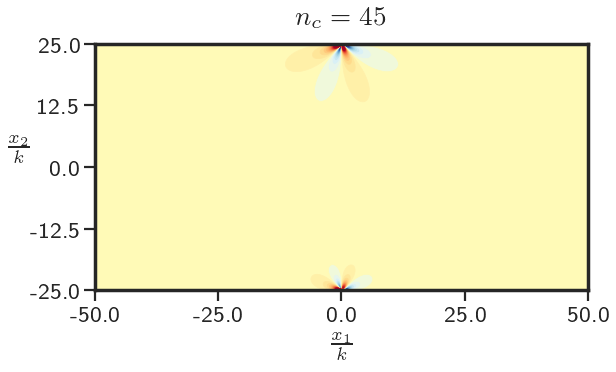}}
	\end{subfigure}%
	\begin{subfigure}[b]{0.33\textwidth}
		\centering
		{\includegraphics[width=\linewidth]{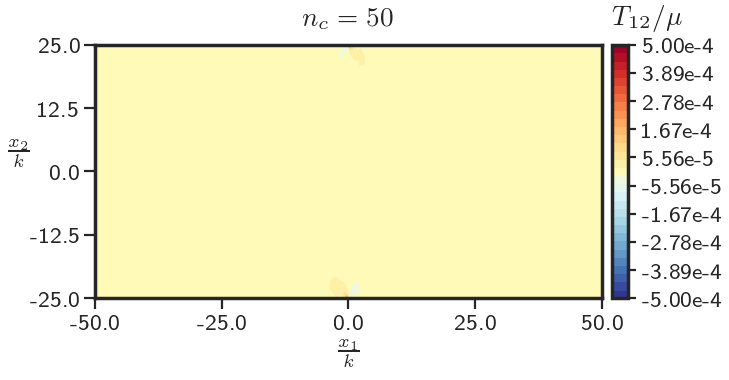}}
	\end{subfigure}\\
	\begin{subfigure}[b]{0.6\textwidth}
		\centering
		{\includegraphics[width=.6\linewidth]{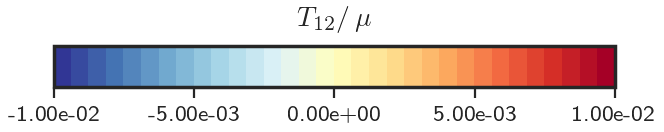}}
	\end{subfigure}%
	
	\caption{$T_{12}$ distribution in the domain for different number of dislocation cores. The colorbar for the cases $n_c < 50$ is shown at the bottom.} 
	\label{fig:dw_stress_fields_45}
\end{figure}

\begin{figure}
	\centering
	\begin{subfigure}[b]{0.33\textwidth}
		\centering
		{\includegraphics[width=\linewidth]{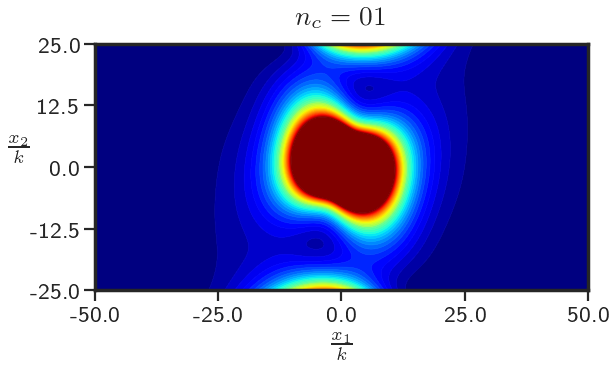}}
	\end{subfigure}%
	\begin{subfigure}[b]{0.33\textwidth}
		\centering
		{\includegraphics[width=\linewidth]{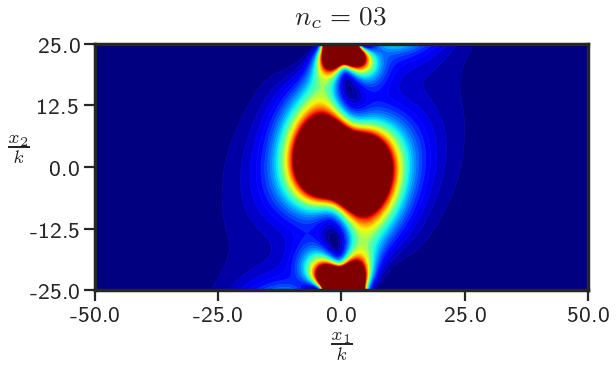}}
	\end{subfigure}%
	\begin{subfigure}[b]{0.33\textwidth}
		\centering
		{\includegraphics[width=\linewidth]{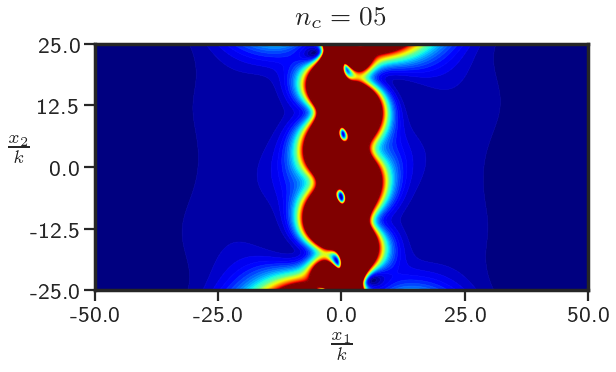}}
	\end{subfigure}\\
	\begin{subfigure}[b]{0.33\textwidth}
		\centering
		{\includegraphics[width=\linewidth]{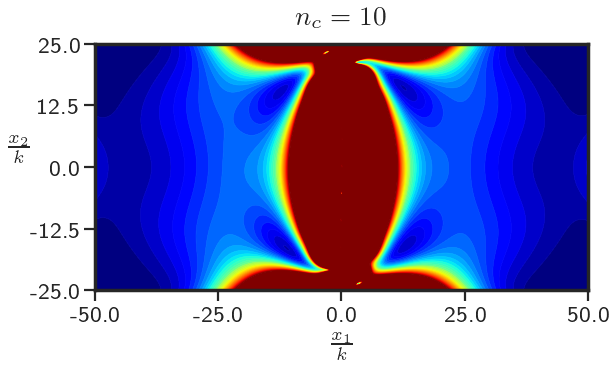}}
	\end{subfigure}%
	\begin{subfigure}[b]{0.33\textwidth}
		\centering
		{\includegraphics[width=\linewidth]{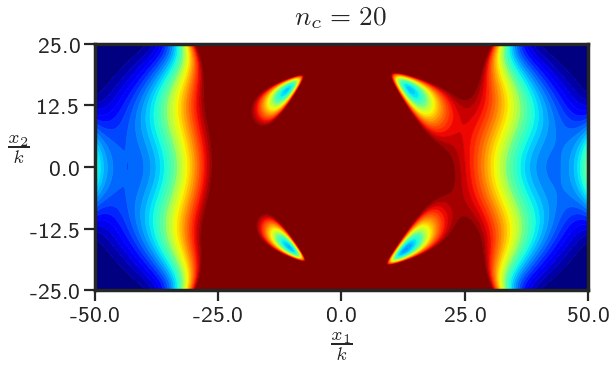}}
	\end{subfigure}%
	\begin{subfigure}[b]{0.33\textwidth}
		\centering
		{\includegraphics[width=\linewidth]{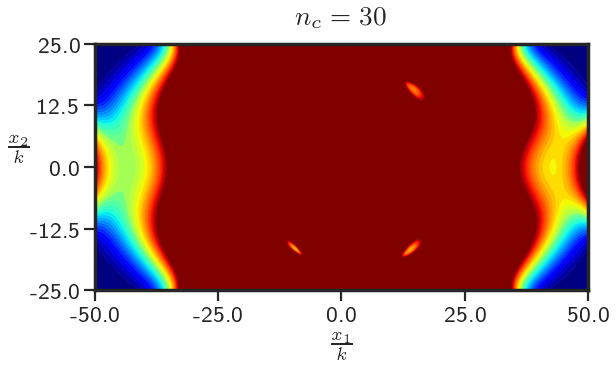}}
	\end{subfigure}\\
	\begin{subfigure}[b]{0.33\textwidth}
		\centering
		{\includegraphics[width=\linewidth]{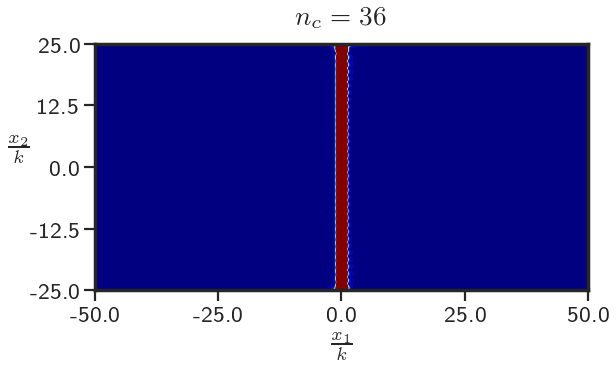}}
	\end{subfigure}%
	\begin{subfigure}[b]{0.33\textwidth}
		\centering
		{\includegraphics[width=\linewidth]{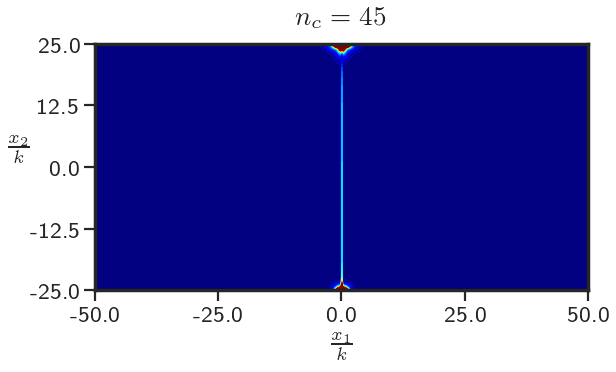}}
	\end{subfigure}%
	\begin{subfigure}[b]{0.33\textwidth}
		\centering
		{\includegraphics[width=\linewidth]{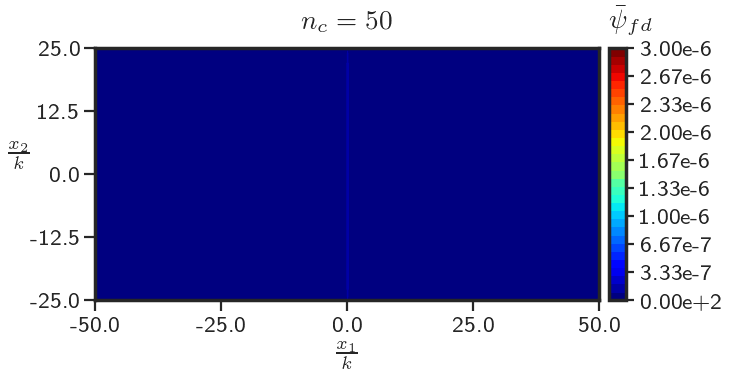}}
	\end{subfigure}\\
	\begin{subfigure}[b]{0.6\linewidth}
	\centering
	{\includegraphics[width=.6\linewidth]{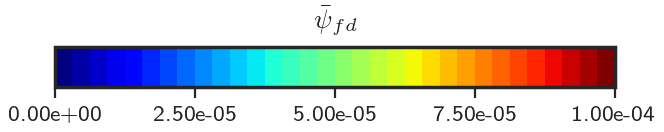}}
\end{subfigure}
	\caption{$\bar{\psi}_{fd}$ distribution in the domain for different number of dislocation cores. The colorbar for the cases $n_c < 50$ is shown at the bottom.}
	\label{fig:dw_sefd_45}
\end{figure}

Figures \ref{fig:dw_stress_fields_45} and \ref{fig:dw_sefd_45} show the distribution of the $T_{12}$ stress component and the finite deformation strain energy density $\bar{\psi}_{fd}$ fields in the body for different number of dislocation cores. An interesting feature of our calculations displayed in Figures \ref{fig:dw_stress_fields_45} and \ref{fig:dw_sefd_45} is that there is a drastic change in the stress and the strain energy distributions when the number of cores $n_c \ge 36$. The variation (for both $T_{12}$ and $\overline{\psi}_{fd}$) which was spread out in the entire domain suddenly becomes localized near the wall and the boundary when $n_c \ge 36$. It is clear that when the number of (same sign) dislocations in the body increases from 0, the stress and energy have to increase. It is also clear that the full wall must represent a stress and energy-free configuration. Whether the transition between these behaviors has to be an abrupt `phase transition' (w.r.t. the number of cores), as indicated by our calculations, is an interesting question for further study. When $\theta_0$ is reduced, the magnitudes for the stress and energy fields become less pronounced with the qualitative conclusions remaining the same.

\begin{figure}
	\centering
		{\includegraphics[width=.50\linewidth]{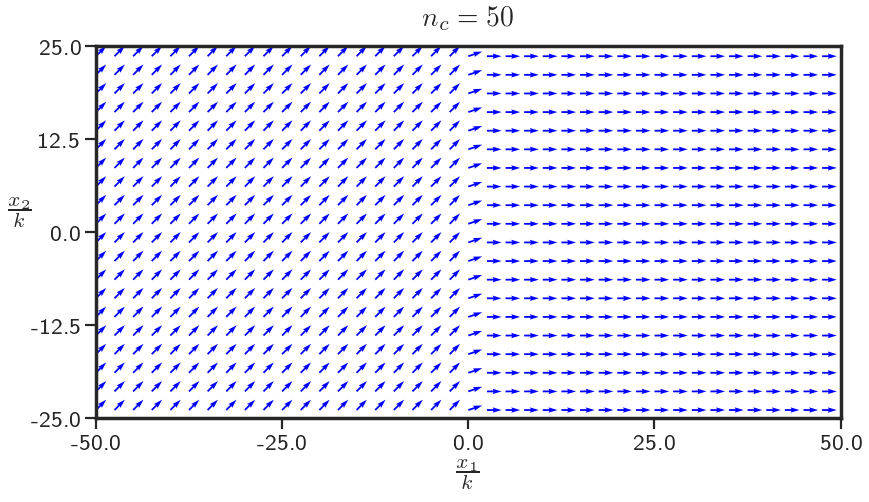}}
		\caption{Mapping of $\bfe_1$ vector by the calculated elastic distortion field $\bfF^e$.}
		\label{fig:dw_Fee1_45}
\end{figure}

Figure \ref{fig:dw_Fee1_45} shows the image of the constant $\bfe_1$ vector field under mapping by the elastic distortion field on the current configuration for the full dislocation wall.  We can see that the lattice on the left of the dislocation wall is rigidly rotated w.r.t.~the lattice on the right by the prescribed misorientation angle $\theta_0$. We also calculate the change in volume  (per unit length of the domain in the $\bfe_3$ direction, i.e. the change in area) of the body due to the presence of the dislocation wall as described in Sec.~\ref{sec:res_vc}. The $\%$ change in volume evaluates to $0$ up to machine precision, a necessary condition for the elastic distortion field to be a rotation tensor field (which is spatially inhomogeneous in this instance).

\subsubsection{Contact with the mathematical literature}
In closing this section, we make contact with the mathematical results of \cite{DW:muller, ginsterb}. A summary of the main result of \cite{DW:muller} adequate for the present context is as follows: Given a frame-indifferent elastic energy density function $W$ and \emph{$\veps$ a nondimensional measure of the magnitude of the Burgers vector of a  dislocation} proportional to a lattice constant of a crystal, 
\begin{equation}\label{MSZ}
 \  \frac{1}{\veps^2 |\log \veps|^2} \int_\Omega W(\bfF^e_\veps) \, dV \longrightarrow \half \int_\Omega \bfbeta : \mathbb{C}\, \bfbeta \, dV + \int_\Omega \varphi(\bfR^e, \bfmu) \, dV,
\end{equation}
where $\bfF^e_\veps$ is a sequence of elastic distortions consistent with a corresponding sequence of discrete dislocation distributions in the body given by $\bfmu_\veps$ through $curl \, \bfF^e_\veps = \bfmu_\veps$, 
 $\bfmu_\veps \to \bfmu$, there exists a sequence of \emph{spatially constant rotation fields} $\bfR^e_\veps \to \bfR^e$ such that $\bfR^{eT}_\veps \bfF^e_\veps - \bfI \to \bfbeta$ with $|\bfR^{eT}_\veps \bfF^e_\veps - \bfI| = \mathcal{O}(\veps |\log\veps|)$, 
 and $curl \, \bfbeta = \bfR^{eT} \bfmu$, $\mathbb{C}$ is the standard linear elastic moduli (with minor (and major) symmetries), $\varphi$ is a function that is deduced, and all arrows represent appropriate (scaled) convergences as $\veps \to 0$. Furthermore, an important hypothesis, for this discussion, on the `well-separated'ness of admissible discrete dislocation distributions is that for any two dislocations comprising $\bfmu_\veps$ located at $\bfx$ and $\bfy$, $|\bfx - \bfy| > 2\rho_\veps$, where ``$\rho_\veps \gg \veps$, with $\rho_\veps \to 0$ as $\veps \to 0$.'' That the sequence of rotation tensors $\bfR^e_\veps$ with limit $\bfR^e$ exists for the sequence $\bfF^e_\veps$ (consistent with admissible $\bfmu_\veps$) is predicated on the assumption that the energy of the sequence as $\veps \to 0$ is bounded, i.e.,
\begin{equation}\label{scaling}
\sup_{\veps \to 0}\  \frac{1}{\veps^2 |\log \veps|^2} \int_\Omega W(\bfF^e_\veps) \, dV  <  \infty.
\end{equation}

Clearly, for the case of polygonization involving a single dislocation wall as we have considered, a representative sequence $\bfF^e_\veps$ would have to converge to a rotation field that is not spatially uniform. Consequently, such a sequence would not be within the purview of the result of \cite{DW:muller}, since the $\bfbeta$ for this sequence is neither skew-symmetric nor $\bfzero$ so that the first term on the RHS of \eqref{MSZ} does not vanish whereas the limit of the LHS of \eqref{MSZ} does for this sequence. Thus, this failure has to be related to the assumptions behind the \cite{DW:muller} analysis. Since such hypotheses, and assumptions of the same ilk, are commonplace in the mathematical literature on dislocations \cite{Garroni_etal, ginstera, ginsterb, lauteri} and has begun to find prominence in the mechanics/engineering literature as well \cite{reina_etal_2016, reina_etal_2018}, we discuss them here and provide our perspective on the matter.
\begin{figure}
\centering
{\includegraphics[width=0.7\linewidth]{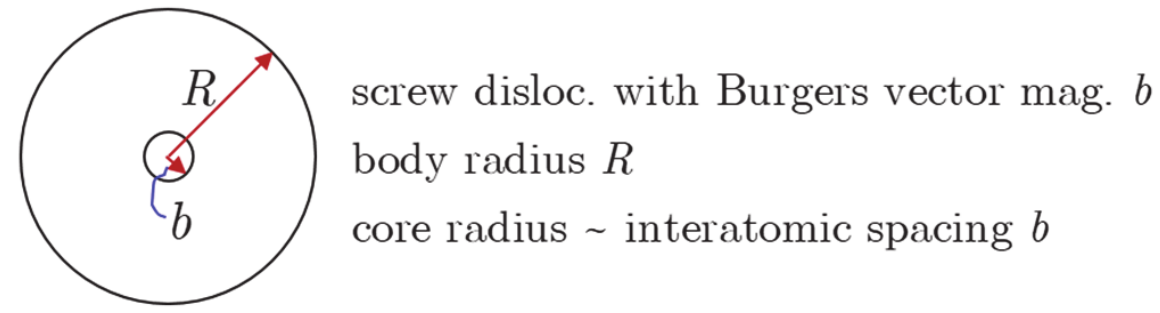}}
	\caption{Physical length scales for a single dislocation in a finite body}
\label{fig:disloc_scaling}
\end{figure}

First, we clarify an issue that is very rarely addressed in the literature (cf.~\cite{reina_etal_2016, reina_etal_2018}), which, nevertheless, is crucial for understanding the physical setting of the mathematical results. This relates to the use of a fixed, bounded configuration containing dislocations whose strengths are assumed to scale with $\veps \to 0$. With reference to Fig.~\ref{fig:disloc_scaling}, we first note that the total energy of the cylinder of length $L$ with a single screw dislocation is given by $E = \int_0^L \int_0^{2\pi} \int_b^R \mu b^2 r^{-2}\, r dr d\theta dl$, and with 
the definition $\veps = bR^{-1}$,
\[
\frac{E}{\mu 2 \pi L} = R^2 \left(\frac{b}{R}\right)^2 |\log \veps|.
\] 
The energy content of the body can be interpreted in two different non-dimensional limits as $\veps \to 0$, corresponding to the expressions
\[
\frac{E}{\mu 2 \pi L R^2} = \veps^2 |\log \veps| \qquad \mbox{and} \qquad \frac{E}{\mu 2 \pi L b^2} = |\log \veps|.
\]
The first
 considers the body to be of fixed radius $R$ with the interatomic spacing of the crystal $b \to 0$ in which case the \emph{limiting total energy of a dislocation in the body vanishes}. With this physical interpretation of $\veps \to 0$, the $\veps^2 |\log \veps|^2$ scaling employed by \cite{DW:muller} may be interpreted, but not necessarily, to correspond to a `weakly unbounded' population of dislocations in the body growing as $|\log \veps|$ as $\veps \to 0$ \cite{Garroni_etal, ginstera, ginsterb} (we discuss a different physical interpretation of this `fixed body scaling' in the next paragraph).
We note, however, that for a physically realistic, non-degenerate scenario where a single dislocation in a body does not result in 0 internal stress and elastic strain energy regardless of its size, the interatomic spacing of a crystal is a finite, well-defined physical length for \emph{a specific} material, as are the dimensions of a body, so choosing $\veps = \frac{b}{R} \to 0$ with $R$ fixed does not appear to be a viable proposition to us (when considering a fixed material). 
 
On the other hand, keeping $b$ fixed and sending $R \to \infty$ seems eminently reasonable (the limit of a progressively large body of the same material), in which case the total energy of the body containing the single dislocation diverges as $|\log \veps| \to  \infty$ as $\veps \to 0$. We believe that it is this physically realistic scaling that should be employed in the analysis of dislocations and their distributions, with the minimum energy scale being $\sup_{\veps \to 0} \frac{1}{b^2 |\log \veps|}\int_{\Omega_\veps} W(\widetilde{\bfF^e_\veps})\, dV < \infty$ (the domain depends on $\veps$ as it grows in size scaled by $b\veps^{-1}$ and we refer to fields on the growing domains with an overhead~ $\widetilde{ }$~). By the nondimensionalization given by $\bfx = \frac{\widetilde{\bfx} \veps}{b}$ with the corresponding domain of \emph{fixed} size represented by $\Omega$, it can be seen that $\int_{\Omega_\veps} W(\widetilde{\bfF^e_\veps})\, d\widetilde{V} = \frac{b^2}{\veps^2} \int_{\Omega} W(\bfF^e_\veps)\, dV$, with  $\bfF^e_\veps (\bfx) = \widetilde{\bfF^e_\veps}( b\veps^{-1} \bfx)$, $\bfmu_\veps (\bfx) = \widetilde{\bfmu_\veps}( b\veps^{-1} \bfx)$, $curl \bfF^e_\veps (\bfx) = \veps^{-1} \bfmu_\veps (\bfx)$\footnote{It is presumably based on this correspondence between the non-dimensionalized problem on the fixed domain and the growing domain problem that \cite{MSZ2,ginsterPC} state that ``From the point of view of physics it is more natural to fix the lattice spacing and to consider domains $\frac{1}{\veps} \Omega$ of increasing size. Upon elasticity scaling both points of view are equivalent and
fixing $\Omega$ rather than the lattice spacing is more convenient for the analysis. Thus $\veps$ really is a dimensionless parameter of the order of lattice spacing divided by the macroscopic dimension of the body. For brevity we will nonetheless often refer to $\veps$ as the lattice spacing.''}. It is in this sense that we interpret all `fixed domain' dislocation-related asymptotic results with discrete dislocation strengths tending to 0, noting, in particular, the large difference between $\int_{\Omega} W(\bfF^e_\veps)\, dV$ and the physical energy content $\int_{\Omega_\veps} W(\widetilde{\bfF^e_\veps})\, d\widetilde{V}$ as $\veps \to 0$.
 
Consider now a `full' dislocation wall described by a piecewise constant rotation field (as constructed in our computational example, but now with a sharp boundary) in a sequence of square domains of size $H \times H$, with $H$ steadily increasing, representing a progressively large domain containing a bicrystal. The magnitude of the Burgers vector of each individual dislocation in the wall is assumed to be $b$. Then the result of the case $\frac{1}{\veps} = \frac{H}{b} \gg 1$ may be approximated by the results of $\frac{H}{b} \to \infty$, \emph{$b$ fixed}. At least within the mathematical context being considered, the entire sequence of domains has $0$ elastic energy (i.e.~the part coming from $W$), which is certainly $\mathcal{O}$ of any of the energy bounds $\veps^2 |\log \veps|$ or  $\veps^2 |\log \veps|^2$ as $\veps \to 0$. Moreover, it can be checked that, on the non-dimensional domain of fixed size, the strength of the Burgers vector of individual dislocations indeed scales like $\veps b$, if $b$ is the magnitude of the Burgers vector of the dislocations on the physical sequence of domains. \emph{Why then does the elastic part of the limit energy in \eqref{MSZ} fail to predict the energy content of such a wall, a commonly observed dislocation configuration, well worthy of prediction?} The answer must lie in the fact that the `full wall,' while being a very low-energy configuration, cannot be achieved as a limit configuration of the admissible dislocation sequences allowed by the \cite{DW:muller} analysis due to the `well-separated'ness assumption -  indeed, it is, in a sense, an `opposite' limit that is considered in our computational example where the core dimensions remains fixed and the inter-core distances reduce to 0. An interesting feature of this specific example is that the number of dislocations in a configuration (when the core radius tends to 0) does not correlate, even roughly, with the total energy content of the configuration, as might be expected based on energy scales related only to self-energy of single dislocations, an argument valid in relatively `dilute' limits \cite{Garroni_etal, ginstera, ginsterb}. Indeed, the defining energetic feature of the `full wall' is that the entire energy of the individual dislocations in the wall gets screened by the `(nonquadratic) interaction energy' between them resulting in a very low-energy state.

Unlike Equation (2.4), Proposition 4.3, and Theorem 4.6 of \cite{DW:muller}, the Generalized Rigidity Estimate (GRE) \cite[Theorem 3.3]{DW:muller} does apply to the `full wall' configuration under discussion. It is instructive to understand the breakdown of that result in validating the elastic part of the limit functional in \eqref{MSZ} for the specific example of the full dislocation wall.
As mentioned earlier in conjunction with \eqref{MSZ}, for \emph{admissible} sequences $(\bfmu_\veps, \bfF^e_\veps)$ satisfying the energy bound \eqref{scaling}, $|\bfR^{eT}_\veps \bfF^e_\veps - \bfI|$ is small so that a quadratic approximation of $W$ estimates it well and it is that quadratic approximation of $W$ with argument $\bfbeta$ that appears in \eqref{MSZ}. For the `wall sequence' (on the growing domains, and recalling that $\frac{b}{H} =: \veps$) with constant unit tangent to the wall in the the direction $\bfe_2$,  $|\widetilde{\bfU^e_\veps} - \bfI|$ is small (actually 0), where $\widetilde{\bfU^e_\veps}$ is the right stretch tensor of the polar decomposition of $\widetilde{\bfF^e_\veps}$ and the question becomes as to whether the spatially non-uniform rotation tensor field of $\widetilde{\bfF^e_\veps}$ can be approximated well by a \emph{constant} rotation $\widetilde{\bfR^e_\veps}$ in the domain $\Omega_\veps$ so that $\widetilde{\bfU^e_\veps}$ can be well approximated by $\widetilde{\bfR^{e}_\veps}^T \widetilde{\bfF^e_\veps}$. We are unable to deduce the necessary control on the point-wise values of  $\widetilde{\bfR^e_\veps} - \widetilde{\bfF^e_\veps}$ from the GRE, given here, for each fixed $\veps$, by
$$
\left|\left|\widetilde{\bfF^e_\veps} - \widetilde{\bfR^e_\veps}\right|\right|_{L^2(\Omega_\veps;  \mathbb{R}^{2 \times2})} \leq C(\Omega_\veps) \left( \left|\left|\mbox{dist}\left(\widetilde{\bfF^e_\veps}, SO(2)\right)\right|\right|_{L^2(\Omega_\veps)} + \left|\widetilde{curl} \, \widetilde{\bfF_\veps}\right|(\Omega_\veps)  \right)
$$
since, although the $\widetilde{\bfF^e_\veps}$ is a rotation tensor field for this specific sequence taking exactly two distinct values, say $\widetilde{\bfR^e_{\veps}}_1 \neq \widetilde{\bfR^e_{\veps}}_2$, the various ingredients of the GRE in this specific example are given by
\begin{align}
& \left|\left|\widetilde{\bfF^e_\veps} - \widetilde{\bfR^e_\veps}\right|\right|_{L^2(\Omega_\veps; \mathbb{R}^{2 \times2})}  = \frac{H}{\sqrt{2}} \sqrt{ \left|\widetilde{\bfR^e_{\veps}}_1 – \widetilde{\bfR^e_\veps}\right|^2 + \left|\widetilde{\bfR^e_{\veps}}_2 – \widetilde{\bfR^e_\veps}\right|^2 } \neq 0 \nonumber\\
& \left|\left|\mbox{dist}\left(\widetilde{\bfF^e_\veps}, SO(2)\right)\right|\right|_{L^2(\Omega_\veps)}  = 0 \nonumber\\
& \left|\widetilde{curl} \, \widetilde{\bfF_\veps}\right|(\Omega_\veps) = H \left|\left(\widetilde{\bfR^e_{\veps}}_1 - \widetilde{\bfR^e_{\veps}}_2\right) \bfe_2\right|, \nonumber
\end{align}
and the constant $C(\Omega_\veps)$ depends on the domain; for the corresponding problem on the fixed-domain, the point-wise squared magnitude of the difference  $\bfR^e_\veps - \bfF^e_\veps$ is therefore bounded at most by an $\mathcal{O}(1)$ quantity independent of $\veps$ (it is physically obvious that no theorem can prove that the rotation field of a high-angle, symmetric tilt boundary can be well-approximated by a constant rotation everywhere). In case such an approximation were to actually fail, then a `quadratization' of $W$ (about $\bfzero$) is not a good approximation for it, and the limit functional in \eqref{MSZ} cannot be the correct one for this wall sequence. Predictions of such a model would be similar in spirit to what is shown, e.g., in Fig.~\ref{fig:dw_sesd_45_poly}. Interestingly, however, the limit elastic functional in \eqref{MSZ} is invariant under superposed rigid deformations.

The above arguments also seem to suggest that for small enough energy scales a better target for the elastic part of the limit functional is $\frac{1}{2}\int_\Omega (\bfU^e - \bfI): \mathbb{C}(\bfU^e - \bfI)\, dV$, where $\bfU^e_\veps \to \bfU^e$ as $\veps \to 0$, which is invariant under superposed rigid deformation as well as succeeds for the case of the dislocation wall. However, it should be noted that in the presence of large `infinities' of dislocations of sufficient strength, control on the energy (and hence the elastic strain field) does not control the rotation field, this being expected since such control is a hallmark of \emph{compatibility} of deformations, as in (non)linear elasticity theory without line defects.

Finally, in the context of energy scales of dislocation configurations for asymptotic analysis, the highest energy scales that have been considered, to our knowledge, are
\begin{equation*}
\sup_{\veps \to 0} \frac{1}{ \veps}\int_{\Omega} W(\bfF^e_\veps)\, dV < \infty \  \cite{lauteri} \qquad \mbox{and} \qquad  \sup_{\veps \to 0}  \int_{\Omega} W(\bfF^e_\veps)\, dV < \infty \ \cite{reina_etal_2016, reina_etal_2018}\footnote{To appreciate the difference between the energy scales implied by the various scalings considered here (constant, $\veps$, 
$\veps^2|\log\veps|$, $\veps^2|\log\veps|^2$), it is instructive to choose the value $\veps = 10^{-10}$ corresponding to an Angstrom scale interatomic spacing in a body of nominal dimension $1m.$};
\end{equation*}
in both cases, no limit energy functional is deduced as in the other references mentioned, and no  (approximate) methods for computing stress fields of the dislocation distributions are devised. 

\subsection{Volume change due to dislocations}
\label{sec:res_vc} 
A fundamental question that was asked by Toupin and Rivlin \cite{toupin1960dimensional} concerns the change in volume of a body when dislocations are introduced; linear elastic theory is not capable of capturing this volume change, due to the fact that the average value of each of the infinitesimal strain components of a self-equilibrated stress field in a body vanishes. However, it has been observed experimentally that the volume of the body changes upon the introduction of dislocations \cite{vc_zener} and the prediction by linear elastic theory (of no volume change) does not seem to be in agreement with experimental observations. Toupin and Rivlin \cite{toupin1960dimensional} used  a second-order approximation of nonlinear elasticity to give explicit expressions for the changes in average dimensions of elastic bodies resulting from the introduction of dislocations.
 
Here, we use finite deformation FDM to capture this volume change. The problem is set up in a $2$-d plane strain setting as follows: Edge dislocations are assumed to be present in a rectangular body of dimensions $[-10b, 10b] \times [-10b, 10b]$.  An edge dislocation, with core centered at point $\bfp = (p_1, p_2)$, is modeled by prescribing a dislocation density of the form 
\begin{align}
	\alpha_{13}(x_1, x_2)  = \begin{cases}
		\phi_0 & |x_1 - p_1| \le \frac{b}{2} ~\text{and}~ |x_2 - p_2| \le \frac{b}{2}\\
		0 & ~\text{otherwise} \\
	\end{cases},
	~~\alpha_{ij} = 0 ~\text{if}~ i\neq 1 ~\text{and}~ j \neq 3.
\end{align}

The constant $\phi_0$ is evaluated by making the
	Burgers vector of the dislocation core equal to $b\bfe_1$, i.e. $\int_{\mOmega} \alpha_{13} da = b$. The external boundaries are assumed to be traction-free. The body is assumed to behave as an isotropic Saint-Venant-Kirchhoff elastic material with $E = 200$ GPa and $\nu = 0.3$.

The volume change is calculated as follows: The ECDD system \eqref{eq:ECDD}-\eqref{eq:ECDD_bc} is solved on the current configuration to obtain the inverse elastic distortion field for a given dislocation density in the domain.  The $\%$ volume change is then calculated from \eqref{eq:vc_all} where $V_{ref}$ and $V_{curr}$ denote the volume of the reference and the current configurations, respectively:
\begin{subequations}
	\begin{align}
		{V_{curr}} = \int_{\mOmega}\, dV ~~;&~~
		{V_{ref}} = \int_{\mOmega} det(\bfW) \, dV \label{eq:vc_vcurrref}\\
		\Delta V = V_{ref} - V_{curr}~~;& ~~
		\% \Delta V = \frac{|\Delta V|}{V_{curr}} \times 100. \label{eq:vc_pcnt_vc}
	\end{align}
	\label{eq:vc_all}
\end{subequations} 

\begin{figure}
	\centering
	\begin{subfigure}[b]{0.15\linewidth}
		\centering
		{\includegraphics[width=\linewidth]{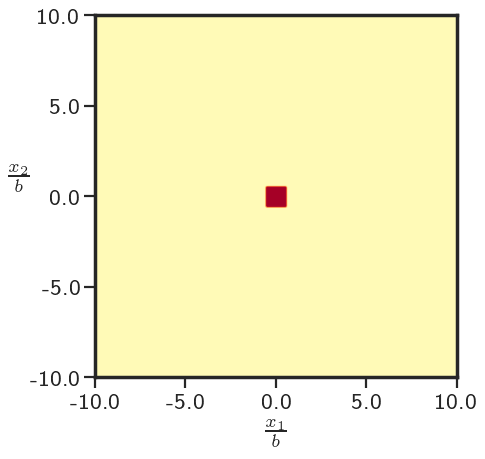}}
		\caption{}
		\label{fig:vc_pbv_alpha_1}
	\end{subfigure}%
	\begin{subfigure}[b]{0.15\linewidth}
		\centering
		{\includegraphics[width=\linewidth]{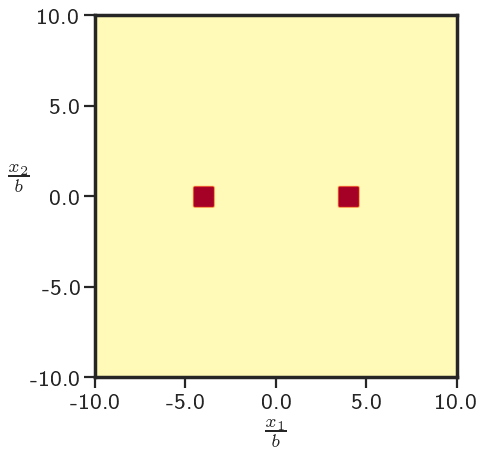}}
		\caption{}
	\end{subfigure}%
	\begin{subfigure}[b]{0.15\linewidth}
		\centering
		{\includegraphics[width=\linewidth]{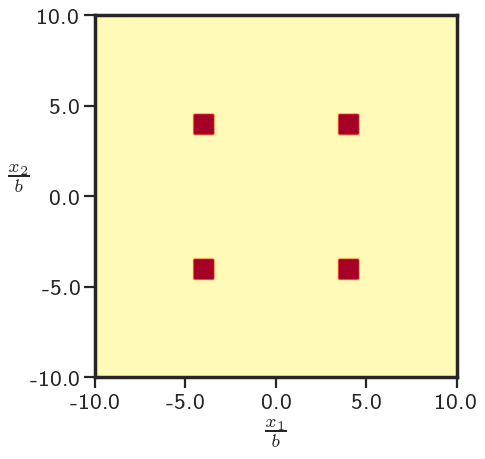}}
		\caption{}
	\end{subfigure}%
	\begin{subfigure}[b]{0.15\linewidth}
		\centering
		{\includegraphics[width=\linewidth]{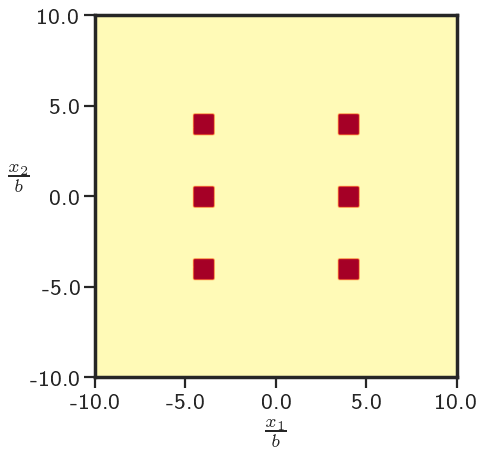}}
		\caption{}
	\end{subfigure}%
	\begin{subfigure}[b]{0.15\linewidth}
		\centering
		{\includegraphics[width=\linewidth]{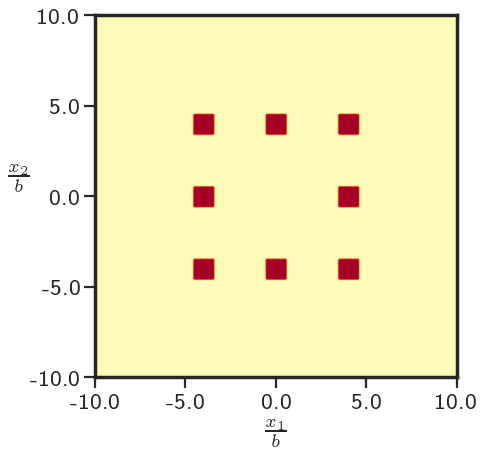}}
		\caption{}
		\label{fig:vc_pbv_alpha_8}
	\end{subfigure}\\
	\begin{subfigure}[b]{0.15\linewidth}
		\centering
		{\includegraphics[width=\linewidth]{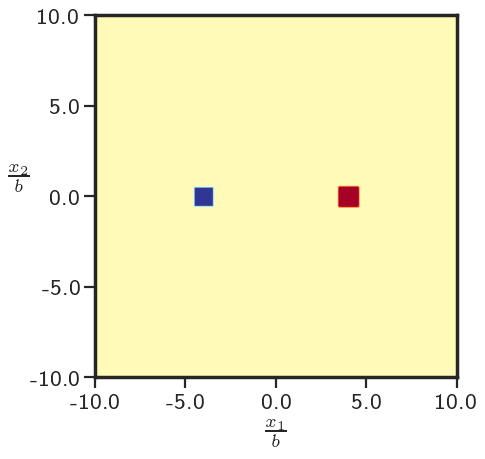}}
		\caption{}    
		\label{fig:vc_zbv_alpha_2}
	\end{subfigure}%
	\begin{subfigure}[b]{0.15\linewidth}
		\centering
		{\includegraphics[width=\linewidth]{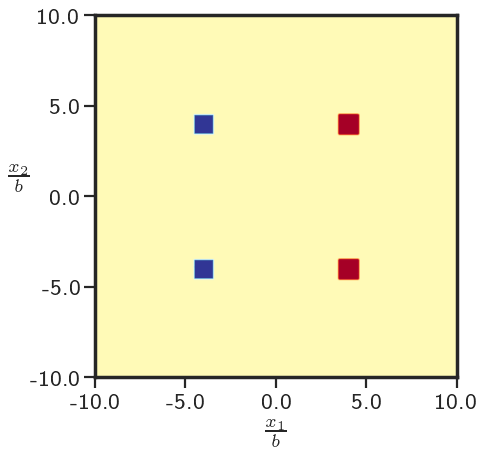}}
		\caption{}
	\end{subfigure}%
	\begin{subfigure}[b]{0.15\linewidth}
		\centering
		{\includegraphics[width=\linewidth]{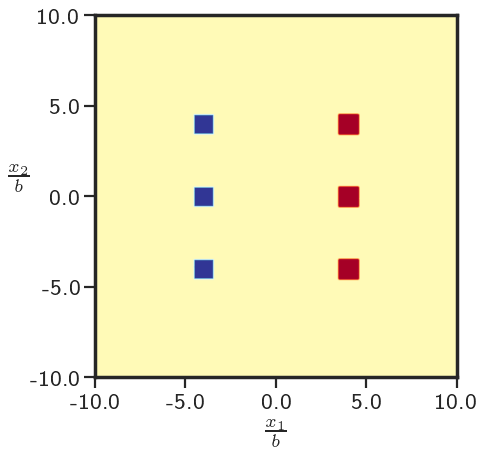}}
		\caption{}    
	\end{subfigure}%
	\begin{subfigure}[b]{0.15\linewidth}
		\centering
		{\includegraphics[width=\linewidth]{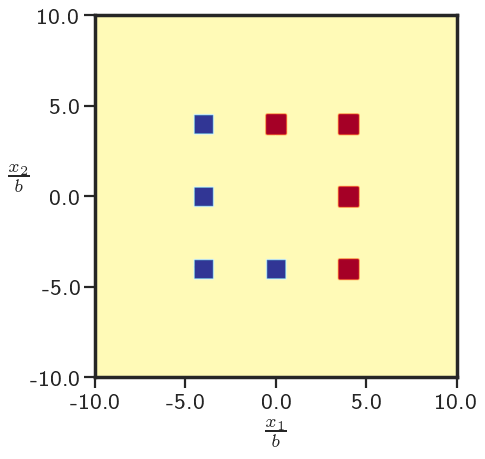}}
		\caption{}    
		\label{fig:vc_zbv_alpha_8}
	\end{subfigure}%
	\begin{subfigure}[b]{0.15\linewidth}
		\centering
		{\includegraphics[scale=.55]{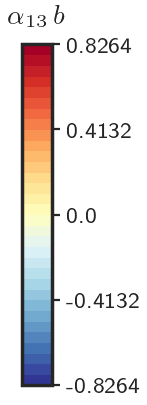}}
	\end{subfigure}
	\caption{Figures a-e show the $\bfalpha$ distributions for the case when the total strength of all dislocations is positive: a) $1$ core	b) $2$ cores c) $4$ cores d) $6$ cores e) $8$ cores. Figures f-i show the $\bfalpha$ distributions for the case when the total Burgers vector is zero: f) $2$ cores
		g) $4$ cores h) $6$ cores i) $8$ cores.  The legend colorbar is common to all the plots.}
	\label{fig:vc_bv_alpha}
\end{figure}
 
In the following, we present the volume change (per unit length along the $\bfe_3$ axis) approximation for two cases corresponding to the total Burgers vector of all the dislocations being a) positive and b) zero.  Table \ref{tab:vc_pbv_vc} shows the $\%$ change in the volume of the body consequent upon the introduction of multiple dislocations of the same sign, distributed in the body  as shown in Figures \ref{fig:vc_pbv_alpha_1} - \ref{fig:vc_pbv_alpha_8}. Table \ref{tab:vc_zbv_vc} shows the $\%$ change in the volume of the body due to the introduction of multiple pairs of dislocations, distributed in the body as shown in Figures \ref{fig:vc_zbv_alpha_2} - \ref{fig:vc_zbv_alpha_8}, such that the total Burgers vector of all dislocations is $\bf0$. The configurations for zero resultant Burgers vector utilize positive and negative straight edge dislocation.

\begin{table}
	\begin{subtable}{0.45\textwidth}
		\centering
		\begin{tabular}{ | c | c | }
			\hline
			Cores  &   $\% \Delta V$  \\ \hline 
			$1$ &   $.1019$ \\ \hline
			$2$ &   $.2064$ \\ \hline
			$4$ &   $.3013$ \\ \hline
			$6$ &   $.3902$ \\ \hline
			$8$ &   $.6025$ \\ \hline        
		\end{tabular}
		\caption{}
		\label{tab:vc_pbv_vc}
	\end{subtable}
	\hfill
	\begin{subtable}{0.45\textwidth}
		\centering
		\begin{tabular}{ | c | c | }
			\hline
			Cores  &   $\% \Delta V$  \\ \hline
			$2$ &   $.1793$ \\ \hline
			$4$ &   $.2938$ \\ \hline
			$6$ &   $.3523$ \\ \hline
			$8$ &   $.5873$ \\ \hline
		\end{tabular}
		\caption{}
		\label{tab:vc_zbv_vc}
	\end{subtable}
	\caption{Volume change in body when the resultant strength of all the dislocations is a) positive b) zero.}
\end{table}

Tables \ref{tab:vc_pbv_vc} and \ref{tab:vc_zbv_vc} show that the change in volume upon introduction of dislocations as quantified by finite deformation FDM is very small for both the cases. Moreover, we see that this volume change is not linear w.r.t the number of cores i.e., the change in volume in the presence of $6$ dislocations is not the same as $3$ times the change in volume when $2$ dislocations are present. The nonlinearity of the ECDD system and interaction among the dislocations  cause the deviation from linear response to give smaller values.

This study suggests that for an isotropic Saint-Venant-Kirchhoff material, the volume change calculated with nonlinear theory is very small and therefore the linear theory prediction of no volume change is a very good approximation.

\subsection{Non-uniqueness of inverse deformation in classical finite elasticity with dislocations}
\label{sec:reference_configurations}
Consider the case when a dislocation with Burgers vector $\bfb$ is assumed to be present in the body. The solution to the ECDD system \eqref{eq:ECDD}-\eqref{eq:ECDD_bc} on the current configuration $\mOmega$ gives the inverse elastic distortion field $\bfW$ in the domain. It can be shown that if the dislocation core is removed from the  body and a cut  is made that runs from the boundary of the core to the external boundary of $\mOmega$ to produce a simply connected body $\mOmega_s$, then there exists a deformation $\bfy$ of this $\mOmega_s$ such that 
\begin{equation}\label{non_unique}
	grad_s\bfy = \bfW,
\end{equation}
where $\bfW^{-1}=:\bfF^e$ and $grad_s$ denotes the gradient on $\mOmega_s$. The reference configuration is then obtained by mapping this `hollowed and cut'  configuration $\mOmega_s$ by the field $\bfy$. Moreover, the field $\bfy$ has two important properties: i) the jump in the value of $\bfy$ (denoted by $\llbracket\bfy\rrbracket$) along the cut surface is equal to the Burgers vector $\bfb$ of the  embedded dislocation in the original body $\mOmega$ and, ii) $\llbracket\bfy\rrbracket$ is independent of the cut surface chosen as well \cite{czhang_gdiscl_theory}. 

Section \ref{sec:reference_configurations_b100} shows that the above topological properties are preserved in the framework of (computational) FDM. More interestingly, we are easily able to do similar calculations for a body containing multiple dislocations when the current configuration cannot be rendered simply-connected by a single cut, as shown in Sec.~\ref{sec:mult_disloc_ref} -- in this case we show that the  jump in $\bfy$ is not constant along the boundary, corresponding to the cuts, of the simply-connected domain, in contrast to the single dislocation, single-cut case.

\subsubsection{Inverse deformation for a single dislocation}

\label{sec:reference_configurations_b100}
\begin{figure}
	\centering
	\begin{subfigure}[b]{0.25\textwidth}
		\centering
		{\includegraphics[width = 0.9\linewidth]{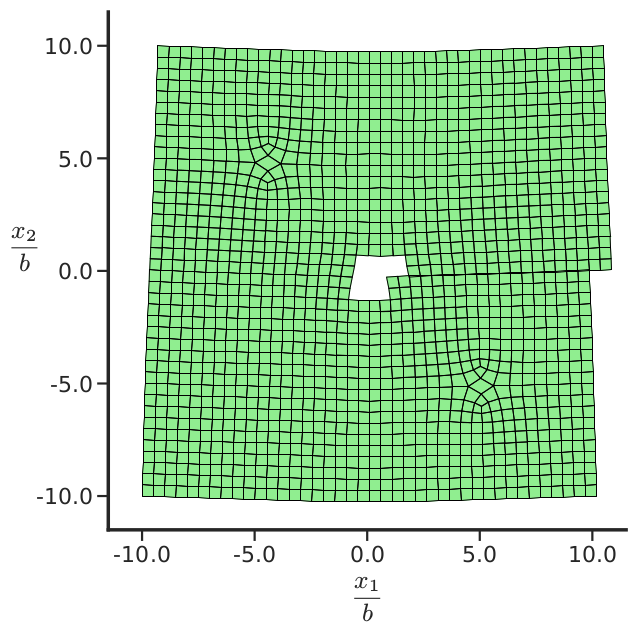}}
		\caption{}
		\label{fig:vc_b100_0angle}
	\end{subfigure}%
	\centering
	\begin{subfigure}[b]{0.25\textwidth}
		\centering
		{\includegraphics[width = 0.9\linewidth]{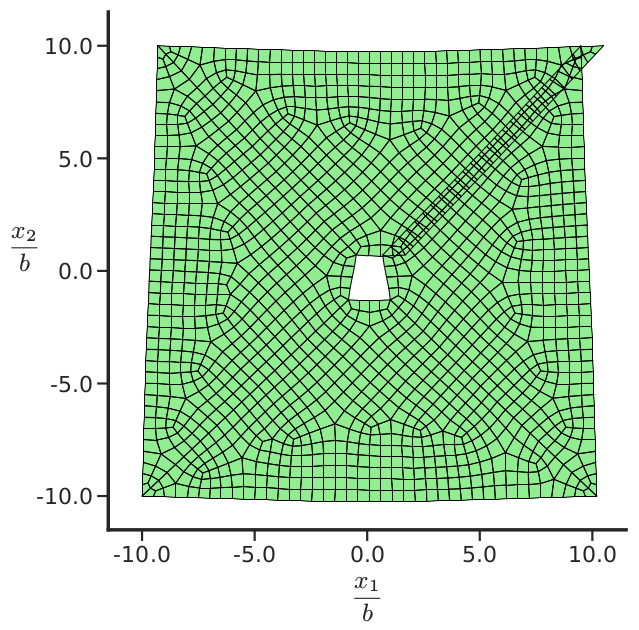}}
		\caption{}
		\label{fig:vc_b100_45angle}
	\end{subfigure}%
	\begin{subfigure}[b]{0.25\textwidth}
		\centering
		{\includegraphics[width = 0.9\linewidth]{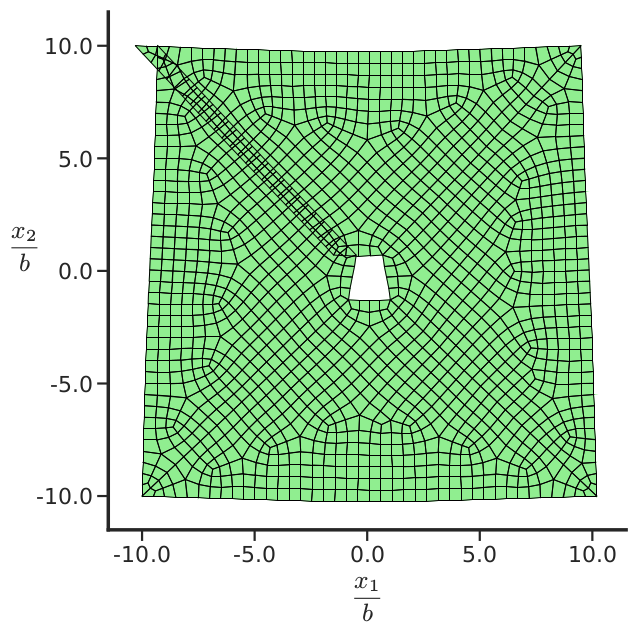}}
		\caption{}
		\label{fig:vc_b100_135angle}
	\end{subfigure}\\
	\centering
	\begin{subfigure}[b]{0.25\textwidth}
		\centering
		{\includegraphics[width = 0.9\linewidth]{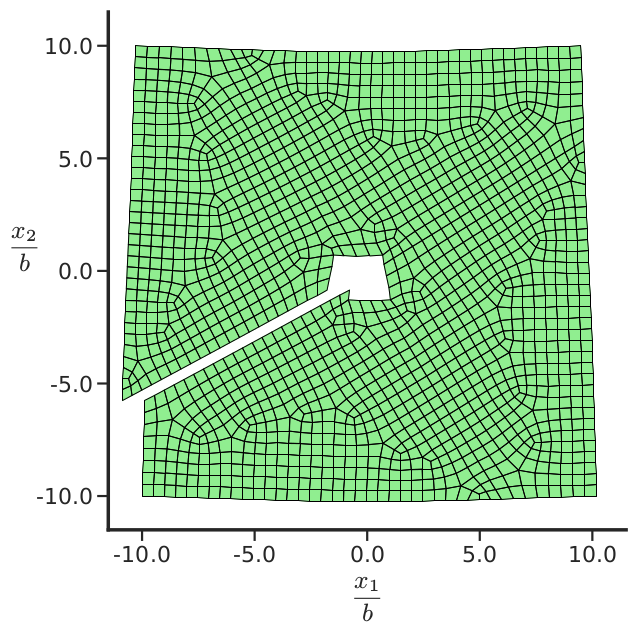}}
		\caption{}
		\label{fig:vc_b100_210angle}
	\end{subfigure}%
	\centering
	\begin{subfigure}[b]{0.25\textwidth}
		\centering
		{\includegraphics[width = 0.9\linewidth]{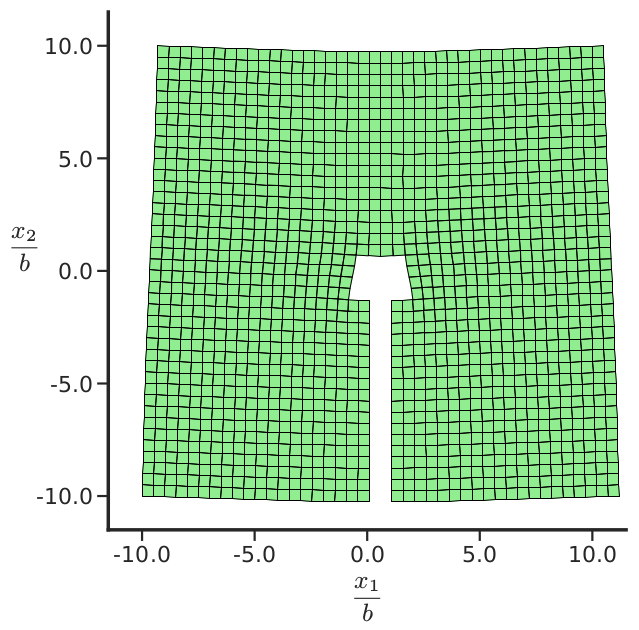}}
		\caption{}
		\label{fig:vc_b100_270angle}
	\end{subfigure}%
	\centering
	\begin{subfigure}[b]{0.25\textwidth}
		\centering
		{\includegraphics[width = 0.9\linewidth]{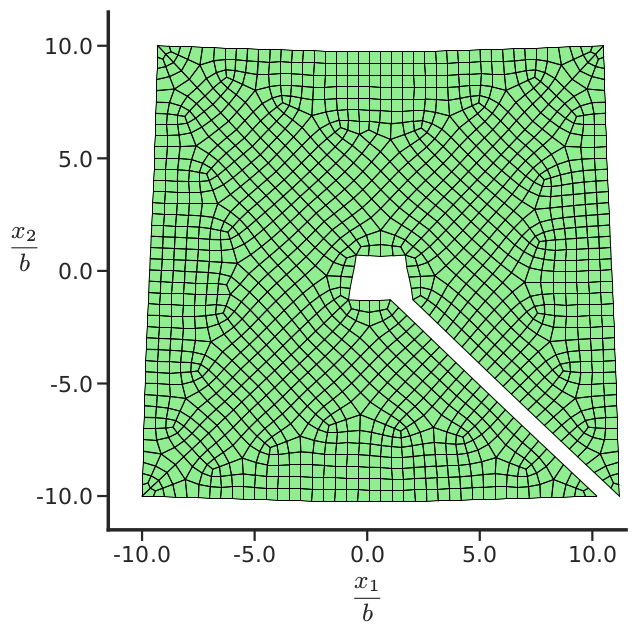}}
		\caption{}
		\label{fig:vc_b100_315angle}
	\end{subfigure}
	\caption{Reference configurations when the cut surface is chosen at varying angles from the $\bfe_1$ axis: a) $\mdeg{0}$ b) $\mdeg{45}$ c) $\mdeg{135}$ f) $\mdeg{210}$ i) $\mdeg{270}$ j) $\mdeg{315}$. The Burgers vector is identical for all the cases, while the overall reference configurations are non-trivially different.}
	\label{fig:vc_non_uniqueness_b100}
\end{figure}

We obtain non-unique reference configurations of a body $\mOmega$ with a single dislocation with Burgers vector $\bfb = b\bfe_1$.  The body $\mOmega$ is made simply connected by removing the core  and making a straight cut, at some angle from the $\bfe_1$ axis, that runs from the boundary of the core to the external boundary. The field $\bfy$ on this simply
connected configuration $\mOmega_s$ is calculated by using the Least-Squares FEM. Figure \ref{fig:vc_non_uniqueness_b100} shows the reference configurations for the cut surface chosen at varying angles w.r.t.~the $\bfe_1$ axis. 

The reference configurations self-penetrate when the cut makes an angle between $\mdeg{0}$ and $\mdeg{180}$ with the $\bfe_1$ axis. When the angle lies between $\mdeg{180}$ and $\mdeg{360}$, the reference configurations show detachment. Also, the reference configurations shown in Figures \ref{fig:vc_b100_45angle} and \ref{fig:vc_b100_135angle}, and \ref{fig:vc_b100_0angle} and \ref{fig:vc_b100_270angle} are entirely different, even  though the corresponding simply connected configurations for these cases are related by a rigid rotation of $-/+\mdeg{90}$ about the $\bfe_3$ axis. This is   a consequence of the constraint that the (vectorial) jump in $\bfy$ has to remain constant along the cut surface. Moreover, this jump is exactly equal to the Burgers vector $b\bfe_1$ of the dislocation in the original configuration $\mOmega$. The jump in the value of the field $\bfy$ is identical for all the cuts showing that the jump is independent of the cut surface, thus verifying the expected topological property.

\subsubsection{Inverse deformations for multiple dislocations}\label{sec:mult_disloc_ref}
\begin{figure}
	\centering
	\begin{subfigure}[b]{0.24\textwidth}
		\centering
		{\includegraphics[width = \linewidth]{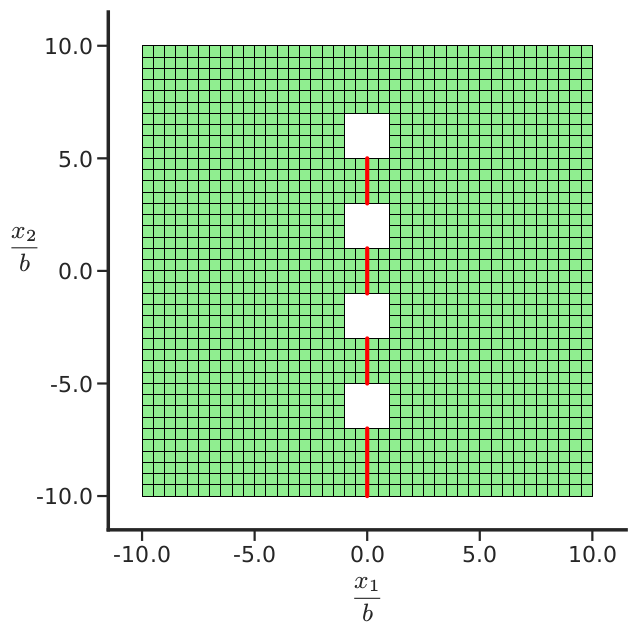}}
		\caption{}
		\label{fig:vc_nu_cutfig1}
	\end{subfigure}%
	\begin{subfigure}[b]{0.24\textwidth}
		\centering
		{\includegraphics[width = \linewidth]{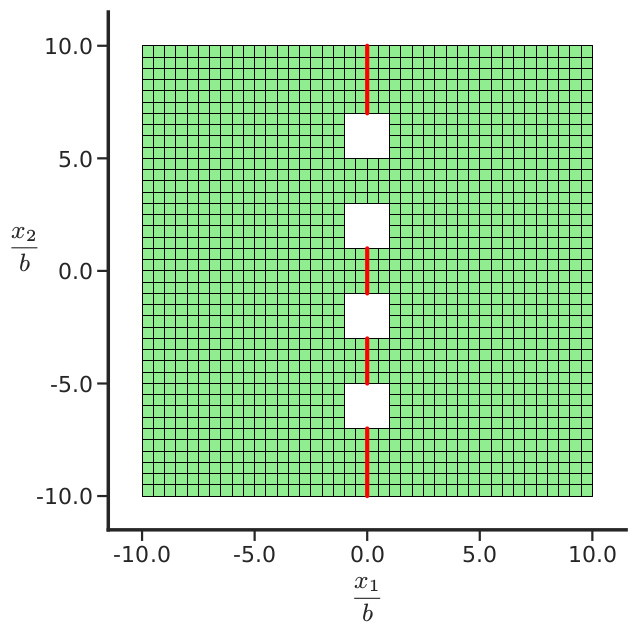}}				
		\caption{}
		\label{fig:vc_nu_cutfig2}
	\end{subfigure}%
	\begin{subfigure}[b]{0.24\textwidth}
		\centering
		{\includegraphics[width = \linewidth]{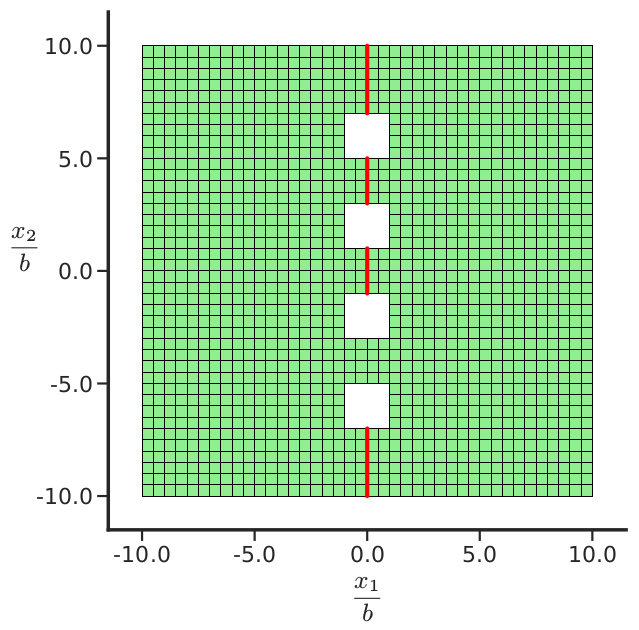}}
		\caption{}
		\label{fig:vc_nu_cutfig4}
	\end{subfigure}%
	\begin{subfigure}[b]{0.24\textwidth}
		\centering
		{\includegraphics[width = \linewidth]{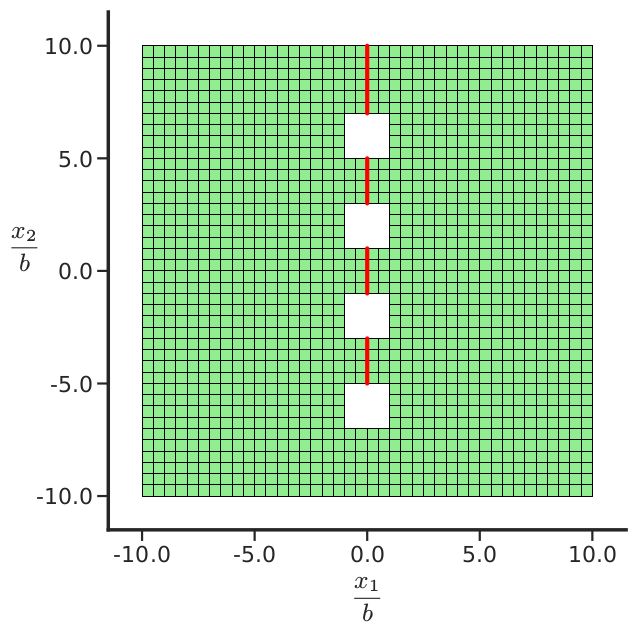}}
		\caption{}
		\label{fig:vc_nu_cutfig5}
	\end{subfigure}\\
	\begin{subfigure}[b]{0.24\textwidth}
		\centering
		{\includegraphics[width = \linewidth]{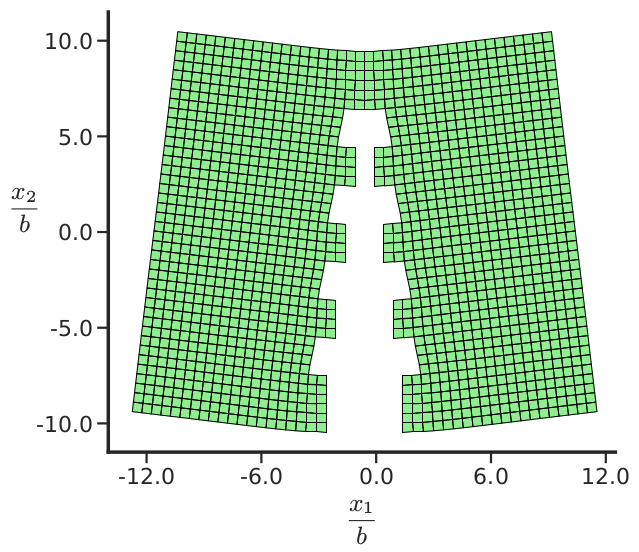}}
		\caption{}
		\label{fig:vc_nu_cutref1}
	\end{subfigure}%
	\begin{subfigure}[b]{0.24\textwidth}
		\centering
		{\includegraphics[width = \linewidth]{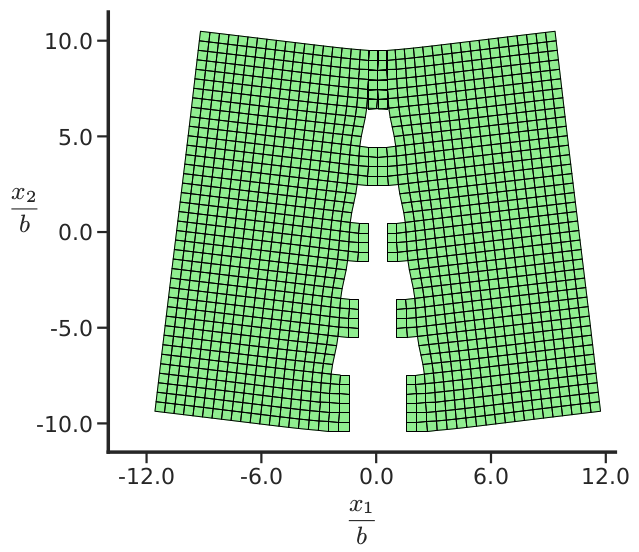}}
		\caption{}
		\label{fig:vc_nu_cutref2}
	\end{subfigure}%
	\begin{subfigure}[b]{0.24\textwidth}
		\centering
		{\includegraphics[width = \linewidth]{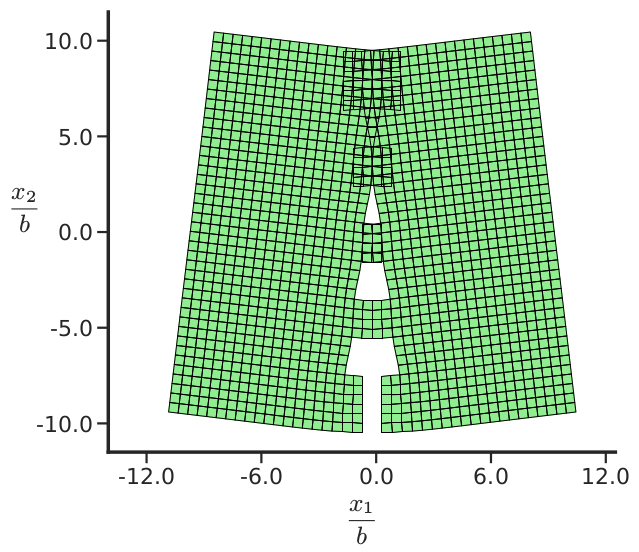}}
		\caption{}
		\label{fig:vc_nu_cutref4}
	\end{subfigure}%
	\begin{subfigure}[b]{0.24\textwidth}
		\centering
		{\includegraphics[width = \linewidth]{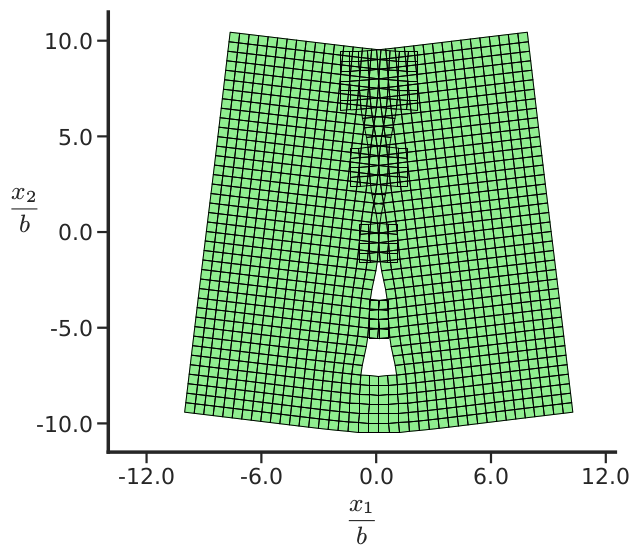}}
		\caption{}
		\label{fig:vc_nu_cutref5}
	\end{subfigure}%
	\caption{Non unique reference configurations (on the bottom) obtained for different simply connected configurations (on the top).}
	\label{fig:vc_nu_ref}
\end{figure}

This section explores the possibility of obtaining non-unique reference configurations of a body with multiple dislocations. Figure \ref{fig:vc_nu_ref} shows such a scenario when 4 dislocations are assumed to be present in the body and reference
 configurations are obtained by solving for $\bfy$ using the Least-Squares FEM for different simply connected configurations.

On the top of Figure \ref{fig:vc_nu_ref}, we show different simply connected configurations, each obtained by making multiple cuts (shown in red) in $\mOmega$. The corresponding reference configurations are shown on the bottom in Figure \ref{fig:vc_nu_ref}. As can be seen from these figures, the reference configuration is non-unique - for a given self-equilibrated $\bfW$ - as it depends on the cuts made in the body to make it simply connected. Moreover, even when the cut-configurations differ by rigid rotations (for example \ref{fig:vc_nu_cutfig1} and \ref{fig:vc_nu_cutfig5}, and  \ref{fig:vc_nu_cutfig2} and \ref{fig:vc_nu_cutfig4}), the corresponding reference configurations (\ref{fig:vc_nu_cutref1} and  \ref{fig:vc_nu_cutref5}, and \ref{fig:vc_nu_cutref2} and  \ref{fig:vc_nu_cutref4},) are entirely different.  Another important thing to note here is that unlike the case of the single dislocation, the jump in $\bfy$ is not constant along the cut in the presence of multiple dislocations,  but is constant along the cut between any two cores,  in accord with the analytical values they should take in these examples of  configurations with multiple cuts.

The above examples highlight an important feature of the ECDD formulation -- if the dislocation stress problem for specified dislocation density is formulated classically, i.e., involving only an inverse deformation field $\bfy$ on the `deformed' configuration with the dislocation considered given, it is clear that there is massive non-uniqueness of solutions $\bfy$ for the same stress field corresponding to $\bfW$\footnote{In the absence of dislocations this corresponds to the classical inverse problem of nonlinear elasticity where the deformed, stressed configuration is assumed given with specified traction boundary conditions and enough kinematic constraints to prevent superposed rigid (inverse) deformation, and the question is to determine the stress-free elastic reference by solving for the inverse deformation.}; ECDD suffers from no such ambiguity and lends itself to robust computation.

\section{Conclusion}
\label{sec:conclusion}
A  partial differential equation based model of finite deformation plasticity coupled to dislocation mechanics has been presented, with no restriction on material and geometric nonlinearities. The model fundamentally accounts for the stress field of arbitrary dislocation distributions and melds elastic dislocation theory at finite deformation with $J_2$ plasticity in a practical manner suitable for application. This paper along with \cite{arora_acharya_ijss, arora_xia_acharya_cmame} form a trio that solves some key problems of current and classical significance in the fields of plasticity and dislocation mechanics, utilizing a finite element based computational framework developed in \cite{arora_xia_acharya_cmame}. As well, Lauteri and Luckhaus \cite{lauteri} state ``It is worthwhile to compare our result with the differential geometric description of dislocation
structures, introduced by Kondo, Kr\"{o}ner and Bilby at al. (see \cite{kazuo1964analytical,kroner1959allgemeine,bilby1955continuous}) and also with $\mathrm{\Gamma}$-limit results in the context of linear elasticity where implicitly or explicitly a volume density of
dislocations is assumed (see \cite{Garroni_etal}). It remains to be investigated if these models remain valid as averaged limits if on an intermediate scale there exists a Cosserat structure of micrograins." - up to the intended meaning of ``averaged limits"  above, we believe that we have answered this latter question in the affirmative, albeit with a far-reaching extension of the pioneering differential geometric works, going beyond simply kinematic considerations at finite deformations. We believe that our work provides a practical pathway for exploring many problems at the intersection of finite deformation plasticity and dislocation mechanics at realistic length and time scales. Further work will involve probing problems of shear localization in $2$ and $3$ dimensional bodies of rate-dependent and rate-independent materials, among others. Finally, improving the constitutive assumptions inherent in $\bfL^p$ based on averaging of dislocation dynamics at the individual defect scale remains a fundamental pursuit.

\subsubsection*{Acknowledgments}
This research was funded by the Army Research Office grant number ARO-W911NF-15-1-0239. This work also used the Extreme Science and Engineering Discovery Environment (XSEDE) \cite{xsede1}, through generous XRAC grants of supercomputing resources, which is supported by National Science Foundation grant number ACI-1548562.  We gratefully acknowledge the Pittsburgh Supercomputing Center (PSC), and Prof.~Jorge Vin\~als and the Minnesota Supercomputing Institute (URL: http://www.msi.umn.edu) for providing computing resources that contributed to the research results reported within this paper. It is a great pleasure to acknowledge insightful discussions with Janusz Ginster and Reza Pakzad.

\begin{appendices}
	
	\section{(Mesoscale) Field Dislocation Mechanics (M)FDM}\label{app:MFDM}
	Significant portions of this section are common with \cite{arora_acharya_ijss,arora_xia_acharya_cmame}, papers developed concurrently with this work.  We include this material here for the sake of being self-contained and since the theory being discussed is quite recent.
	
	Field Dislocation Mechanics (FDM) was developed in \cite{acharya2001model, acharya2003driving, acharya2004constitutive, acharya2007jump} building on the pioneering works of Kr\"oner \cite{kroner1981continuum}, Willis \cite{willis1967second},  Mura \cite{mura1963continuous}, and Fox \cite{fox1966continuum}.  The theory utilizes a tensorial description of dislocation density \cite{nye1953, bilby1955continuous}, which is related to special gradients of the (inverse) elastic distortion field.  The governing equations of FDM at finite deformation are presented below:
	\begin{subequations}
		\begin{align}
		&~\mathring{\bfalpha}\equiv tr(\bfL)\,\bfalpha+\dot{\bfalpha}-\bfalpha\bfL^T = -curl\left(\bfalpha\times \bfV \right)
		\label{eq:fdm_alpha}\\[1mm]
		&~\bfW = \bfchi+grad\!\bff ; \quad\bfF^{e} := \bfW^{-1} \nonumber\\[1.25mm]
		&\left.\begin{aligned}
		& curl\bfW = curl{\bfchi} = -\bfalpha\\
		&div{\bfchi} = \bf0
		\label{eq:fdm_chi}
		\end{aligned}~~~~\qquad\qquad\right\}\\[1.25mm]
		&~div\left(grad\!\dot{\bff}\right) = div\left(\bfalpha\times \bfV - \dot{\bfchi}-\bfchi\bfL\right)\label{eq:fdm_fevol}\\
		&~div\,[\bfT(\bfW)]  = \begin{cases}
		\bf 0 & \text{quasistatic} \\
		\rho\dot{\bfv} & \text{dynamic}. \\
		\end{cases}
		\label{eq:fdm_eqb}
		\end{align}
		\label{eq:fdm}
	\end{subequations}
	
	Here, $\bfF^e$ is the elastic distortion tensor, ${\bfchi}$ is the incompatible part of $\bfW$, $\bff$ is the plastic position vector \cite{acharya2006size}, $grad\!\bff$ represents the compatible part of $\bfW$, $\bfalpha$ is the dislocation density tensor, $\bfv$ represents the material velocity field, $\bfL=grad\bfv$ is the velocity gradient, and $\bfT$ is the (symmetric) Cauchy stress tensor. The dislocation velocity, $\bfV$, at any point is the instantaneous velocity of the dislocation complex at that point relative to the material; at the microscopic scale, the dislocation complex at most points consists of single segment with well-defined line direction and Burgers vector. At the same scale, the mathematical model assigns a single velocity to a dislocation junction, allowing for a systematic definition of a thermodynamic driving force on a dislocation complex that consistently reduces to well-accepted notions when the complex is a single segment, and which does not preclude dissociation of a junction on evolution.
	
	The statement of dislocation density evolution \eqref{eq:fdm_alpha} is derived from the fact that the rate of change of Burgers vector content of any arbitrary area patch has to be equal to the flux of dislocation lines into the area patch carrying with them their corresponding Burgers vectors. Equation \eqref{eq:fdm_chi} is the fundamental statement of elastic incompatibility and relates the dislocation density field to the incompatible part of the inverse elastic distortion field $\bfW$. It can be derived by considering the closure failure of the image of any closed loop in the current configuration on mapping by $\bfW$. Equation \eqref{eq:fdm_fevol} gives the evolution equation for the compatible part of the inverse elastic distortion field. It can be shown to be related to the permanent deformation that arises due to dislocation motion \cite{acharya2004constitutive}. The field $grad\!\bff$ can also be viewed as the gradient of the inverse deformation for purely elastic deformations. Equation \eqref{eq:fdm_eqb} is the balance of linear momentum (in the absence of body forces). Balance of mass is assumed to hold in standard form, and balance of angular momentum is satisfied by adopting a symmetric stress tensor.

The equations of FDM  \eqref{eq:fdm} can also be succinctly reformulated as
	\begin{equation}\label{eqn:HJ}
	\begin{split}
	\dot{\bfW} & = -\bfW \bfL - (curl \bfW) \times \bfV\\
	~div\,[\bfT(\bfW)]  &= \begin{cases}
	\bf 0 & \text{quasistatic} \\
	\rho\dot{\bfv} & \text{dynamic}, \\
	\end{cases}
	\end{split}
	\end{equation} 
	but since the \emph{system} of Hamilton-Jacobi equations in \eqref{eqn:HJ}$_1$ is somewhat daunting, we work with \eqref{eq:fdm} instead, using a Stokes-Helmholtz decomposition of the field $\bfW$ and the evolution equation for $\bfalpha$ in the form of a conservation law. 	Equation \eqref{eq:fdm} is augmented with constitutive equations for the dislocation velocity $\bfV$ and the stress $\bfT$ in terms of $\bfW$ and $\bfalpha$ \cite{acharya2004constitutive, zhang2015single} to obtain a closed system.

	FDM  is a model for the representation of dislocation mechanics at a scale where individual dislocations are resolved. In order to develop a model of plasticity that is applicable to mesoscopic scales, a space-time averaging technique utilized in the study of multiphase flows (see e.g. \cite{babic1997average}) is applied to microscopic FDM  \cite{acharya2006size, acharya2011microcanonical}. For any microscopic field $m$ given as a function of space and time, the weighted, space-time running average field $\overline{m}$ is given as
	\begin{align*}
	\overline{m}(\bfx, t) := \dfrac{1}{\int_{B(\bfx)} \int_{I(t)} w(\bfx-\bfx', t-t')d\bfx' dt'  } {\int_{\mLambda} \int_{\mOmega} w(\bfx-\bfx', t-t') \,m(\bfx',t') d\bfx' dt'},
	\end{align*} where $\mOmega$ is the body and $\mLambda$ is a sufficiently large interval of time. $B(\bfx)$ is a bounded region within the body around the point $\bfx$ with linear dimension of the spatial resolution of the model to be developed, and $I(t)$ is a  bounded interval contained in  $\mLambda$. The weighting function $w$ is non-dimensional and assumed to be  smooth in the variables $\bfx, \bfx', t, t'$. For fixed $\bfx$ and $t$, $w$ is only non-zero in $B(\bfx) \times I(t)$ when viewed as a function of $\bfx'$ and $t'$. 
	
Mesoscale Field Dislocation Mechanics (MFDM) is obtained by applying the above space-time averaging filter to the FDM equations \eqref{eq:fdm} with the assumption that all averages of products are equal to the product of averages except for $\overline{\bfalpha \times \bfV}$. The governing equations of MFDM \cite{acharya2006size, acharya2011microcanonical, arora_acharya_ijss} at finite deformation (without body forces) are written as 
	\begin{subequations}
		\begin{align}
		&~\mathring{\overline\bfalpha}\equiv tr(\overline\bfL)\,\overline\bfalpha+\dot{\overline\bfalpha}-\overline\bfalpha\overline\bfL^T = -curl\left(\overline\bfalpha\times \overline\bfV + \bfL^p\right)\label{eq:mfdm_a_app}\\[1mm]
		&~\overline\bfW = \overline\bfchi+grad\overline\bff \nonumber\\[1.25mm]
		&\left.\begin{aligned}
		&curl{\overline{\bfW}} = curl{\overline\bfchi} = -\overline\bfalpha\\
		&div{\overline\bfchi} = \bf0
		\label{eq:mfdm_chi_app} 
		\end{aligned}\right\}\\[1.25mm]
		&~div\left(grad\dot{\overline\bff}\right) = div\left(\overline\bfalpha\times \overline\bfV + \bfL^p - \dot{\overline\bfchi}-\overline\bfchi\overline\bfL\right)\label{eq:mfdm_fevol_app}\\
		&~div\,[\overline\bfT(\overline\bfW)]  = \begin{cases}
		\bf 0 & \text{quasistatic} \\
		\overline\rho\,\dot{\overline\bfv} & \text{dynamic}, \\
		\end{cases}
		\label{eq:mfdm_f_app}
		\end{align}
		\label{eq:mfdm_app}
	\end{subequations}
	where $\bfL^p$ is defined as
	\begin{align}\label{eqn:Lp}
	\bfL^p(\bfx,t) := \overline{(\bfalpha - \overline{\bfalpha}(\bfx,t)) \times \bfV}(\bfx,t) = \overline{\bfalpha \times \bfV}(\bfx,t) - \overline{\bfalpha}(\bfx,t) \times \overline{\bfV}(\bfx,t).
	\end{align}
	
	The barred quantities in \eqref{eq:mfdm_app} are simply the weighted, space-time, running averages of their corresponding microscopic fields (defined in \eqref{eq:fdm}). The field $\overline{\bfalpha}$ is  the Excess Dislocation Density (ED). The microscopic density of Statistical Dislocations (SD)  at any point is defined as the difference between the microscopic dislocation density $\bfalpha$ and its  averaged field $\overline{\bfalpha}$:
	\begin{equation*}
	\bfbeta(\bfx,\bfx',t,t') = \bfalpha (\bfx',t') - \overline{\bfalpha}(\bfx,t),
	\end{equation*}
	which implies
	\begin{align}
	\label{eqn:tot_gnd_ssd}
	\begin{split}
	\rho_t &= \sqrt{\rho_g^2 + \rho_s^2}\\
	\rho_t(\bfx, t) := \sqrt{ \overline{ \left( \dfrac{|\bfalpha  |}{b}\right)^2} (\bfx,t)} \ \ ; \  \rho_g(\bfx, t) &:=    \dfrac{|\overline{\bfalpha}(\bfx,t)|}{b} \ \  ; \ \ \rho_s(\bfx, t) := \sqrt{ \overline{ \left( \frac{ |\bfbeta|}{b} \right)^2 }(\bfx, t) },
	\end{split}
	\end{align}
	with $b$ the magnitude of the Burgers vector of a dislocation in the material, $\rho_t$ the \emph{total dislocation density}, $\rho_g$ the magnitude of ED (commonly referred to as the geometrically necessary dislocation density), and $\rho_s$ is, up to a scaling constant, the root-mean-squared SD. We refer to $\rho_s$ as the scalar statistical dislocation density (\emph{ssd}).   It is important to note that spatially unresolved dislocation loops below the scale of resolution of the averaged model do not contribute to the ED ($\overline\bfalpha$) on space time averaging of the microscopic dislocation density, due to sign cancellation. Thus, the magnitude of the ED is an inadequate approximation of the total dislocation density. Similarly, a consideration of `symmetric' expansion of unresolved dislocation loops shows that the plastic strain rate produced by SD, $\bfL^p$ \eqref{eqn:Lp}, is not accounted for in $\overline{\bfalpha} \times \overline{\bfV}$, and thus the latter is not a good approximation of the total averaged plastic strain rate $\overline{\bfalpha \times \bfV}$.
	
	In MFDM, closure assumptions are made for the field $\bfL^p$ and the evolution of $\rho_s$, as is standard in most, if not all, averaged versions of nonlinear microscopic models, whether of real-space or kinetic theory type. As such, these closure assumptions can be improved as necessary (and increasingly larger systems of such a hierarchy of nonlinear pde can be formally written down for MFDM). In this paper, we adopt simple and familiar closure statements from (almost) classical crystal   and $J_2$ plasticity theories and present the finite element formulation for the model.  Following the works of Kocks, Mecking, and co-workers \cite{mecking1981kinetics, estrin1984unified} we describe the evolution of $\rho_s$ through a statement, instead, of evolution of material strength $g$ described by \eqref{eq:softening}; $\bfL^p$ is defined by \eqref{eq:Lp_crystal} (or \eqref{eq:Lp_j2}) following standard assumptions of crystal/$J_2$ plasticity theory and thermodynamics. A significant part of the tensorial structure of \eqref{eq:Lp_crystal}  and \eqref{eq:Lp_j2}  can be justified by  elementary averaging considerations of dislocation motion on a family of slip planes under the action of their Peach-K\"{o}ehler driving force \cite{acharya2012elementary}. 
	
	\emph{Below, and in system \eqref{eq:mfdm} as well as the rest of the paper, we drop the overhead bars for convenience in referring to averaged quantities}. 
	
	As shown in \cite{acharya2015dislocation}, \eqref{eq:mfdm_a} and \eqref{eq:mfdm_chi} imply
	\begin{equation}
	\dot{\bfW} + \bfW\bfL = \bfalpha \times \bfV + \bfL^p
	\label{eq:vel_grad}
	\end{equation} up to the gradient of a vector field, which is re-written as
	\begin{equation*}
	\bfL = \dot{\bfF^{e}} {\bfF^{e-1}} + \bfF^{e}(\bfalpha\times\bfV+\bfL^p),
	\end{equation*}
	where $\bfF^e := \bfW^{-1}$. This can be interpreted as the decomposition of the velocity gradient into an elastic part, given by $\dot{\bfF^{e}} {\bfF^{e-1}}$, and a plastic part given by  $\bfF^{e}(\bfalpha\times\bfV+\bfL^p)$. The plastic part is defined by the motion of dislocations, both resolved and unresolved, on the current configuration and \emph{no notion of any pre-assigned reference configuration is needed}. Of  significance is also the fact that \emph{MFDM involves no notion of a plastic distortion tensor and yet produces (large) permanent deformation}.

\section{A brief review of prior work}

\label{sec:literature_review_jmps}

Here, we briefly review some of the vast literature on strain gradient plasticity theories. An exhaustive review of the subject is beyond the scope of this paper.

Given the significance of an accurate prediction of size-effect for industries involved in the manufacturing of small scale electro-mechanical systems, and the need for a continuum scale model that enables a predictive study of dislocation mediated plastic deformation, several extensions of conventional continuum models, differing in origin and mathematical structure, have been proposed that incorporate one or many in-built length scales \cite{aifantis1984microstructural, 
	acharya1996non,
	acharya2000lattice,  
	acharya2004boundary, 
	tang2004effects, tang2005directional, acharya2000grain, arsenlis2004evolution,evers2004non, geers2006second,kuroda2008finite,rudraraju2014three,fleck2001reformulation,niordson2003non, fleck1994strain, gurtin2002gradient, yefimov2004comparison, gurtin2008finite,gurtin2014gradient, ma2006dislocation, niordson2004size,niordson2005instabilities,lynggaard2019finite, po2019continuum}.

One class of continuum models  produce size-effects  by modifying  the hardening law to take into account the hardening due to  the presence of incompatibility in elastic distortion \cite{acharya2000lattice, bassani2001plastic, acharya2004boundary, acharya1996non,tang2004effects, tang2005directional}. These elastic incompatibilities correspond to `Geometrically Necessary Dislocations' (GNDs) \cite{ashby1970deformation}. The advantage of these `lower order gradient' theories is that the classical structure of the underlying incremental boundary-value-problem (bvp) with its natural boundary conditions  remains unchanged, retaining all associated results on the uniqueness of  solutions  to the bvp of incremental equilibrium for the rate-independent (and rate-dependent) material \cite{hill1958general}, as observed in \cite{acharya1995thermodynamic}. This is in contrast with many strain gradient theories which involve higher order stresses or additional boundary conditions.

The second class of models  \cite{fleck1994strain, fleck2001reformulation, niordson2003non, gudmundson2004unified} incorporate hardening effects due to plastic strain gradients by defining measures of the effective plastic strain which depend on invariants of gradients of plastic strain. These extensions of the conventional theories involve higher order stresses and require additional boundary conditions on appropriate variables.

The third class of models are based on the framework of Gurtin \cite{gurtin2000plasticity, gurtin2002gradient, gurtin2008finite, gurtin2014gradient}. In these models, the classical force balance is supplemented with a statement of microscopic force balance, one for each slip system, which also form the nonlocal flow rules of the theory when combined with thermodynamically consistent constitutive equations.
 
The fourth class of models \cite{arsenlis2004evolution, geers2006second, yefimov2004comparison, kuroda2008finite, ma2006dislocation, po2019continuum} augment conventional elasto-plastic theories with additional equations that aim to model  dislocation motion and evolution in the domain, resulting in plastic flow. This is in contrast with the other three class of models wherein plastic flow results through constitutive assumptions without explicitly characterizing dislocation motion and evolution. All these models  \cite{arsenlis2004evolution, geers2006second, kuroda2008finite, gurtin2008finite, ma2006dislocation, po2019continuum} involve the prescription of boundary conditions on scalar dislocation densities at grain boundaries in  multi/poly-crystalline aggregates; this is an ambiguous, but necessary requirement,  as slip system  dislocation densities in different grains have different physical meanings  due to the change in lattice orientation between the grains.

We now review some of the continuum models of strain gradient plasticity theory mentioned above.

Arsenlis et al.~\cite{arsenlis2004evolution} developed a micromechanical model of single crystal plasticity that incorporates a material length scale dependence in its framework. The authors developed a set of pdes for the evolution of slip system dislocation densities, one for each system. This requires constitutive assumptions for dislocation velocity, cut-off radii for annihilation, and segment-length interaction matrices for each component and slip system. Evolution equations including SSDs are also proposed.  The framework is applied to predict size effects in simple shear for thin films using constrained and quasi-free boundary conditions for an idealized crystalline geometry.

Geers et al.~\cite{geers2006second} present a strain-gradient crystal plasticity theory that has additional equations which govern the evolution of GNDs and SSDs in the domain. The back stresses due to the presence of dislocations are accounted for by using analytical integral expressions for stress fields in an infinite medium.  Aspects of modeling backstresses by such assumptions are  assessed in \cite{acharya2008counterpoint}.

Kuroda et al.~\cite{kuroda2008finite} present a finite deformation higher order strain gradient crystal plasticity formulation without introducing higher order stresses.  The conventional theory is supplemented by a length scale dependent back-stress which in-turn depends on the gradients of scalar edge and screw GND densities whose evolution is governed by their respective pdes. Po et al.~\cite{po2019continuum} present a finite deformation continuum dislocation-based plasticity formulation to model dislocation microstructure that is observed in the wedge micro-indentation experiment of  \cite{kysar2010experimental}.

A finite strain generalization of the  strain gradient plasticity formulation of Fleck et al.~\cite{fleck2001reformulation} is proposed and implemented in a finite element framework by Niordson et al.~\cite{niordson2004size, niordson2005instabilities}. Numerically, the formulation requires the equivalent plastic strain rate to be one of the nodal variables which requires  additional boundary conditions to be applied at the boundary separating parts of the body loading elastically and plastically,  along with boundary conditions in higher order  stresses. The model is used to study necking under plane strain tension and compression \cite{niordson2005instabilities}, and size effects in plane strain necking of thin sheets \cite{niordson2004size}. %

\cite{yefimov2004bending} propose a nonlocal version of crystal plasticity motivated by elementary statistical-mechanics  descriptions of collective behavior of dislocations  \cite{groma1999investigation,groma1997link} wherein $2$ coupled pdes govern the evolution of the scalar dislocation density and GNDs. The model is applied to study the shearing of a model composite material in single slip with constrained boundary conditions \cite{yefimov2004bending}  and bending of a single-crystal strip in plane strain \cite{yefimov2004bending}. The model and computational implementation is limited to single slip deformations in $2$-$d$ problems.

Gurtin \cite{gurtin2008finite} develops a finite deformation gradient theory of crystal plasticity involving configurational microforce balances for each slipsystem \cite{gurtin2002gradient}. A central ingredient of the theory involves inclusion of an additional term based on GND density in the free energy that characterizes an 
energetic hardening mechanism associated with the accumulation of GNDs.

Rudraraju  et al.~\cite{rudraraju2014three}  present stress fields of dislocations using Toupin's gradient elasticity theory \cite{toupin1962elastic} at finite strains wherein the defects  are represented by force dipole distributions. The theory involves $4^{th}$ order derivatives of displacement and therefore requires the use of isogeometric analysis \cite{hughes2005isogeometric, cottrell2009isogeometric} for its implementation. The representation of a complex network of line defects can be onerous by dipole forces. Moreover, it is not very straightforward to postulate evolution of defects  coupled to the underlying stress field, given its representation in the form of dipole forces. 

A hybrid method to model dislocation core widths and their mutual interactions under quasi-static and dynamic loadings was developed in \cite{denoual2004dynamic, denoual2007modeling}. The approach involves a combination of Pierls-Nabarro model and Galerkin methods.  However, finite deformation frameworks for these approaches have not been demonstrated.

\end{appendices}

\clearpage
\newpage

\bibliographystyle{alpha}
\bibliography{gen_bib}

\end{document}